\documentclass{article}
\usepackage{geometry}
\usepackage{amsmath,amssymb,amsfonts,amsthm,amsopn,dsfont, mathrsfs}
\usepackage{bm}
\usepackage{graphicx}
\usepackage{float}

\usepackage{standalone}
\usepackage{tikz}
\usepackage{pgfplots}
\usepackage{pgfplotstable}
\pgfplotsset{compat=newest}
\usepackage{subcaption}
\usepackage{colortbl}
\usepackage{booktabs}

\usepackage[T1]{fontenc}

\usepackage{xcolor}

\usepackage[ruled,vlined,linesnumbered]{algorithm2e}
\usepackage{multirow}

\newcommand{\hsp}{\mathcal{H}^{\Gamma_k}}
\newcommand{\extg}{\mathcal{E}}
\newcommand{\exts}{\Xi}
\newcommand{\trg}{\gamma}
\newcommand{\Vhatz}{{{}\hat{V}_0^k}}
\newcommand{\Vhatc}{{{}\hat{V}_c^k}}
\newcommand{\Vhat}{{{}\hat{V}^k}}
\newcommand{\VD}{V^k}
\newcommand{\Vtilde}{\widetilde{V}^k}

\newcommand{\hpsid}{\hat{\psi}^D}
\newcommand{\hpsis}{\hat{\psi}^\Sigma}

\newcommand{\D}{\mathcal{D}^k}
\newcommand{\partD}{\partial \mathcal{D}}
\newcommand{\Deltaik}{\Delta \mathcal{I}_k}

\begin{document}

\title{An optimization based 3D-1D coupling strategy for tissue perfusion and chemical transport during tumor-induced angiogenesis}
\author{Stefano Berrone$^{1,a}$,
Chiara Giverso$^{2,b}$,
Denise Grappein$^{3,a}$,\\
Luigi Preziosi$^{4,b}$,
Stefano Scial\`o$^{5,a}$\smallskip\\{\footnotesize Dipartimento di Scienze Matematiche, Politecnico di Torino,
	Corso Duca degli Abruzzi 24,}\\{\footnotesize 10129 Torino, Italy.}\\{\footnotesize $^1$stefano.berrone@polito.it, $^2$chiara.giverso@polito.it, $^3$denise.grappein@polito.it,}\\{\footnotesize $^4$luigi.preziosi@polito.it, $^5$stefano.scialo@polito.it }\medskip\\{\footnotesize ${}^a$Member of INdAM research group GNCS}\\{\footnotesize ${}^b$Member of INdAM research group GNFM}}
\date{}

\maketitle

\begin{abstract}
A new mathematical model and numerical approach are proposed for the simulation of fluid and chemical exchanges
between a growing capillary network and the surrounding tissue, in the context of tumor-induced angiogenesis.
Thanks to proper modeling assumptions the capillaries are reduced to their
centerline: a well posed mathematical model is hence worked out, based on the
coupling between a three-dimensional and a one-dimensional equation (3D-1D
coupled problem). Also the application of a PDE-constrained optimization formulation is here proposed for the first time for angiogenesis simulations. Under this approach no mesh conformity is required, thus making the method particularly suitable for this kind of application, since no remeshing is required as the capillary network grows. In order to handle both the evolution of the quantities of interest and the changes in the geometry, a discrete-hybrid strategy is adopted, combining a continuous modeling of the tissue and of the chemicals with a discrete tip-tracking model to account for the vascular network growth. The tip-tracking strategy, together with some proper rules for branching and anastomosis, is able to provide a realistic representation of the capillary network.
\end{abstract}
\subsection*{Keywords} 3D-1D coupling, domain-decomposition, non conforming mesh, optimization methods for PDE problems, mathematical model of angiogenesis, fluid and chemical transport in evolving networks\medskip\\
\textbf{MSC[2010]} 65N30, 65N50, 68U20, 86-08, 35Q92, 92B05, 92C17
	
\section{Introduction}

Angiogenesis is the crucial process that leads to the formation of new blood vessels from an existing vasculature, with the aim of providing the correct amount of nutrients and oxygen to the tissue and warranting metabolic waste removal from it  \cite{HillenFbio}.
This process occurs in many different conditions, either physiologically (e.g., embryogenesis, wound-healing, and female cycle) \cite{Graham, Arnold_wound} or pathologically (e.g., rheumatoid and inflammatory disease, duodenal ulcers, abnormal vascolarization in the eye, and the initiation and progression of cancers) \cite{Walsh, Gimbrone, Folkman_1985, Folkman_1992, Folkman_1995}. 
In all cases, angiogenesis entails a well-organized sequence of events, comprising the rearrangement, migration, and proliferation of endothelial cells (ECs) forming the capillary wall. These processes should coordinate with the establishment of blood flow inside the new capillaries in order to reach a properly functional vessel network \cite{Folkman_1987}.
The whole process is orchestrated by biochemical stimuli, released both by neighbouring cells and by the ECs themselves, and by the mechanical interaction between the ECs and the surrounding environment (refer to \cite{Risau, Carmeliet, Hudlicka_1991, Jain_2003} for an extensive review of the key chemical and mechanical cues in angiogenesis).

In this paper, we specifically focus on tumour-induced angiogenesis, occurring when an avascular tumour reaches a critical diameter of approximately 1-2 mm, above which nutrients and oxygen diffusion from the existing vasculature are no longer sufficient to sustain cancer progression and cells inside the tumour experience hypoxia \cite{Folkman_1987, Paweletz}. The hypoxic condition triggers the secretion by tumor cells of a number of chemicals, collectively called \emph{tumour angiogenic factors} (TAFs) \cite{HillenFbio, Folkman_1987}. 
These substances diffuse through the nearby tissue and, when they reach the vasculature, they activate ECs by binding to the transmembrane cell receptors and activating specific molecular pathways. This initiates the first step of angiogenesis, in which ECs change shape and adhesion properties, leading to a weakening of the junctions within the endothelial layer and a subsequent increase in the permeability of the blood vessel \cite{Nagy2008}. In the next phase, ECSs produce proteolytic enzymes, which degrade the basal lamina of the vessel and the surrounding extracellular matrix \cite{Kabelic}, enabling the formation of the first protrusion (\emph{sprout}) and the subsequent cell migration towards the source of the TAFs (\emph{chemotaxis}) \cite{Ausprunk, Eilken2010}. 
At this stage, it is possible to distinguish between \emph{tip} cells, i.e., the specialized cells at the extremity of newly formed capillaries that guide vessel outgrowth through chemotactic motion, and \emph{stalk} cells, which are highly proliferative cells that follow the tip cell motion, establishing tight junctions to ensure the stability of the new sprout and the formation of the nascent capillary \cite{Siekmann}. 
During the development of the network, tip-sprouts may undergo \emph{branching} or they can merge when two capillaries encounter each other (\emph{anastomosis}) \cite{Ausprunk, Bentley}. 
As the sprout approaches the tumour, the branches noticeably increase in number \cite{Muthukkaruppan} and the whole process culminates with the penetration of the new capillaries inside the tumour. Once the tumour becomes vascularized, the cancer cells gain access to an almost unlimited supply of nutrients and oxygen and they can eventually enter inside the vasculature and form metastasis also in distant sites \cite{Carmeliet}.  

Tumor-induced angiogenesis has been extensively studied through \emph{in vitro} and \emph{in vivo} biological experiments \cite{Madri, Miura, Staton, Ribatti}, that pointed out the main biomechanical factors and pathways involved in both physiological and pathological vascular progression. However, many aspects are still under investigation and a comprehensive biological set-up able to study the process as a whole, considering the different spatio-temporal scales and all the components involved in this complex mechanism is yet to be developed.
In this regard, \emph{in silico} models may be an efficient way to study and replicate selected features of the experimental system and forecast the evolution of the entire process in biologically relevant conditions.
Therefore, in the last decades, different mathematical models have been proposed, with different approaches, from continuous \cite{Balding, Liotta_1977, Chaplain_Sleeman, Chaplain_Stuart, Byrne_1995, Orme, Gomez2017, GiversoCiarletta} to discrete/hybrid \cite{Stokes, Gerhardt, Alarcon_2003, Gomez, Milde} models, with either deterministic or stochastic rules for cell branching (see \cite{Mantzaris, SciannaPreziosi, Gomez_review} for a review on mathematical models on angiogenesis). 
Continuous models \cite{Balding, Chaplain_Sleeman, Chaplain_Stuart, andersonchaplain}, typically encapsulates systems of coupled nonlinear partial differential equations (PDEs) describing the migration of ECs from the parent vessel towards the solid tumour in response to the TAFs and other chemicals dispersed in the ECM or eventually released by the ECs themselves.
Despite the ability of such PDEs continuous models to capture important angiogenic features at the macroscopic scale (i.e, average ECs and sprout density, average vessel growth and network expansion rates), they are not able to explicitly represent the geometry of the developing capillary network. Even in those cases in which the boundary between the capillary network and the surrounding tissue is tracked through a diffuse interface approach \cite{Gomez2017, Gomez_review}, the proposed models are not suitable to evaluate the inner blood flow.
At the same time, in those continuous models in which the morphology of nascent vessels is captured through a sharp interface, whose evolution is controlled both by chemical and mechanical cues \cite{GiversoCiarletta}, the vascularization of the network is not described, since it would require the definition of the complex fluid-structure interaction in an evolving 3D network.
Conversely, the description of the vessel networks morphology has been widely reproduced with discrete models operating at the scale of single ECs \cite{Bentley, Markus}, coupled with a continuous representation for the chemicals \cite{Stokes, andersonchaplain, Nekka, Levine_distance, Gomez, sun, Milde}. Although hybrid models have the advantage of describing the motion of individual ECs for simulating a realistic capillary network, the number of vessels that can be considered is limited by the computational costs of the discrete representation of each EC. Furthermore, the simulation of the blood flow inside the expanding geometry is generally disregarded. Indeed, despite its renowned importance in tumour growth and drug transport, the study of blood flow through the new vessel network and the mechanisms of fluid transport inside the surrounding tissue have only recently gained attention in mathematical modeling.

Starting from the preliminary works of Baxter and Jain presenting a macroscopic model for fluid and macromolecules transport inside a tumour with a distributed vascular source \cite{Baxter_1, Baxter} and a microscopic model describing flow transport around a single vessel \cite{Baxter4}, some mathematical models to describe the blood flow in non-evolving vascular networks have been proposed \cite{cattzun0, cattzun}. Recently some attempts to couple discrete models of angiogenesis, with continuous models of blood flow have been done. Specifically, in \cite{Secomb}, the flow  into each node of a network of vessel, evolving accordingly to the discrete angiogenesis model proposed in \cite{Tong}, is expressed in terms of nodal pressures and flow resistances of the segment, in order to satisfy the conservation of mass. 
A similar approach to determine blood flow in the vessel was used in \cite{Dougall, Alarcon_2003, Stephanou2005, Stephanou2006}, with different discrete models to reproduce the spatial and temporal evolution of the network. In these works, the simulation of flow was carried out a posteriori, either after generation of an hexagonal hollow vessel network constructed ad-hoc \cite{Alarcon_2003} or after an initial cell migration model that reproduce the vessel network \cite{Dougall, Stephanou2005, Stephanou2006}.
Notably, Rieger et al. developed a 2D and 3D cellular automata model \cite{Rieger2006, Rieger2008, Rieger2013} that considered angiogenesis, vessel co-option and vessel collapse in a time-dependent model of blood flow and tumour growth.
However, in all these models, nascent vessels need to be conformal to the mesh, leading to unnatural geometries of the network and blood flow inside it. Furthermore when cellular automata are used, the number of cells that can be simulated and therefore the extent of the vascular network that can be reproduced are limited in size.

Therefore in this work we propose a hybrid model, coupling a continuous representation of the tissue and of the chemical dispersed inside it with a discrete tip-tracking model that describes the motion of tip cells and the subsequent formation of vessels.
In principle this task would require the definition of a three dimensional domain, representing the healthy tissue, with tubular evolving structures immersed inside it, to represent the vessels. Then, all the equations of the model should be defined in both domains, with proper conditions at the interface between the vessel and the surrounding tissue.
The presence of domains with embedded tubular inclusions with radius much smaller than the length and than the size of the domain itself is challenging from a simulation standpoint. The generation of the computational mesh in the interior of the inclusions, indeed, constraints the mesh-size in the neighbourhood, thus resulting in a linear system with a very large number of unknowns. Further, for complex networks of inclusions, the mesh generation process might result even unfeasible, due to the large number of geometrical constraints. All these complexities are further aggravated if time-dependent simulations are to be performed and when branching and anastomosis occur.

The geometrical reduction of the inclusions to one dimensional objects can be performed, in these situations, to mitigate the computational cost and to allow the treatment of arbitrarily complex configurations. The 3D inclusions are collapsed on their centrelines, and simultaneously the external domain is extended to fill the voids. The resulting problem is composed of a three dimensional bulk domain with an embedded network of one dimensional domains, and a coupling of the solutions of such 3D-1D problems is required. 

The mathematical formulation of 3D-1D coupled problems requires particular care, as there is no bounded trace operator on 1D manifolds for functions in $H^1$-spaces on 3D domains. In \cite{Dangelo2012,DangeloQuarteroni2008} 3D problems with singular source terms on 1D segments are investigated, posing the basis for the analysis of 3D-1D coupled problems. Some authors suggest a regularization of the singular terms, such as in \cite{Tornberg2004,Heltai2019,Koch2020}, or splitting techniques, which treat in a separate manner the regular and irregular part of the solution \cite{Gjerde2018a,Gjerde2020}. Coupled 3D-1D problems are treated by \cite{Zunino2019} through the introduction of suitable averaging operators, and by \cite{Kuchta2021} through the use of Lagrange multipliers in a domain decomposition setting.
Here the method proposed by \cite{BGS3D1D2022,BGS3D1Ddisc} is adopted, in which a well posed problem is derived in suitable function spaces, defined on the basis of some assumptions on the regularity of the solution, and the problem is solved by means of an optimization based domain decomposition method. According to this approach, the 3D problem and the 1D problem are independently written, introducing additional variables at the interfaces, and a cost functional, expressing the error in fulfilling interface conditions is minimized to recover a global solution. The advantages of this approach lie in the possibility of building independent meshes on the various domains and in the direct computation of the interface variables, as, e.g. the flux at the 3D-1D interface, that are of particular interest for the applications. 

In detail, the paper is organized as follows. In Section \ref{sec:notandassum} we introduce the notation and the main modeling assumptions. The governing equations of the angiogenesis hybrid model, described as a 3D-1D coupled problem, are presented in Section \ref{sec:model}, in which the optimization based domain decomposition for 3D-1D coupling is also introduced.  Section \ref{sec:discretization}  is devoted to the discussion of the details concerning the numerical discretization of the model equations, while in Section \ref{sec:simulation} some numerical experiments are presented. Finally, in Section \ref{sec:conclusion} we summarize the main features of the work and present possible directions for further research.

\section{Notation and model assumptions}\label{sec:notandassum}
\begin{figure}
	\centering
	\includegraphics[width=1\textwidth]{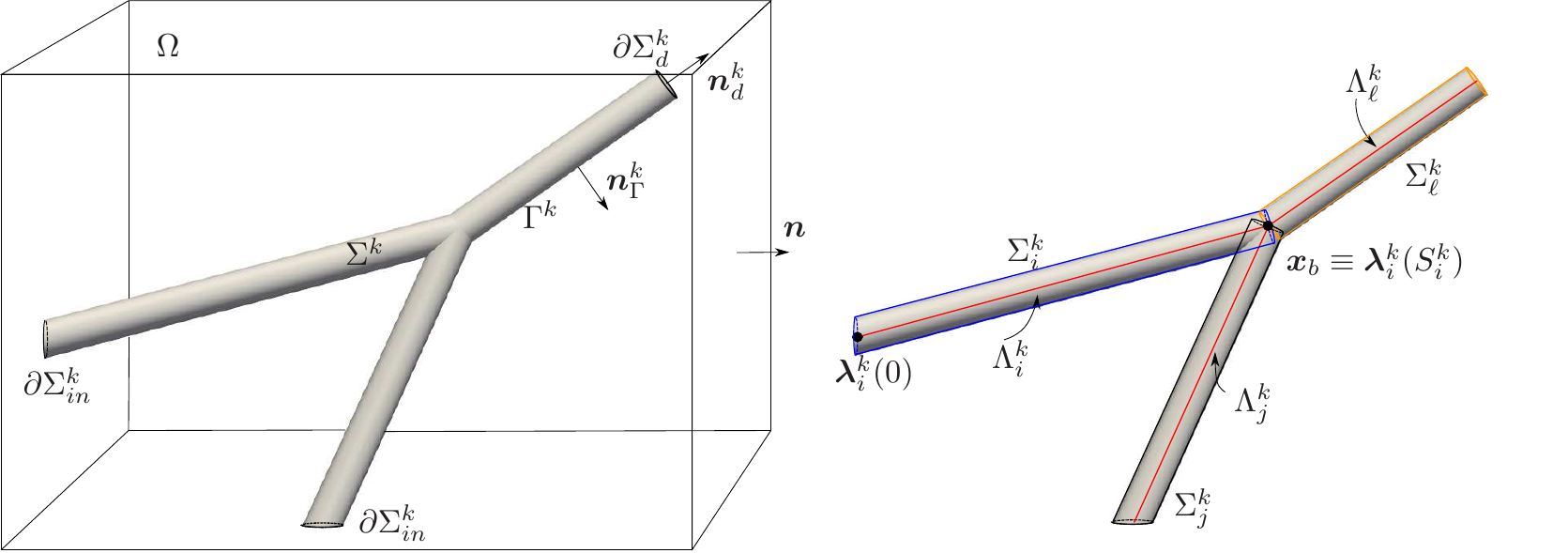}
	\caption{Nomenclature for capillary network representation at a bifurcation point.}
	\label{Fig:SegInters}
\end{figure}

Let us consider the time interval $[0,T]$ and a partition defined as
$$0=t_0<t_1<...<t_k<...<t_K=T,$$ with $\mathcal{I}_k=(t_{k-1},t_k]$ and $\Deltaik=t_k-t_{k-1}$. Let $\Sigma(t_k)=\Sigma^k\subset\Omega\subset \mathbb{R}^3$ denote the capillary network at time $t_k$, such that $\Sigma^0\subseteq\Sigma^1\subseteq...\subseteq \Sigma^k$. More precisely, in a continuous-discrete hybrid framework, $\Sigma^k$ denotes the fixed capillary network which is considered as the quantities of interest vary during the time-interval $\mathcal{I}_k$. We assume that the network $\Sigma^k$ is composed by thin tubular vessels of constant radius $R$, as shown in Figure~\ref{Fig:SegInters}.
The surrounding interstitial volume for $t \in \mathcal{I}_k$ is defined as $\mathcal{D}^k=\Omega\setminus \overline{\Sigma^k}$. The boundary of $\Omega$ is denoted by $\partial \Omega$ whereas $\partial \Sigma^k$ is the boundary of $\Sigma^k$.
Such boundary is split into four parts, i.e. $$\partial \Sigma^k= \Gamma^k\cup\left( \partial\Sigma_{in}^k\cup\partial\Sigma_{out}^k\cup\partial\Sigma_d^k\right),$$ with $\Gamma^k$ denoting the lateral surface of $\Sigma^k$, and $\partial\Sigma_{in}^k\subset \partial \Omega$ and $\partial\Sigma_{out}^k\subset \partial \Omega$ being the union, respectively, of the inlet and outlet cross-sections of $\Sigma^k$, the nomenclature referring to blood velocity. Finally, the set $\partial \Sigma_d^k$ collects the extremal cross sections of $\Sigma^k$ lying in the interior of $\Omega$ and thus having an empty intersection with $\partial \Omega$. We will assume that the extremal cross sections of $\Sigma^k$ are either completely lying on $\partial \Omega$, or inside $\Omega$, for all $k=0,...,K$. For clarity of exposition, we also suppose the set of the inflow and outflow sections to be fixed in time, i.e. we exclude the possibility for any growing vessel to reach the boundary of the domain. For this reason the sets $\partial\Sigma_{in}$ and $\partial \Sigma_{out}$ will actually not depend on time, so that the superscript $k$ is dropped. On the other hand, $\partial\Sigma_d^k$ is allowed to change in time.  Vectors $\tilde{\bm{n}}_{out}$ and $\tilde{\bm{n}}_d^k$ are used to denote the unit normal vectors to $\partial\Sigma_{out}$ and $\partial \Sigma_d^k$ respectively, both outward pointing from $\Sigma^k$. 
We define $\partD\subset\partial \Omega$ as $$\partial \mathcal{D}=\partial \Omega \setminus (\partial \Sigma_{in}\cup \partial \Sigma_{out})$$ and we denote by $\bm{n}$ the unit normal vector to $\partD$ outward pointing from $\D$.

We assume that the radius $R$ is much smaller than the inclusion length and than the characteristic size of $\Omega$, and thus that the original 3D-3D equi-dimensional problems can be approximated by 3D-1D problems. First, in each time interval $\mathcal{I}_k$, the original network $\Sigma^k$ is covered by a set of thin cylindrical vessels $\Sigma_i^k$, eventually cut by domain borders. We denote the centerline of each cylindrical vessel by $\Lambda^k_i=\lbrace \bm{\lambda}_i^k(s), ~s\in (0, S_i^k)\rbrace$, for $i\in Y^k$, being $Y^k$ the set if vessel indexes in $\mathcal{I}_k$. We also call $\Gamma_i^k$ the lateral surface of each vessel. Vessel centrelines are possibly connected at their endpoints, and we denote by $\lbrace\bm{x}_b\rbrace_{b \in B^k}$ the set of points at which vessel centrelines join or bifurcate as a consequence of branching or anastomosis events. The subset $Y_b^k\subset Y^k$ of centreline indexes is introduced, such that segments $\Lambda_j^k$, $j\in Y_b^k$ are connected in $\bm{x}_b$, and we call $S_{i,b}$ the curvilinear abscissa such that $\bm{\lambda}_i^k(S_{i,b})=\bm{x}_b$, $i \in Y_b^b$ (clearly $S_{i,b}$ is either 0 or $S_i^k$). We assume that up to three centrelines can meet at each junction point. 
The domain $\Sigma^k$ is thus replaced, in the derivation of the 3D-1D problems, by $\bigcup_{i\in Y^k} \Sigma_i^k$, while $\Lambda^k=\bigcup_{i\in Y^k}\Lambda_i^k \cup \lbrace\bm{x}_b\rbrace_{b \in B^k} $. We remark that, in general, according to the above definition, the union of the $\Sigma_i^k$ might be different from $\Sigma^k$, but, as $R$ is small such difference can be considered negligible.

Under the same assumption of small vessel radius, we consider negligible the variations of a generic quantity $\tilde{u}(\bm{x},t)$ on the transversal sections of each $\Sigma^k_i$. According to this, we can replace quantity $\tilde{u}(\bm{x},t)$ with the uniform extension to the whole cross-section of the value on the centreline $\hat{u}_i(s,t)$. In cylindrical coordinates, for $s \in (0,S_i^k)$,
\begin{equation}
\tilde{u}_{|{\Sigma_i^k}}(r,\theta,s,t)=\hat{u}_i(s,t) \quad \forall r \in [0,R], ~\forall \theta \in [0,2\pi),~t \in \mathcal{I}_k.\label{fund_assum}
\end{equation}

\section{Models and methods}\label{sec:model}
The process of oxygen delivery by blood flow during angiogenesisis is here described by means of three partial differential equation systems and an ordinary differential equation.

The first PDE problem describes the distribution of fluid pressure in the considered domain. This problem is first formulated as a coupled problem between a 3D tissue and a 3D domain given by the union of the vessels. To ease the meshing process and to reduce the computational cost of simulations, the 3D-3D equi-dimensional problem is re-formulated as a 3D-1D coupled problem by dimensionally reducing the cylindrical vessels to one dimensional domains coinciding with their centrelines, adopting assumption \eqref{fund_assum}. Further, a domain decomposition method, based on numerical optimization \cite{BGS3D1D2022,BGS3D1Ddisc}, is used to solve the 3D-1D coupled problem. The advantages of this approach lie in the possibility of using completely  independent meshes on the various subdomains and on the interfaces. Further, interface-internal variables, such as the trace of the pressure at vessel-tissue boundary, and consequently the flux, are directly computed, without any post process, and thus avoiding any loss of accuracy. The pressure problem is solved in a quasi-stationary framework, i.e. an equilibrium problem is solved each time the geometry is updated.

The same optimization based 3D-1D coupled strategy is adopted for the second PDE problem, which concerns the transport and diffusion of oxygen in the tissue and inside the vessel network. The velocity field for the transport term is computed according to the gradient of fluid pressure, previously obtained. The equations are in this case parabolic, i.e. although the geometry is fixed $\forall t \in \mathcal{I}_k$, we consider the continuous variation of oxygen distribution in the given time-interval.

The third PDE problem models instead the distribution of a chemotactic growth factor in the 3D tissue. The absorption of the chemotactic agent by the capillary network is modeled as a singular sink term located at vessel centrelines, while no equation is solved inside the vessel network. Even in this case a time-dependent advection-diffusion-reaction equation is considered.

Finally the growth of the capillary network is described by an ODE for the evolution of the position of vessel free ends, with given rules for preferential growth paths, branching and anastomosis.

In the remaining of this Section, we first propose the three PDE problems, both in the equi-dimensional form, and in the 3D-1D reduced form. Subsequently the optimization based domain decomposition method is proposed, for brevity, only for the oxygen concentration problem, which given its parabolic nature is more complex. The extension to the pressure problem is however similar.
Finally, the model of network growth is described in details.

\subsection{The pressure problem}
Let us denote by $p(\bm{x},t) $ and $\tilde{p}(\bm{x},t)$ respectively the interstitial fluid pressure in the tissue $\mathcal{D}^k$ and the blood pressure in the capillary network $\Sigma^k$, with $t\in\mathcal{I}_k=(t_{k-1},t_k]$. Assuming that the tissue can be suitably represented by a saturated porous medium \cite{Preziosi}, with the cell and the extracellular matrix representing the solid skeleton, the Darcy's law can be used to model the interstitial flow. The solid skeleton is assumed to be rigid and the growth of cells in the tissue surrounding the capillary and the degradation/deposition of the ECM are neglected. Therefore the cell volume fraction of the interstitial fluid and the solid phase can be assumed constant.
	Inside the capillaries, the motion of the blood, described as an incompressible viscous fluid can be appropriately described by the Poiseuille's law for laminar stationary flow \cite{Baxter_1, cattzun0}. Therefore, the following 3D-3D quasi-stationary coupled problem completely describe the motion of the fluid inside the capillary and the interstitium $\forall t \in \mathcal{I}_k$:
\begin{align}
-&\nabla \cdot \Big(\frac{\kappa}{\mu} \nabla p(\bm{x},t)\Big)+\beta_p^{LS}\frac{S}{V}(p(\bm{x},t)-p_{LS})=0 & \qquad \bm{x} \in \D\label{eq1_pressure}\\[0.2em]
&\frac{\kappa}{\mu}\nabla p(\bm{x},t)\cdot \bm{n}^k_\Gamma(\bm{x})=\beta_p(\tilde{p}(\bm{x},t)-p(\bm{x},t)-\Delta p_{onc}) &\qquad \bm{x} \in \Gamma^k\label{coup1}\\[0.2em]
&\frac{\kappa}{\mu}\nabla p(\bm{x},t)\cdot \bm{n}(\bm{x})=\beta_p^{ext}(p_{ext}-p(\bm{x},t)) &\qquad \bm{x} \in \partD\label{Robin_pressure}\\[0.2em]
&\nabla p(\bm{x},t)\cdot \bm{n}_d^k(\bm{x})=0  &\qquad \bm{x} \in \partial \Sigma_{d}^k\\[0.8em]
&-\nabla \cdot \Big(\frac{R^2}{8\mu} \nabla \tilde{p}(\bm{x},t)\Big)=0 & \qquad \bm{x} \in \Sigma^k\label{eq1_pressure_tilde}\\[-0.4em]
&\frac{R^2}{8\mu}\nabla \tilde{p}(\bm{x},t)\cdot \tilde{\bm{n}}_\Gamma^k(\bm{x})=\beta_p(p(\bm{x},t)-\tilde{p}(\bm{x},t)+\Delta p_{onc}) &\qquad \bm{x} \in \Gamma^k\label{coup2}\\
&\tilde{p}(\bm{x},t)=\tilde{p}_{in} &\qquad \bm{x} \in \partial\Sigma_{in}\\[0.2em]
&\tilde{p}(\bm{x},t)=\tilde{p}_{out} &\qquad \bm{x} \in \partial\Sigma_{out}\\[0.2em]
&\nabla\tilde{p}(\bm{x},t)\cdot \tilde{\bm{n}}_d^k(\bm{x})=0  &\qquad \bm{x} \in \partial \Sigma_{d}^k\label{eq_end_pressure}
\end{align}	
Parameters $\kappa$ and $\mu$ are positive scalars, denoting respectively the hydraulic permeability of the tissue and blood viscosity. Also $\beta_p$ and $\beta_p^{ext}$ are positive scalars, representing respectively the permeability of the capillary wall $\Gamma^k$ and the conductivity of the external boundary, while $p_{ext}$ is the basal pressure. In numerical simulations we will consider $\tilde{p}_{out}=p_{ext}$. Vector $\bm{n}_\Gamma^k=-\tilde{\bm{n}}_\Gamma^k$ is the unit normal vector to $\Gamma^k$ outward pointing from $\D$, while $\bm{n}_d^k=-\tilde{\bm{n}}_d^k$.

Equations \eqref{eq1_pressure} and \eqref{eq1_pressure_tilde} are coupled through \eqref{coup1} and \eqref{coup2}, which impose flux conservation across the surface $\Gamma^k$. The flux is defined through the Kedem-Katchalsky condition, assuming it is proportional to the pressure jump. The term $\Delta p_{onc}$ accounts for the contribute of oncotic pressure, whose gradient is relied to the different concentration of the chemicals on the two sides of the vessel wall. The most significant contribute to the oncotic pressure gradient is given by proteins, in particular by albumin, whose concentration is considered constant. For this reason we handle the oncotic pressure as a known correction term $\Delta p_{onc}=\xi(\tilde p_{onc}-{p}_{onc})$, with $p_{onc}$ and $\tilde{p}_{onc}$ denoting the oncotic pressure of albumin inside $\D$ and $\Sigma^k$ respectively ($\forall k$) and $\xi$ being the departure of the membrane from perfect permeability (see \cite{cattzun} for further details). Finally the term $\beta_p^{LS}\frac{S}{V}(p(\bm{x},t)-p_{LS})$ accounts for the absorption of the fluid in excess by the lymphatic system. This contribution is treated as a distributed sink term, with  $\beta_p^{LS}$ denoting the permeability of the lymphatic wall, $\frac{S}{V}$ the surface area of lymphatic vessels per unit of tissue volume and $p_{LS}=p_{ext}$ the pressure inside the lymphatic system. 
Once the pressure distribution is computed inside $\D$ and $\Sigma^k$ the fluid velocity can be defined as
\begin{align}
&\bm{v}(\bm{x},t)=-\frac{\kappa}{\mu}\nabla p(\bm{x},t) \quad \bm{x}\in \D,~t\in \mathcal{I}_k \label{vel3D}\\[0.3em]
&\tilde{\bm{v}}(\bm{x},t)=-\frac{R^2}{8\mu}\nabla \tilde{p}(\bm{x},t) \quad \bm{x} \in \Sigma^k,~t\in \mathcal{I}_k\label{vel1D}
\end{align}
Under assumption \eqref{fund_assum} we have that, in cylindrical coordinates, for $s \in (0,S_i^k)$ and $t \in \mathcal{I}_k$
\begin{align}
&\tilde{p}_{|{\Sigma_i^k}}(r,\theta,s,t)=\hat{p}_i(s,t),&\quad \forall r\in [0,R],~\forall \theta \in [0,2\pi)\label{assum_pr}\\[0.3em]
&p_{|{\Gamma_i^k}}(R,\theta,s,t)=\check{p}_i(s,t),&\quad \forall \theta \in [0,2\pi),
\end{align}
Thus, defining function
\begin{equation}
\hat{f}_p^i(s,t)=2\pi R \beta_p(\hat{p}_i(s,t)-\check p_i(s,t)-\Delta p_{onc}), \quad s \in (0,S_i^k),~ t \in \mathcal{I}_k
\end{equation}
equations \eqref{eq1_pressure}-\eqref{coup1} and \eqref{eq1_pressure_tilde}-\eqref{coup2} can be rewritten as a 3D-1D coupled system. For $t \in \mathcal{I}_k$:
\begin{align}
-&\nabla \cdot \Big(\frac{\kappa}{\mu} \nabla p(\bm{x},t)\Big)+\beta_p^{LS}\frac{S}{V}(p(\bm{x},t)-p_{LS})=\sum_{i \in Y^k}\hat{f}_p^i\delta_{\Lambda_{i}^k} & \bm{x} \in \D\label{pr1D_1}\\
-&\frac{\partial }{\partial s}\Big(\frac{\pi R^4}{8\mu}\frac{\partial \hat{p}_i(s,t)}{\partial s}\Big)=-\hat{f}_p^i(s,t) & \forall i\in Y^k,~ s \in (0,S_i^k) \\[0.4em]
&\sum_{j \in Y_b} \frac{\partial \hat{p}_j}{\partial s}(S_{j,b},t)=0 & \forall b\in B^k\\
&\hat{p}_i(S_{i,b},t)=\hat{p}_j(S_{j,b},t) &\forall i\neq j \in Y_b,~\forall b \in B^k,\label{pr1D_end}
\end{align}	
where the last two equations express respectively flux balance and pressure continuity at bifurcation points. Let us observe that according to \eqref{assum_pr}, the velocity $\tilde{\bm{v}}(\bm{x},t)$ is always parallel to $\Lambda$ and its magnitude depends only on the curvelength $s$. For this reason in the following we will adopt the notation $\hat{v}_i(s,t)=\tilde{\bm{v}}(\bm{x},t)\cdot \bm{\tau}_{\Lambda_i^k}$, with $\bm{\tau}_{\Lambda_i^k}$ denoting the direction tangential to $\Lambda_i^k$. 

\subsection{The oxygen concentration problem}\label{Concentration}
Denoting by $c(\bm{x},t)$ and $\tilde{c}(\bm{x},t)$ the concentration of oxygen respectively in $\D$ and in the capillary network $\Sigma^k$, we can write, for $t \in \mathcal{I}_k$ the following 3D-3D reaction-diffusion-convection coupled problem:
\begin{align}
&\frac{\partial c(\bm{x},t)}{\partial t}=\nabla \cdot \left(D_c \nabla c(\bm{x},t)\right)-\bm{v}(\bm{x},t)\cdot\nabla c(\bm{x},t)-m_c(c(\bm{x},t)) &\bm{x} \in \D\label{eq1_concentration}\\[0.2em]
&D_c\nabla c(\bm{x},t)\cdot \bm{n}_\Gamma^k(\bm{x})=\beta_c(\tilde{c}(\bm{x},t)-c(\bm{x},t))&\bm{x} \in \Gamma^k\label{coup1_conc}\\[0.2em]
&D_c\nabla c(\bm{x},t)\cdot \bm{n}(\bm{x})=\beta_c^{ext}(c_{ext}-c(\bm{x},t)) & \bm{x} \in \partD \label{Robin_oxygen}\\[0.2em]
&\nabla c(\bm{x},t)\cdot \bm{n}_d^k(\bm{x})=0 & \bm{x} \in \partial \Sigma_{d}^k\\[0.8em]
&\frac{\partial \tilde{c}(\bm{x},t)}{\partial t}=\nabla \cdot (\tilde{D}_c \nabla \tilde{c}(\bm{x},t))-\tilde{\bm{v}}(\bm{x},t)\cdot\nabla \tilde{c}(\bm{x},t) & \bm{x} \in \Sigma^k\label{eq_concentration_tilde}\\[0.2em]
&\tilde{c}(\bm{x},t_{k-1})=0 & \bm{x} \in \Sigma^k\setminus \Sigma^{k-1} \label{init0_oxy}\\
&\tilde{D}_c\nabla \tilde{c}(\bm{x},t)\cdot \tilde{\bm{n}}_\Gamma^k(\bm{x})=\beta_c(c(\bm{x},t)-\tilde{c}(\bm{x},t)) & \bm{x} \in \Gamma^k\label{coup2_conc}\\[0.2em]
&\tilde{c}(\bm{x},t)=c_{in}& \bm{x} \in \partial \Sigma_{in}\\[0.2em]
&\tilde{D}_c\nabla\tilde{c}(\bm{x},t)\cdot \tilde{\bm{n}}_{out}(\bm{x})=0 & \bm{x} \in \partial\Sigma_{out}\\[0.2em]
&\nabla\tilde{c}(\bm{x},t)\cdot \tilde{\bm{n}}_d^k(\bm{x},t)=0 & \bm{x} \in \partial \Sigma_{d}^k\label{eq_end_concentration}
\end{align}	
For $k=0$ we define the initial conditions
\begin{align*}
&c(\bm{x},0)=c_{0}(\bm{x}) & \bm{x} \in \mathcal{D}^0 \\
&\tilde{c}(\bm{x},t_{0})=\tilde{c}_0(\bm{x}) & \bm{x} \in \Sigma^0
\end{align*}
while for $k>0$ the concentrations at time $t_{k-1}$ are available from the final concentrations computed in $\mathcal{I}_{k-1}$ and only the amount of oxygen in the newborn capillaries has to be initialized, as in equation \eqref{init0_oxy}.
Parameters $D_c$ and $\tilde{D}_c$ are positive scalars denoting the diffusivity respectively in $\D$ and $\Sigma^k$. Also $\beta_c$ and $\beta_c^{ext}$ are positive scalars, which denote the permeability to oxygen respectively of the blood vessel wall and of the boundary of $\D$. Finally, function $m_c(c(\bm{x},t))$ accounts for the oxygen metabolization by cells. In particular we choose $m_c(c(\bm{x},t))=M_cc(\bm{x},t)$, with $M_c$ positive scalar.

Exploiting again assumption \eqref{fund_assum} we have that, in cylindrical coordinates, for $s \in (0,S_i^k)$ and $t \in \mathcal{I}_k$
\begin{align}
&\tilde{c}_{|{\Sigma_i^k}}(r,\theta,s,t)=\hat{c}_i(s,t),&\quad \forall r\in [0,R],~\forall \theta \in [0,2\pi)\label{chat}\\
&c_{|{\Gamma_i^k}}(R,\theta,s,t)=\check{c}_i(s,t),&\quad \forall \theta \in [0,2\pi)\label{ccheck}
\end{align}
Defining the quantity
\begin{equation}
\hat{f}_c^i(s,t)=2\pi R \beta_c(\hat{c}_i(s,t)-\check c_i(s,t)), \quad s\in(0,S_i^k),~t\in \mathcal{I}_k
\end{equation}
equations \eqref{eq1_concentration}-\eqref{coup1_conc} and \eqref{eq_concentration_tilde}-\eqref{coup2_conc} can now be rewritten as the 3D-1D coupled system, for $t \in \mathcal{I}_k$:
\begin{align}
&\frac{\partial c(\bm{x},t)}{\partial t}-\nabla \cdot \Big(D_c \nabla c(\bm{x},t)\Big)+\bm{v}(\bm{x},t)\cdot\nabla c(\bm{x},t)+m_c(c(\bm{x},t))=\sum_{i \in Y^k}\hat{f}^i_c\delta_{\Lambda_i^k}\label{oxy1D_1}\\[-0.3em]
&\hspace{10cm} \bm{x} \in \D\nonumber  \\
&\pi R^2\frac{\partial \hat{c}_i(s,t)}{\partial t}-\frac{\partial}{\partial s_i}\Big(\pi R^2 \tilde{D}_c\frac{\partial \hat{c}_i(s,t)}{ds}\Big)+\pi R^2\hat{v}_i(s,t)\frac{\partial \hat{c}_i(s,t)}{\partial s_i}=-\hat{f}_c^i(s,t)\\[-0.2em]
&\hspace{8cm}\forall i\in Y^k,~s \in (0,S_i^k) \nonumber \\
&\sum_{j \in Y_b} \frac{\partial \hat{c}_j}{\partial s}(S_{j,b},t)=0 \hspace{6.8cm} \forall b\in B^k\\
&\hat{c}_i(S_{i,b},t)=\hat{c}_j(S_{j,b},t) \hspace{4.5cm}\forall i\neq j \in Y_b,~\forall b \in B^k.\label{oxy1D_end}
\end{align}

\subsubsection{The optimization based domain decomposition for 3D-1D coupling}\label{sec:opt}
The problems of pressure and oxygen concentration require a coupling between a 3D and a 1D problem. Providing a well-posed mathematical formulation for this kind of coupling is not trivial, since no bounded trace operator is defined when the dimensionality gap between the interested manifolds is higher than one. However it is possible to define suitable subspaces of the Sobolev spaces typically employed for the variational formulation of partial differential equations, in which the definition of such a trace operator is never required. After having defined the proper spaces in which to look for the solutions we aim at applying the optimization based 3D-1D coupling strategy presented in \cite{BGS3D1Ddisc}. This choice is related to the fact that this approach has no mesh conformity requirements and, hence, it allows to easily handle complex time-varying geometries as the ones characterizing angiogenesis simulations. We refer to \cite{BGS3D1D2022}-\cite{BGS3D1Ddisc} for a wider presentation of the method, while trying to give the main ideas in the following.
In this section we focus on problem \eqref{oxy1D_1}-\eqref{oxy1D_end} in which time derivation, advection and reaction contributions are present. The same considerations hold, however, for more simple elliptic problems as \eqref{pr1D_1}-\eqref{pr1D_end}.

Let us again consider $t \in \mathcal{I}_k=(t_{k-1},t_k]$ and let us define the space $$H^1(\Lambda^k)=\prod_{i \in Y^k}H^1(\Lambda_i^k) \cap \mathcal{C}^0(\Lambda^k),$$ being nothing but the space of continuous functions on $\Lambda^k$ whose restriction to $\Lambda_i^k$ is in $H^1(\Lambda_i^k)$. Each function $\hat{u}\in H^1(\Lambda^k)$ can be written as $$\hat{u}=\prod_{i \in Y^k}\hat{u}_i, \quad \hat{u}_i \in H^1(\Lambda^k_i).$$ 
More in general, in the remaining of this section, we denote by $w_i$ the restriction of a sufficiently regular function $w$ to $\Lambda_i^k$. We then define a trace operator $$\trg_i^k:H^1(\D)\cup H^1(\Sigma^k_i)\rightarrow H^{\frac{1}{2}}(\Gamma^k_i)$$ which, given $u \in H^1(\D)\cup H^1(\Sigma^k)$, returns $\trg^k_iu=u_{|_{\Gamma^k_i}}$ $i \in Y^k$, and an extension operator
$$\extg_i^k: H^1(\Lambda^k_i) \rightarrow H^{\frac{1}{2}}(\Gamma^k_i)$$
which, given $\hat{u}_i\in H^1(\Lambda^k_i)$ uniformly extends the value $\hat{u}_i(s)$ to the boundary $\Gamma^k_i(s)$ of the transversal section $\Sigma^k_i(s)$, i.e. $\extg^k_i\hat{u}_i(s)=\tilde{u}(\bm{x})~\forall \bm{x} \in \Gamma^k_i(s)$.
Then, we define
\begin{equation*}
\Vhatz=\left\lbrace \hat{u} \in H^1(\Lambda^k):~\hat{u}_{|_{\Lambda_{in}}}=0 \right\rbrace, \quad \Vhat=\left\lbrace \hat{u} \in H^1(\Lambda^k):~\hat{u}_{|_{\Lambda_{in}}}=\hat{c}_{in} \right\rbrace,
\end{equation*}
with $\Lambda_{in}$ collecting the centers of the inflow sections $\partial \Sigma_{in}$, and
\begin{equation*}
\hsp_i=\lbrace u\in H^{\frac{1}{2}}(\Gamma^k_i): u =\extg^k_i\hat{u}_i, ~\hat{u}\in \Vhat\rbrace
\end{equation*} 
\begin{equation*}
\VD=\left\lbrace u \in H^1(\D): \trg^k_iu \in \hsp_i,~\forall i \in Y^k\right\rbrace.
\end{equation*}
Also, the space $\Vtilde_i$ is introduced as:
$$\Vtilde_i=\lbrace u \in H^1(\Sigma^k_i): u =\exts^k_i\hat{u}_i, ~\hat{u}\in \Vhat\rbrace,$$
where $\exts^k_i: H^1(\Lambda^k_i) \rightarrow H^1(\Sigma^k_i)$ is an extension operator which uniformly extends the value $\hat{u}_i(s)$ to the cross section $\Sigma^k_i(s)$ of the cylinder, i.e. $\exts^k_i\hat{u}_i(s)=\tilde{u}(\bm{x})~\forall \bm{x}\in \Sigma^k_i(s)$.

Exploiting assumption \eqref{fund_assum}, we can choose $$c(t)\in\VD \quad \text{and} \quad \tilde{c}(t)=\prod_{i \in Y^k}\tilde{c}_i(t),~\tilde{c}_i(t)\in \Vtilde_i.$$ We denote by $\check{c}(t)\in\Vhat$ the function such that $\trg^k_ic(t)=\extg^k_i\check{c}_i(t)$, as in \eqref{ccheck}, and similarly by $\hat{c}(t) \in\Vhat$ the function such that $\tilde{c}(t)(\bm{x})=\exts^k_i\hat{c}_i(t)$ $\forall \bm{x} \in \Sigma^k_i$.

 The 3D-1D variational formulation of problem \eqref{oxy1D_1}-\eqref{oxy1D_end} can now be written as:
 $\forall t \in \mathcal{I}_k $, \textit{find} $c(t) \in \VD$, $\hat{c}(t)\in \Vhat$ \textit{such that}
 \begin{align}
 &\Big(\frac{\partial c }{\partial t},\eta\Big)_{L^2(\D)}+\big(D_c\nabla c,\nabla \eta\big)_{L^2(\D)}+\big(\bm{v}\cdot\nabla c, \eta\big)_{L^2(\D)}+\big(M_c c,\eta\big)_{L^2(\D)}+\nonumber \\
 &\quad +(\beta_c^{ext}c,\eta)_{L^2(\partD )}+\sum_{i \in Y^k}(2\pi R\beta_c(\check{c}_i-\hat{c}_i),\check{\eta}_i)_{L^2(\Lambda^k_i)}=(\beta_c^{ext}c_{ext},\eta)_{L^2(\partD )}\label{variational1} \\[-0.5em]&\hspace{6.5cm}\forall \eta \in \VD: \trg^k_i\eta=\extg^k_i\check{\eta}_i, \check{\eta}\in \Vhatz \nonumber \\[0.7em]
 &\sum_{i \in Y^k}\Bigg[\Big(\pi R^2\frac{\partial \hat{c}_i }{\partial t},\hat{\eta}_i\Big)_{L^2(\Lambda^k_i)}+\Big(\pi R^2\tilde{D}_c\frac{\partial\hat{c}_i}{\partial s},\frac{\partial\hat{\eta}_i}{\partial s}\Big)_{L^2(\Lambda^k_i)}+\Big(\pi R^2\hat{v}_i\frac{\partial \hat{c}_i}{\partial s},\hat{\eta}_i \Big)_{L^2(\Lambda^k_i)}+\nonumber \\[-0.5em]
 &\hspace{1cm}+\Big(2\pi R\beta_c(\hat{c}_i-\check{c}_i),\hat{\eta}_i)_{L^2(\Lambda^k_i)}\Bigg]=0, \qquad \forall \hat{\eta}\in \Vhatz \label{variationalend}
 \end{align}
 
As aforementioned, we now aim at applying the optimization based 3D-1D coupling approach presented in \cite{BGS3D1Ddisc}. The method resorts to a domain decomposition strategy, in which two auxiliary variables are introduced at the interface in order to decouple the problems defined in the vascular network and in the surrounding tissue. We denote such variables by $\hpsid(t)$ and $\hpsis(t)$ and we rewrite problem \eqref{variational1}-\eqref{variationalend} as:\\ $\forall t \in \mathcal{I}_k $, \textit{find} $c(t) \in \VD$, $\hat{c}(t)\in \Vhatc$, $\hpsid(t)\in \Vhat$, $\hpsis(t) \in \Vhat$ \textit{such that}
\begin{align}
&\Big(\frac{\partial c }{\partial t},\eta\Big)_{L^2(\D)}+\big(D_c\nabla c,\nabla \eta\big)_{L^2(\D)}+\big(\bm{v}\cdot\nabla c, \eta\big)_{L^2(\D)}+\big(M_c c,\eta\big)_{L^2(\D)}+\nonumber \\
&\quad +(\beta_c^{ext}c,\eta)_{L^2(\partD )}+\sum_{i \in Y^k}(2\pi R\beta_c\check{c}_i,\check{\eta}_i)_{L^2(\Lambda^k_i)} - \sum_{i \in Y^k}(2\pi R\beta_c\hpsis_i,\check{\eta}_i)_{L^2(\Lambda^k_i)}=\nonumber \\
&\quad =(\beta_c^{ext}c_{ext},\eta)_{L^2(\partD )},\qquad \forall \eta \in \VD: \trg^k_i\eta=\extg^k_i\check{\eta}_i, \check{\eta}\in \Vhatz \label{var_concentration1} \\[0.7em]
&\sum_{i \in Y^k}\Bigg[\Big(\pi R^2\frac{\partial \hat{c}_i }{\partial t},\hat{\eta}_i\Big)_{L^2(\Lambda^k_i)}+\Big(\pi R^2\tilde{D}_c\frac{\partial\hat{c}_i}{\partial s},\frac{\partial\hat{\eta}_i}{\partial s}\Big)_{L^2(\Lambda^k_i)}+\Big(\pi R^2\hat{v}_i\frac{\partial \hat{c}_i}{\partial s},\hat{\eta}_i \Big)_{L^2(\Lambda^k_i)}+\nonumber \\[-0.5em]
&+\Big(2\pi R\beta_c\hat{c}_i,\hat{\eta}_i\Big)_{L^2(\Lambda^k_i)}-(2\pi R\beta_c\hpsid_i,\hat{\eta}_i)_{L^2(\Lambda^k_i)}\Bigg]=0, \qquad \forall \hat{\eta}\in \Vhatz \label{var_concentration2}
\end{align}
with interface conditions, $\forall i \in Y^k$,
\begin{align}
&\left\langle\check{c}_i(t)-\hpsid_i(t) ,\hat{\mu}_i\right\rangle_{\Vhatz,{\Vhatz}'}=0&~\forall \hat{\mu}_i \in  \Vhatz',~t\in \mathcal{I}_k,\label{condpsi_u}\\
&\left\langle \hat{c}_i(t)-\hpsis_i(t),\hat{\mu}_i\right\rangle_{\Vhatz,\Vhatz'}=0&~\forall \hat{\mu}_i \in  \Vhatz',~t\in \mathcal{I}_k.\label{condpsi_hat}
\end{align}

The final step is to recast our problem into a PDE-constrained optimization problem.  We hence  introduce a cost functional, which mimics the error committed by approximating $\check{c}(t)$ and $\hat{c}(t)$ by $\hpsid(t)$ and $\hpsis(t)$ respectively:
\begin{equation}
J^k(\hpsid(t),\hpsis(t))=\cfrac{1}{2}\sum_{i \in Y^k}\Big( ||\check{c}_i(t)-\hpsid_i(t)||_{L^2(\Lambda^k_i)}^2+||\hat{c}_i(t)-\hpsis_i(t)||_{L^2(\Lambda^k_i)}^2\Big).	\label{functional}
\end{equation}
The variational PDE-constrained optimization formulation of problem \eqref{eq1_concentration}-\eqref{eq_end_concentration} finally reads: $\forall t \in \mathcal{I}_k $
\begin{equation}
\min_{\hpsid(t),\hpsis(t)\in \Vhat}J^k(\hpsid(t),\hpsis(t))  \text{ subject to \eqref{var_concentration1}-\eqref{var_concentration2}}. \label{minJ}
\end{equation}
Handling the 3D-1D coupled problem as an optimization problem ends up in a method for which no mesh conformity is required. This represents a great advantage in angiogenesis simulations, since we will never need to remesh the tissue as the vascular network grows. We refer to Section \ref{sec:discretization} for details about the meshes and for the discretization of problem \eqref{minJ}.

\subsection{The chemotactic growth factor problem}\label{CGF_section}
Let us denote by $\mathcal{C}\subset \mathbb{R}^3$ the portion of space occupied by the tumor. We suppose that $\partial \Omega\cap\partial\mathcal{C}\neq \emptyset$ but $\mathcal{C}\nsubseteq \Omega$, i.e. our computational domain does not account for the tumor region, but a portion of its boundary is chosen as an interface with the tumor itself. Let us then denote by $g(\bm{x},t)$ the concentration of a vascular endothelial growth factor (VEGF) and let us describe its evolution for $t \in \mathcal{I}_k$ by the following set of equations 
\begin{align}
&\frac{\partial g(\bm{x},t) }{\partial t}=\nabla\cdot \left(D_g\nabla g(\bm{x},t)\right)-\bm{v}(\bm{x},t)\cdot \nabla g(\bm{x},t)-\sigma g(\bm{x},t) \quad \bm{x} \in \D\label{eq_g}
\end{align}\vspace{-0.6cm}
\begin{align}
&D_g\nabla g(\bm{x},t)\cdot \bm{n}_\Gamma^k(\bm{x})=-\tilde{\sigma}g(\bm{x},t)  &\qquad \bm{x} \in \Gamma^k \label{interface_g} \\
&g(\bm{x},t)=g^{\mathcal{C}}(\bm{x}) &\qquad \bm{x} \in \partial \mathcal{C}\cap \partD\\
&\nabla g(\bm{x},t)\cdot \bm{n}(\bm{x})=0  &\qquad \bm{x} \in \partD\setminus  (\partial\mathcal{C}\cap \partD) \\
&\nabla g(\bm{x},t)\cdot \bm{n}_d^k(\bm{x})=0 &\bm{x} \in \partial\Sigma_d^k\label{eq_g_fin}
\end{align}
For $k=0$ we set an initial condition
\begin{equation*}
g(\bm{x},0)=g_0(\bm{x}) \qquad \bm{x} \in \mathcal{D}^0.
\end{equation*}
At the interface between the tissue sample and the tumor a constant Dirichlet boundary condition is imposed, modelling a constant distribution $g^{\mathcal{C}}$ of growth factor at the tumor boundary. Parameters $D_g$ and $\sigma$ are positive scalars denoting respectively the diffusivity of the VEGF and its natural decay rate, $\tilde{\sigma}$ is the rate of consumption of VEGF by endothelial cells. The vector field $\bm{v}$ is computed according to \eqref{vel3D}.
 
Exploiting assumption \eqref{fund_assum}, we can write, in cylindrical coordinates, for $s \in (0,S_i^k)$
$$g_{|\Gamma_i^k}(R,\theta,s,t)=\check{g}_i(s,t),\quad \forall \theta \in [0,2\pi),~ t \in \mathcal{I}_k.$$
Equations \eqref{eq_g}-\eqref{interface_g} can thus be reduced to a 3D equation with a singular reaction term
\begin{align}
&\frac{\partial g(\bm{x},t)}{\partial t}-\nabla \cdot \big(D_g \nabla c(\bm{x},t)\big)+\bm{v}(\bm{x},t)\cdot\nabla g(\bm{x},t)+\\[-0.4em]
\nonumber&\hspace{4cm}+\sigma g(\bm{x},t)=-\sum_{i \in Y^k}2\pi R\tilde{\sigma}\check{g}(s,t)\delta_{\Lambda_i^k}, \quad \bm{x} \in \D,~t \in \mathcal{I}_k \nonumber
\end{align}
The consumption of VEGF by endothelial cells is to be intended in terms of receptor mediated binding, i.e. there is not a flux of VEGF through the vessel wall. For this reason no equation is defined for the concentration of VEGF inside the capillaries.

Let us introduce the spaces
\newcommand{\Vzero}{V_{0^{\mathcal{C}}}^k}
\newcommand{\Vg}{V_{g^{\mathcal{C}}}^k}
\begin{equation*}
\Vzero=\left\lbrace u \in H^1(\D): \trg^k_i u \in\hsp_i \text{ and } u_{|_{\partD\cap\partial \mathcal{C}}}=0\right\rbrace,
\end{equation*}
\begin{equation*}
\Vg=\left\lbrace u \in H^1(\D): \trg^k_i u \in \hsp_i \text{ and } u_{|_{\partD\cap\partial \mathcal{C}}}=g^{\mathcal{C}}\right\rbrace.
\end{equation*}
Functions in $\Vg$ assume a constant value on section boundaries $\Gamma^k_i(s)$, thus satisfying assumption \eqref{fund_assum}. The variational problem then  reads: \textit{Find $g(t) \in \Vg$ such that }
\begin{equation}
\begin{cases}
\left(\partial_t g,\eta\right)_{L^2(\D)}+\left(D_g\nabla g,\nabla \eta\right)_{L^2(\D)}+(\bm{v}\cdot\nabla g,\eta)_{L^2(\D)}+(\sigma g,\eta)_{L^2(\D)}+\\\hspace{1.5cm}+\sum_{i \in Y^k}\limits\left(2\pi R\tilde{\sigma}\check{g}_i,\check{\eta}_i\right)_{L^2(\Lambda^k_i)}=0 \quad \forall \eta \in \Vzero: \trg^k_i\eta=\extg^k_i\check{\eta}, \check{\eta}\in H^1(\Lambda^k) \label{var_g}\\g(0)=g_0.\end{cases}\end{equation}
where $\trg^k_ig(t)=\extg^k_i\check{g}_i(t)$.

\subsection{The growth of the vascular network}\label{growth_sec}
Given assumption \eqref{fund_assum}, in the following we identify the vascular network at time $t_k$ with its centerline $\Lambda^k$ and we extend the domain $\D$ to the whole $\Omega$, with embedded subdomains $\Lambda^k$. We denote by $\mathcal{P}^k$ the set of capillary tips at time $t_k$, being nothing but the centers of the sections $\partial \Sigma_d^k$. We assume that the endothelial cells respond chemotactically to VEGF gradients, undergoing mitosis and producing sprout extension. We model this phenomenon as a displacement of the tip cells, thus monitoring their number and position. In particular, the position $\bm{x}_{P}$ of a generic tip cell $P\in \mathcal{P}^k$ evolves according to
\begin{equation}
\frac{d\bm{x}_P}{dt}=\bm{w}(g(\bm{x}_P,t),\bm{x}_P)\label{grow}
\end{equation}
with $\bm{w}$ denoting the tip velocity and defined, according to \cite{sun}, as
\begin{equation}
\bm{w}(g,\bm{x})=\begin{cases}\cfrac{l_e}{t_c(g)}\cfrac{\bm{K}_{ECM}(\bm{x})\nabla g}{||\bm{K}_{ECM}(\bm{x})\nabla g||} &\text{ if }g\geq g_{lim}\\
0 &\text{otherwise}\end{cases} \label{growth_vel}
\end{equation}
In the above definition, $g_{lim}$ represents the minimum VEGF concentration for endothelial cell proliferation, i.e. for tip displacement, $l_e$ is the endothelial cell length and $t_c$ is a cell cycle division time, modeled as \cite{sun}
\begin{equation}
t_c(g)=\tau \left(1+e^{\left(\frac{\bar{g}}{g}-1\right)}\right),\label{tc}
\end{equation}
where $\tau$ is a cell proliferation parameter, while $\overline{g}$ is the concentration at which $t_c=2\tau$. In Figure \ref{fig:Pbr_tc_branching} (top left) function $t_c(g)$ is plotted for $\tau=12~\rm h$ and $\overline{g}=1\cdot 10^{-13}~ \rm kg/mm^3$. The local orientation of the extracellular matrix (ECM) fibers is modeled through matrix $\bm{K}_{ECM}$, which is assumed constant in time but variable in space. When no data on the ECM orientation is available, the matrix is defined as $\bm{K}_{ECM}=\bm{K}_{ECM}^{rand}$, with
\begin{equation}
\bm{K}_{ECM}^{rand}(\bm{x})=\bm{I}+k_{an}(\bm{x}) \bm{K}_{ECM}^{an}(\bm{x})\label{kecm}.
\end{equation}
Matrix $\bm{K}_{ECM}^{an}(\bm{x})$ represents deviations from isotropicity and it takes the form
\begin{equation*}
\bm{K}_{ECM}^{an}(\bm{x})=
\begin{bmatrix}
-k_2(\bm{x})^2-k_3(\bm{x})^2 & k_1(\bm{x})k_2(\bm{x}) &k_1(\bm{x})k_3(\bm{x}) \\ k_1(\bm{x})k_2(\bm{x}) & -k_1(\bm{x})^2-k_3(\bm{x})^2 & k_2(\bm{x})k_3(\bm{x}) \\ k_1(\bm{x})k_3(\bm{x}) & k_2(\bm{x})k_3(\bm{x}) & -k_1(\bm{x})^2-k_2(\bm{x})^2
\end{bmatrix}.
\end{equation*}
The parameters $k_i(\bm{x})$, $i=1,...,3$ and the weight of the anisotropic contribution $k_{an}(\bm{x})$ are randomly chosen for each position in space, such that $\sum_{i=1}^3k_i(\bm{x})^2=1$ and $k_{an}(\bm{x})\in [0,1]$. We refer to Section~\ref{TestSphere} for a case in which random perturbation is exploited together with some known features on the ECM orientation.
\subsubsection{Branching}
In our model the generation of new sprouts from an existing sprout tip (branching) can occur only when the following conditions are both satisfied (\cite{sun}):
\begin{enumerate}
	\item the age of the current sprout is greater than a threshold-age $\tau_{br}$. This means that a time $\tau_{br}$ has passed since the sprout last branched;
	\item the ratio between the norm of the orthogonal projection of $\bm{w}$ on the plane perpendicular to the current sprout orientation and the norm of $\bm{w}$ is greater than a threshold value $\alpha^w_{br}$.
\end{enumerate}
\begin{figure}
\begin{minipage}{.6\textwidth}
	\includegraphics[width=0.6\linewidth]{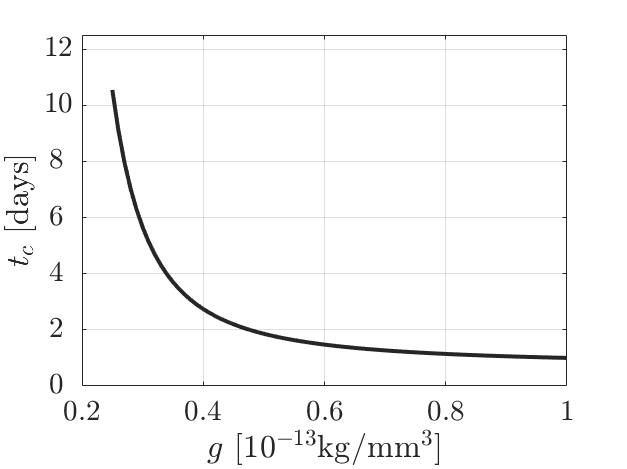}\\
	\includegraphics[width=0.6\linewidth]{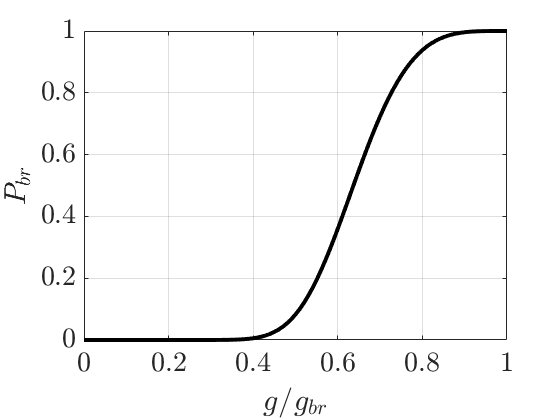}
\end{minipage}\hspace{-1cm}%
\begin{minipage}{.6\textwidth}
	\includegraphics[width=0.7\textwidth]{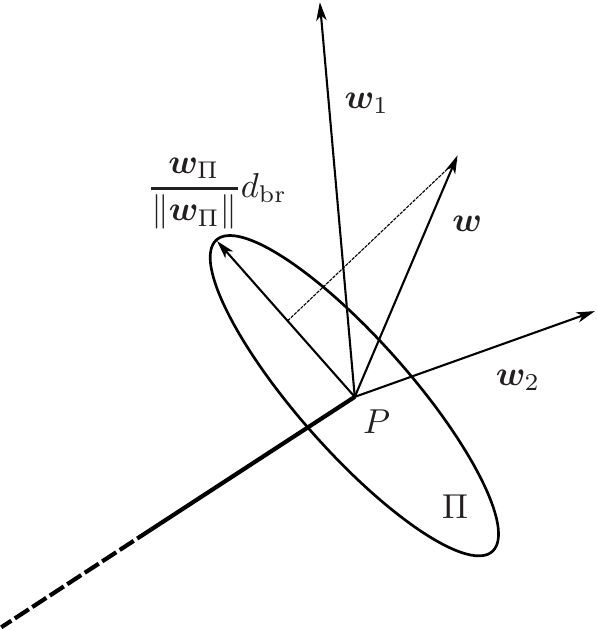}
	\end{minipage}
\caption{Top left, cell cycle division time \eqref{tc} for $g \in [0.25,1]\cdot 10^{13}~\rm kg/mm^3$, $\tau=12h$ and $\bar{g}=1\cdot 10^{-13}$; bottom left,  branching probability \eqref{br_prob} for $g_{br}=\bar{g}=1\cdot 10^{-13}$ and $\tau_{br}=48 ~\rm h$. On the right representation of the branching directions}
\label{fig:Pbr_tc_branching}
\end{figure}
We introduce also a \textit{branching probability} $P_{br}(g)$, such that branching occurs more probably when the concentration of VEGF is high. The aim of the branching probability function is also to avoid the formation of new branches at a rate which could be incoherently higher than the proliferation rate. According to \eqref{tc}, the cell cycle division time is in fact related to VEGF concentration so that we impose a very low probability of branching when $\tau_{br}<t_c$. In particular we define
\begin{equation}
P_{br}(g)=\begin{cases}
e^{-a\Big(\frac{g}{g_{br}}-1\Big)^4} &\text{if } g<g_{br}\\
1 &\text{otherwise}
\end{cases}\label{br_prob}
\end{equation}
with $g_{br}$ corresponding to the VEGF concentration at which the probability of branching is $100\%$. We choose $a$ such that the probability of branching is $5\%$ for a value of $g$ corresponding to $t_c=\tau_{br}$.
More or less restrictive choices are of course possible, possibly supported by biological experiments. The branching probability function for $g_{br}=\bar{g}=1\cdot 10^{-13}$ and $\tau_{br}=48 ~\rm h$ is reported in Figure~\ref{fig:Pbr_tc_branching} (bottom left).

Let $P\in \mathcal{P}^k$ being a sprout tip at time $t_k$. Let us denote by $\Pi$ the plane perpendicular to the sprout orientation and by $\bm{w}_{\Pi}$ the orthogonal projection of $\bm{w}(g(\bm{x}_P,t_k),\bm{x}_P)$ onto $\Pi$. Let us suppose that $\frac{||\bm{w}_{\Pi}||}{||\bm{w}||}>\alpha^w_{br}$ and that the sprout is old enough to branch. If, according to the probability $P_{br}(g(\bm{x}_P,t_k))$, branching actually occurs, then the directions $\bm{w}_1$ and $\bm{w}_2$ of the two new sprouts are obtained as
\begin{equation*}
\bm{w}_1=\bm{w}+\frac{\bm{w}_{\Pi}}{||{\bm{w}_{\Pi}}||}d_{br}, \qquad
\bm{w}_2=\bm{w}-\frac{\bm{w}_{\Pi}}{||{\bm{w}_{\Pi}}||}d_{br}
\end{equation*} with $d_{br}$ denoting the diameter of a single capillary and corresponding to the distance by which the new sprout tips will be separated (see Figure \ref{fig:Pbr_tc_branching}-right).

\subsubsection{Anastomosis}
Anastomosis, i.e. the formation of loops as a consequence of the fusion of two vessels, is supposed to occur when a sprout tip meets another sprout tip or a portion of a sprout that is not older than a certain threshold-age $\tau_{an}$. Using the nomenclature in \cite{sun}, the first configuration is called tip-to-tip anastomosis and produces the deactivation of both tips, while the second is called tip-to-sprout and leads only one tip to become inactive. In the simulations, anastomosis is forced when the new sprout tip lies within a distance $d_{an}$ from another tip or from a sufficiently young sprout. 

\section{Problem discretization}\label{sec:discretization}
Here the discretization of the previous problems is reported. We start from the problem for oxygen concentration, resulting from the application of the optimization method, then we move to the VEGF problem and to the network growth. The discretization of the pressure problem can be easily derived generalizing the steps made for oxygen concentration.

\subsection{The discrete optimization problem}
As already mentioned in Section~\ref{growth_sec}, for the discretization of the 3D-1D coupled problems, the 3D domain is extended to the whole $\Omega$. Let us consider a tetrahedral mesh $\mathcal{T}$ on domain $\Omega$, independent from the position of the vessel network. Denoting by $N$ the number of DOFs for oxygen inside $\Omega$, let us define the Lagrangian basis functions $\left\lbrace \varphi_j\right\rbrace_{j=1}^{N}$ such that the discrete approximation of $c(t)$ is $C(t)=\sum_{j=1}^{N}C_j(t)\varphi_k$. For what concerns the 1D variables, we build on $\Lambda^k$ three different partitions $\hat{\mathcal{T}}^k$, $\tau_{D}^k$ and $\tau_{\Sigma}^k$, independent from each other and from $\mathcal{T}$. Such meshes could change at each time-step, but, for computational efficiency we choose to incrementally add mesh elements as the network grows. To ensure mesh uniformity, a minimum element size can be fixed, such that new mesh elements are created only if larger than the minimum size. We define the basis functions $\left\lbrace\hat{\varphi}_{j} \right\rbrace _{j=1}^{\hat{N}^k}$ on $\hat{\mathcal{T}}^k$, $\left\lbrace \theta_{j}^D\right\rbrace_{j=1}^{\hat{N}_{D}^k}$ on $\tau_{D}^k$ and $\left\lbrace \theta_{j}^\Sigma\right\rbrace_{j=1}^{\hat{N}_{\Sigma}^k}$ on $\tau_{\Sigma}^k$, with $\hat{N}^k$, $\hat{N}_{D}^k$ and $\hat{N}_{\Sigma}^k$ denoting the number of DOFs at time $t_k$ of the discrete approximations of the variables $\hat{c}(t)$, $\hpsid(t)$ and $\hpsis(t)$. We remark that the basis functions do not depend on time, only their number can vary with time. The approximations of $\hat{c}(t)$, $\hpsid(t)$ and $\hpsis(t)$ on the vascular network $\Lambda_k$ are defined as:
 \begin{equation*}
 \hat{C}(t)=\sum_{j=1}^{\hat{N}^k}\hat{C}_{j}(t)~\hat{\varphi}_{j}, \quad
  \Psi^D(t)=\sum_{j=1}^{\hat{N}_{D}^k}\Psi_{j}^D(t)~\theta_{j}^D, \quad \Psi^\Sigma(t)=\sum_{j=1}^{\hat{N}_{\Sigma}^k}\Psi_{j}^\Sigma(t) ~\theta_{j}^\Sigma.
 \end{equation*}
 Let us now define the matrices
 \begin{align*}
 &\bm{A}^k \in \mathbb{R}^{N\times N} \text{ s.t. } A_{lj}^k=\int_{\Omega}\left( D_c\nabla\varphi_j\nabla\varphi_l+(\bm{v}\cdot\nabla\varphi_j)\varphi_l+m_c\varphi_j\varphi_l\right) ~d\omega+\\[-0.2em] &\hspace{4cm}+\int_{\partial \Omega}\beta_c^{ext}{\varphi_j}_{|_{\partial \Omega}}{\varphi_l}_{|_{\partial \Omega}}d\sigma+\int_{\Lambda^k}2\pi R \beta_c{\varphi_j}_{|_{\Lambda^k}}{\varphi_l}_{|_{\Lambda^k}} ds,\\[0.8em]
 &\bm{\hat{A}}^k \in \mathbb{R}^{\hat{N}^k\times \hat{N}^k} \text{ s.t. } \hat{A}_{lj}^k=\int_{\Lambda^k}\Big(\pi R^2\tilde{D}_c\frac{d\hat{\varphi}_{j}}{ds}\frac{d\hat{\varphi}_{l}}{ds}+\pi R^2\hat{v}\frac{d\hat{\varphi}_j}{ds}\hat{\varphi}_l\Big) ~ds+\int_{\Lambda^k}2\pi R \beta_c\hat{\varphi}_{j}\hat{\varphi}_{l}~ds\\
 &\bm{M} \in \mathbb{R}^{N\times N} \text{ s.t. } M_{lj}=\int_{\Omega}\varphi_j\varphi_l~d\omega\\
 &\bm{\hat{M}}^k\in \mathbb{R}^{\hat{N}\times \hat{N}} \text{ s.t. } \hat{M}_{lj}^k=\int_{\Lambda^k}\pi R^2\hat{\varphi}_j\hat{\varphi}_l~ds,\\[1em]
 &{\bm{\hat{D}}_{\beta}}^k \in \mathbb{R}^{\hat{N}^k\times \hat{N}_{D}^k} \text{ s.t. } ({}\hat{D}_\beta^k)_{lj}=\int_{\Lambda^k}2\pi R \beta_c{\hat{\varphi}_{l}~\theta_{j}^D}~ds,\\&
 \bm{S}_{\beta}^k \in \mathbb{R}^{N\times \hat{N}_{\Sigma}^k} \text{ s.t. } (S_{\beta}^k)_{lj}=\int_{\Lambda^k}2\pi R \beta_c{\varphi_l}_{|_{\Lambda^k}}\theta_{j}^{\Sigma}~ds
 \end{align*}
 and the vector
 $${F}\in \mathbb{R}^{N} \text{ s.t. } F_l=\int_{\partial \Omega}\beta_c^{ext}c_{ext}\varphi_l~d\sigma.$$

The implicit Euler scheme is adopted for time-discretization. At this aim let us define a uniform partition of the time interval $\mathcal{I}_k$ with a step $\Delta t \leq \Delta \mathcal{I}_k$ and $t_{k,q}=t_{k-1}+q\Delta t$, $q\geq 0$. The fully discretized version of equations \eqref{var_concentration1}-\eqref{var_concentration2} then reads:
 \begin{equation}
 \begin{cases}
 \left(\bm{M}+\Delta t\bm{A}^k\right)C(t_{k,q})-\Delta t\bm{S}_\beta^k\Psi_{\Sigma}(t_{k,q})=\bm{M}C(t_{k,q-1})+\Delta t F\\
 \big(\bm{\hat{M}}^k+\Delta t\bm{\hat{A}}^k\big)\hat{C}(t_{k,q})-\Delta t \bm{\hat{D}}_\beta^k\Psi_{D}(t_{k,q})=\bm{\hat{M}}^k\hat{C}(t_{k,q-1})\\
 C(t_{k,0})=\begin{cases}
 C_0 & \text{ if } k=0\\ 
 C(t_{k-1}) &\text{ if } k>0
 \end{cases}\\
 \hat{C}(t_{k,0})=\begin{cases}\hat{C}_0& \text{ if } k=0\\
 \hat{C}_\sharp(t_{k-1}) &\text{ if } k>0
 \end{cases}
 \end{cases}\label{discr_c}
 \end{equation}
 where $\hat{C}_\sharp(t_{k-1})$ is the trivial extension of $\hat{C}(t_{k-1})\in \mathbb{R}^{\hat{N}^{k-1}}$ to $\mathbb{R}^{\hat{N}^k}$ by zero elements in correspondence of DOFs defined on $\Lambda^{k}\setminus\Lambda^{k-1}$.
 
In order to work out also the discrete formulation of functional \eqref{functional} let us build the matrices
\begin{align*}
&\bm{G}^k \in \mathbb{R}^{N \times N} \text{ s.t. } G_{lj}^k=\int_{\Lambda^k}{\varphi_j}_{|_{\Lambda^k}}{\varphi_l}_{|_{\Lambda^k}}ds,\\
&\bm{\hat{G}}^k \in \mathbb{R}^{\hat{N}^k \times \hat{N}^k} \text{ s.t. } \hat{G}_{lj}^k=\int_{\Lambda^k}{\hat{\varphi}}_{j}~{\hat{\varphi}}_{l}~ds,\\
&\bm{{G}_{D}}^k \in \mathbb{R}^{\hat{N}_{D}^k \times \hat{N}_{D}^k}  \text{ s.t. } ({G}_{D}^k)_{lj}=\int_{\Lambda^k}\theta_{j}^D~\theta_{l}^D~ds,\\
&\bm{{G}_{\Sigma}}^k \in \mathbb{R}^{\hat{N}_{\Sigma}^k \times \hat{N}_{\Sigma}^k}  \text{ s.t. } ({G}_{\Sigma}^k)_{lj}=\int_{\Lambda^k}\theta_{j}^\Sigma~\theta_{l}^\Sigma~ds
\end{align*}
\begin{align*}
&\bm{D}^k \in \mathbb{R}^{N\times \hat{N}_{D}^k} \text{ s.t. } D^k_{lj}=\int_{\Lambda^k}{{\varphi_{l}}_{|_{\Lambda^k}}\theta_{j}^D}~ds,\\ 
&\bm{\hat{S}}^k \in \mathbb{R}^{\hat{N}^k\times \hat{N}_{\Sigma}^k} \text{ s.t. } \hat{S}^k_{lj}=\int_{\Lambda^k}{\hat{\varphi}}_{l}~\theta_{j}^\Sigma~ds.
\end{align*}
The discrete cost functional at time $t_{k,q}$ then reads:
 \begin{align}
 &\tilde{J}^{k,q}=\cfrac{1}{2}\big( C(t_{k,q})^T\bm{G}^kC(t_{k,q})-C(t_{k,q})^T\bm{D}^k\Psi_D(t_{k,q})-\Psi_D(t_{k,q})^T(\bm{D}^k)^TC(t_{k,q})+\nonumber\\&+\Psi_D(t_{k,q})^T\bm{G_D}^k\Psi_D(t_{k,q})+\hat{C}(t_{k,q})^T\bm{\hat{G}}^k\hat{C}(t_{k,q})-\hat{C}(t_{k,q})^T\bm{\hat{S}}^k\Psi_\Sigma(t_{k,q})+\nonumber\\& -\Psi_\Sigma(t_{k,q})^T(\bm{\hat{S}}^k)^T\hat{C}(t_{k,q})+\Psi_\Sigma(t_{k,q})^T\bm{G_\Sigma}^k\Psi_\Sigma(t_{k,q})\big)\label{Jtildec}
 \end{align}
 Introducing the matrices 
 $$\bm{\mathcal{G}}^k=\begin{bmatrix}
 \bm{G}^k& \bm{0} & -\bm{D}^k & \bm{0} \\
 \bm{0} & \bm{\hat{G}}^k & \bm{0} &-\bm{\hat{S}}^k \\ 
 -(\bm{D}^k)^T & \bm{0} & \bm{G_D}^k & \bm{0} \\
 \bm{0} &-(\bm{\hat{S}}^k)^T & \bm{0} & \bm{G_\Sigma}^k
 \end{bmatrix}$$
 $$ \bm{\mathcal{A}}^k=\begin{bmatrix} \bm{A}^k & 0\\
 0& \bm{\hat{A}}^k\end{bmatrix}, \quad \bm{\mathcal{M}}^k=\begin{bmatrix} \bm{M} & 0\\
 0& \bm{\hat{M}}^k\end{bmatrix}, \quad \bm{\mathcal{C}_\beta}^k=\begin{bmatrix} 0 & \bm{S}_\beta^k \\
 \bm{D}_\beta^k&0\end{bmatrix},$$
 $$\bm{\mathcal{A}_{\Delta t}}^k=\begin{bmatrix}
 \Delta t\bm{\mathcal{A}}^k+\bm{\mathcal{M}}^k &-\Delta t\bm{\mathcal{C}_\beta^k}\end{bmatrix}$$
first order optimality conditions for the minimization of \eqref{Jtildec} constrained by Equations~\eqref{discr_c} are collected in the saddle-point system
\begin{equation*}
\bm{\mathcal{K}}^k=
\begin{bmatrix}
\bm{\mathcal{G}}^k & (\bm{\mathcal{A}_{\Delta t}}^k)^T\\ 
\bm{\mathcal{A}_{\Delta t}}^k &\bm{0}
\end{bmatrix}
\end{equation*}
\begin{equation}
\bm{\mathcal{K}}^k
\begin{bmatrix}
C(t_k)\\\hat{C}(t_{k,q})\\ \Psi_D(t_{k,q}) \\ \Psi_\Sigma(t_{k,q}) \\ -\Pi(t_{k,q})\\ -\hat{\Pi}(t_{k,q})
\end{bmatrix}=\begin{bmatrix}
0 \\0  \\ 0 \\ 0 \\\bm{M}C(t_{k,q-1})+\Delta t F \\\bm{\hat{M}}^k\hat{C}(t_{k,q-1})
\end{bmatrix}\label{KKT}
\end{equation}
which is solved at each time-step $\delta t$. Vectors $\Pi$ and $\hat{\Pi}$ are the vector of DOFs of Lagrange multipliers.  Let us observe that, by a proper organization of the DOFs, most matrices do not need to be rebuilt completely at each $k$: only the integrals on $\Lambda^k\setminus \Lambda^{k-1}$ have to be computed and properly concatenated to the matrices that were already available at time $t_{k-1}$. Only the matrices $\bm{A}$ and $\hat{\bm{A}}$ need a further update, in order to account for the variation of the velocity field. This update affects however only the advection contribution to such matrices. For the proof of the uniqueness of the solution to \eqref{KKT} we refer to \cite{BGS3D1Ddisc}. 

\subsection{The discrete problem for VEGF and capillary growth}

The finite element discretization of Problem \eqref{var_g} is obtained considering a tetrahedral mesh $\mathcal{T}$ on domain $\Omega$  and defining on it Lagrangian basis functions $\left\lbrace \varphi_j\right\rbrace_{j=1}^{N_G}$, such that $$G(t)=\sum_{j=1}^{N_G}G_k(t)\varphi_j$$ is the discrete approximation of variable $g(t)$ being $N_G$ the number of degrees of freedom. We then define the matrices
\begin{align*}
&\bm{B}^k\in \mathbb{R}^{N_G\times N_G} \text{ s.t. }B_{lj}^k=\int_{\Omega}D_g\nabla\varphi_j\nabla\varphi_l~d\omega+\int_{\Omega}(\bm{v}\cdot\nabla \varphi_j)
\varphi_l~d\omega+\\[-0.4em]
&\hspace{5cm}+\int_{\Omega}\sigma\varphi_j\varphi_l~d\omega+\int_{\Lambda^k}2\pi R\tilde{\sigma}{\varphi_j}_{|{\Lambda^k}}{\varphi_l}_{|{\Lambda^k}}~ds \\
&\bm{H}\in \mathbb{R}^{N_G\times N_G} \text{ s.t. }H_{lj}=\int_{\Omega}\varphi_j\varphi_l~d\omega
\end{align*}
such that the space semi-discretization of problem \eqref{var_g} reads
\begin{equation}
\begin{cases}
\bm{H}G'(t)+\bm{B}^kG(t)=0 &t\in \mathcal{I}_k\\
G(0)=G_0
\end{cases}
\end{equation}
with $G'=\partial_tG$.
For what concerns time discretization we adopt an implicit Euler scheme, considering again a uniform partition of $\mathcal{I}_k$ with $\Delta t \leq \Delta \mathcal{I}_k$ and $t_{k,q}=t_{k-1}+q\Delta t$ we solve at each time step a system in the form
\begin{equation}
\begin{cases}
\left(\bm{H}+\Delta t \bm{B}^k\right)G(t_{k,q})=\bm{H}G(t_{k,q-1}) &k>0\\
G(t_{k,0})=\begin{cases}
G_0 &\text{ if } k=0\\
G(t_{k-1}) &\text{ if } k>0
\end{cases}
\end{cases}\label{discrG}
\end{equation}
For what concerns Equation~\eqref{grow}, discretization is made by the explicit Euler method. Once the VEGF concentration at time $t_k$ has been computed, the position of the tip cells is updated as
\begin{equation}
\bm{x}_P(t_{k+1})=\bm{x}_P(t_{k})+\Delta \mathcal{I}_{k+1}\cdot \bm{w}(G(t_{k}),\bm{x}_P(t_{k})), \quad \forall P \in \mathcal{P}^k\label{discr_grow}
\end{equation}
providing the position of a tip cell in $\mathcal{P}^{k+1}$.
The points $\bm{x}_P(t_{k+1})$ and $\bm{x}_P(t_{k})$ are then connected by a line, such that the capillary network is represented by sets of connected segments in the 3D space. 
In case branching occurs, the orthogonal projection $\bm{w}_{\Pi}$ of $\bm{w}(G(t_{k}),\bm{x}_P(t_{k}))$ onto the plane $\Pi$ perpendicular to the direction $\bm{x}_P(t_{k})-\bm{x}_P(t_{k-1})$ has to be computed in order to obtain the branching directions $\bm{w}_1$ and $\bm{w}_2$. Equation \eqref{discr_grow} is then splitted into
\begin{equation}
\bm{x}_P^1(t_{k+1})=\bm{x}_P(t_{k})+\Delta\mathcal{I}_{k+1}\bm{w}_1,\qquad 
\bm{x}_P^2(t_{k+1})=\bm{x}_P(t_{k})+\Delta\mathcal{I}_{k+1}\bm{w}_2\label{br}
\end{equation}
producing two new sprout tips. Once the new positions of the tip cells have been computed, the network is updated as
$$\Lambda^{k+1}=\Lambda^{k}\cup\bigcup_{\substack{P\in \mathcal{P}^k\\i=1,2}}[x_P^i(t_{k+1}),x_P(t_{k})]$$
which is the fixed geometry on which the other quantities will evolve for $t \in \mathcal{I}_{k+1}$.

\section{Numerical experiments}\label{sec:simulation}
\begin{figure}
	\includegraphics[width=0.69\linewidth]{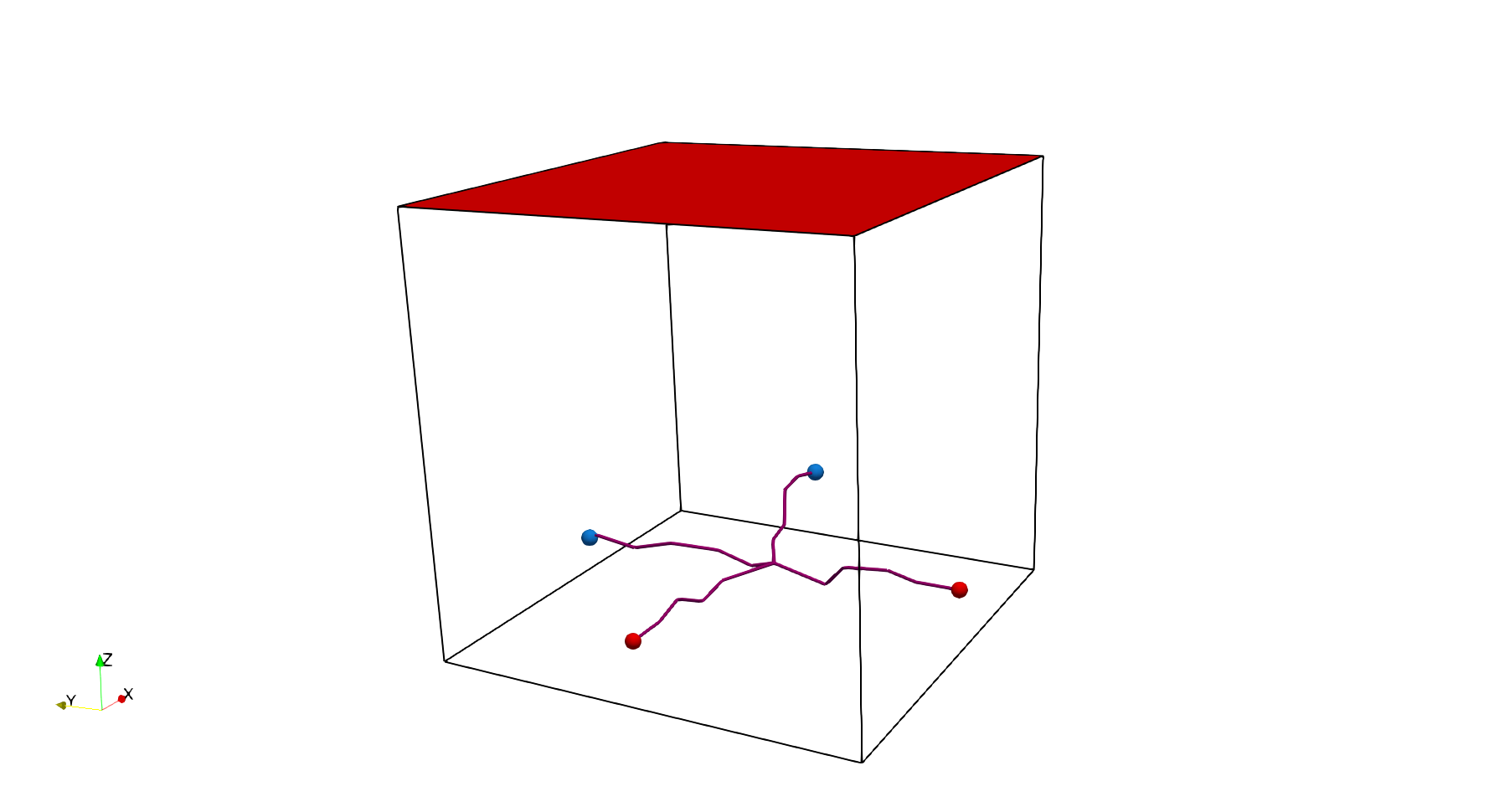}
	\hspace{-2cm}\includegraphics[width=0.39\linewidth]{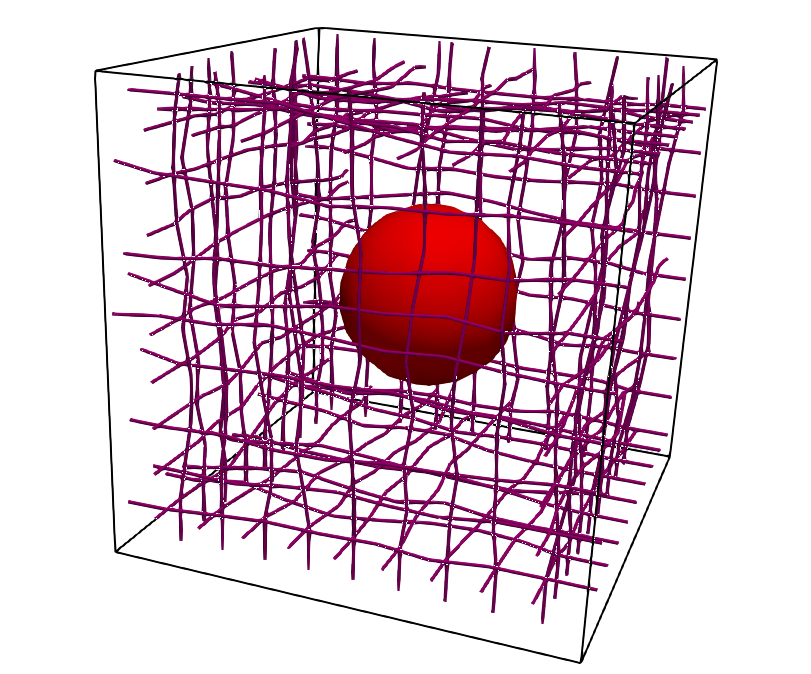}
	\caption{Initial configurations for the proposed numerical experiments, on the left \textit{TestFace} (inlet and outlet extrema in red and blue respectively), on the right \textit{TestSphere}.}
	\label{initialconfig}
\end{figure}

In this section we provide some numerical examples to validate the proposed approach. In particular we present two tests, labeled \textit{TestFace} and \textit{TestSphere}. In the first case the domain $\Omega$ is a cube and the tumor interface corresponds to one of its faces. A small initial vascular network with two inlet and two outlet points is considered, located at the opposite of the tumour interface, as shown in Figure \ref{initialconfig}-left. For the second test the tumor is instead supposed to be spherical and located at the center of a cubic domain. For this case a more complex initial network is considered, as reported in Figure \ref{initialconfig}-right. In the following we choose $\Delta t =\Deltaik$, i.e. we consider the same, uniform, time-stepping for the growth of the network and for the evolution on the quantities defined on it. 

\begin{table}
	\renewcommand*{\arraystretch}{1.1}
	\centering 
	\caption{Default parameters for the geometry}
	\label{table_geom}
	\begin{tabular}{ccclc}
			\hline
		\textbf{Parameter} &\textbf{Value} &\textbf{Unit}& \textbf{Description}&\textbf{Reference}\\
		\hline
		$L$ & $0.5-2.5$ &\small mm & Domain edge length& \cite{Gimbrone, angio_velocity, Cavallo} \\
		$R$ & $5\cdot 10^{-3}$ &\small mm &Vessel radius&\cite{capillary_wall}\\		
	\end{tabular}
\end{table}
\begin{table}
	\renewcommand*{\arraystretch}{1.3}
	\centering 
	\caption{Defualt parameters for pressure}
	\label{table_pr}
	\begin{tabular}{ccclc}
		\hline
		\textbf{Parameter} &\textbf{Value} &\textbf{Unit}& \textbf{Description}&\textbf{Reference}\\
		\hline
		\multirow{1.5}{*}{$\beta_p^0$} & \multirow{1.5}{*}{$2.78\cdot 10^{-10} $} & \multirow{1.5}{*}{$\frac{\rm{mm}^2h}{kg}$} &Hydraulic permeability &\\[-0.5em]&&&of healthy capillary wall&\cite{cattzun}\\
			\multirow{1.5}{*}{$r^\beta_p$} & \multirow{1.5}{*}{$100$} & \multirow{1.5}{*}{$-$} &Increase of wall permeability&\\[-0.5em] &&& for tumor-generated capillaries&\cite{cattzun0}\\
		\multirow{1.5}{*}{$\Delta p_{onc}$} & \multirow{1.5}{*}{$4.82\cdot 10^{7}$ }& \multirow{1.5}{*}{$\rm \frac{kg}{h^2 mm}$} &Oncotic pressure jump &\\[-0.5em]&&& at the capillary wall&\cite{cattzun}\\
		\multirow{1.5}{*}{$\beta_p^{LS}\frac{S}{V}$} &\multirow{1.5}{*}{$2.89\cdot 10^{-7}$} & \multirow{1.5}{*}{$\rm\frac{ mm~h}{kg}$} &Effective permeability& \\[-0.5em]&&& of the lymphatic vessels&	\multirow{-1.5}{*}{\cite{cattzun,Baxter}} \\
		\multirow{1.5}{*}{$\kappa$} & 	\multirow{1.5}{*}{$1.0 \cdot 10^{-12}$}& 	\multirow{1.5}{*}{\small$\rm mm^2$} &  Hydraulic permeability&\\[-0.5em]&&& of the tissue&\cite{cattzun}\\
		$\mu$ & $1.44\cdot 10^{-2}$ & $\rm \frac{kg}{mm ~h}$ & Blood viscosity&\cite{cattzun}\\
		$\tilde{p}_{in}$ & $6.05 \cdot 10^{7}$& $\rm \frac{kg}{h^2 mm}$ & inflow pressure&\cite{cattzun}\\
		$\tilde{p}_{out}$ & $5.83 \cdot 10^{7}$& $\rm \frac{kg}{h^2 mm}$ & outflow pressure&\cite{cattzun}\\
		$\beta_p^{ext}$ & $1.4\cdot 10^{-8}$ & $\rm \frac{mm^2h}{kg}$ & boundary conductivity&\cite{cattzun}\\
	\end{tabular}
\end{table}
\begin{table}
	\renewcommand*{\arraystretch}{1.3}
	\centering 
	\caption{Defualt parameters for oxygen}
	\label{table_oxy}
	\begin{tabular}{ccclc}
		\hline
		\textbf{Parameter} &\textbf{Value} &\textbf{Unit}& \textbf{Description}&\textbf{Reference }\\
		\hline
		\multirow{1.5}{*}{$\beta_c^0$} & \multirow{1.5}{*}{$126$} & \multirow{1.5}{*}{$\rm \frac{mm}{h}$} &Permeability of the healthy&\\[-0.5em] &&&capillary wall&\cite{cattzun}\\
		\multirow{1.5}{*}{$r^\beta_c$} & \multirow{1.5}{*}{$10$} & \multirow{1.5}{*}{$-$} &Increase of wall permeability&\\[-0.5em] &&& for tumor-generated capillaries&\\
		$D_c$ & $4.86 $& $\rm \frac{mm^2}{h}$ &  Diffusivity, tissue &\cite{cattzun}\\
		\multirow{1.5}{*}{$m_c$} & \multirow{1.5}{*}{$3.6$} &\multirow{1.5}{*}{$\rm \frac{1}{h}$} & Decay/metabolization&\\[-0.5em] &&&parameter&\\
		$\tilde{D}_c $ & $1.8\cdot 10^{3}$ & $\rm \frac{mm^2}{h}$ & Vascular diffusivity&\cite{cattzun}\\
		$\tilde{c}_{in}$ & $1.73\cdot 10^{8}$& $\rm \frac{kg}{h^2 mm}$ & inflow concentration&\\
		$\beta_c^{ext}$ & $18$ & $\rm\frac{mm}{h}$ & boundary permeability&\cite{cattzun}\\
		\multirow{1.5}{*}{$c_{ext}$} & \multirow{1.5}{*}{$6.05\cdot 10^6$} & \multirow{1.5}{*}{$\rm \frac{kg}{h^2 mm}$} & External oxygen&\\[-0.5em] &&& concentration&				
	\end{tabular}	
\end{table}

Tables \ref{table_geom}, \ref{table_pr}, \ref{table_oxy} and \ref{table_VEGF} provide a set of parameter values to which we will refer as \textit{default parameters}. Changes to these values will be specified and motivated time by time.
The parameter values for pressure and oxygen concentration are mainly taken from \cite{cattzun}. For what concerns the VEGF, in \cite{scirep_angionetwork} its concentration is assumed to be at a constant value of 20 $\rm ng/ml$, while in \cite{wang} endothelial cells are stimulated to migrate with a VEGF concentration of 50 $\rm ng/ml$. Converting to the units of measure used in the numerical simulations, the order of magnitude is of $10^{-14}~\rm kg/mm^3$. We choose $g_{lim}=2.5\cdot 10^{-14}\rm kg/mm^3$ and, according to \cite{sun} we set $\overline{g}=4g_{lim}=1\cdot 10^{-13}~\rm kg/mm^3$ in Equation \eqref{tc}. At the tumor interface we fix $g^{\mathcal{C}}=10^{-13}~\rm kg/mm^3$, so that the cell proliferation time $t_c$ in the simulation will be in the range $(2 \tau, +\infty)$.  The diffusion coefficient inferred from biological data, for the vast majority of angiogenic growth factors, is in the order of $10-600 \, \rm{\mu m^2/s}$ \cite{Miura, andersonchaplain, Serini}, i.e. between 0.036 and 2.16 $\rm mm^2/h$, while the decay rate of the VEGF in the surrounding tissue is in the order of $0.456 - 0.65 \, \rm{h}^{-1}$ \cite{Serini, Plank}.
\begin{table}
	\renewcommand*{\arraystretch}{1.3}
	\centering 
	\caption{Defualt parameters for VEGF}
	\label{table_VEGF}
	\begin{tabular}{ccclc}
		\hline
		\textbf{Parameter} &\textbf{Value} &\textbf{Unit}& \textbf{Description}&\textbf{Reference}\\
		\hline
		$D_g$ & $0.29$ & $\rm\frac{{mm}^2}{h}$ &VEGF diffusivity&\cite{sun} \\
		$\sigma$ & $0.5$ & $\rm \frac{1}{h}$ &VEGF interstitial decay&\cite{sun}\\
		\multirow{1.5}{*}{$\tilde{\sigma}$} &\multirow{1.5}{*}{$0.7$} & \multirow{1.5}{*}{$\rm\frac{1}{h}$ } &Endothelial cell VEGF&\\[-0.5em]&&&  consumption rate &\\
		\multirow{1.5}{*}{$g^{\mathcal{C}}$ }&	\multirow{1.5}{*}{$1.0\cdot 10^{-13}$} & 	\multirow{1.5}{*}{$\rm \frac{kg}{mm^3}$} & VEGF concentration &\\[-0.5em] &&&at tumor interface&\\
		\multirow{2}{*}{$g_{lim}$} & 	\multirow{2}{*}{$2.5\cdot 10^{-14}$}& 	\multirow{2}{*}{$\rm \frac{kg}{mm^3}$} &  minimum VEGF &\\[-0.5em] &&& concentration &\\[-0.5em] &&& for proliferation & \multirow{-2.5}{*}{\cite{scirep_angionetwork},\cite{wang}}\\
		\multirow{1.5}{*}{$\bar{g}$} & 	\multirow{1.5}{*}{$1.0\cdot 10^{-13}$} & 	\multirow{1.5}{*}{$\rm \frac{kg}{mm^3}$} & VEGF concentration &\\[-0.5em] &&& for $t_c=2\tau$&\\
		\multirow{1.5}{*}{$\tau$} & \multirow{1.5}{*}{$12$} & \multirow{1.5}{*}{\small h} & cell proliferation &\\[-0.5em] &&&  parameter & \multirow{-1.5}{*}{\cite{sun}}\\ 
		$l_e$ & $0.04$ & \small mm & endothelial cell length&\cite{sun}\\
		\multirow{1.5}{*}{$\alpha^w_{br}$} & \multirow{1.5}{*}{$0.3$} & \multirow{1.5}{*}{-} & threshold of $\frac{||\bm{w}_\Pi||}{||\bm{w}||}$ &\\[-0.5em] &&&for branching &\\
		$d_{br}$ & $1.0\cdot 10^{-2}$ & \small mm & branching distance&\\
		\multirow{1.5}{*}{$\tau_{br}$} & \multirow{1.5}{*}{$48$} & \multirow{1.5}{*}{\small h} & threshold age&\\[-0.5em] &&&for branching&\\
		\multirow{1.5}{*}{$g_{br}$} & 	\multirow{1.5}{*}{$1.0\cdot 10^{-13}$} &	\multirow{1.5}{*}{$\rm \frac{kg}{mm^3}$} & VEGF concentration&\\[-0.5em]&&& for $P_{br}=1$ &\\
		\multirow{1.5}{*}{$d_{an}$} & 	\multirow{1.5}{*}{$1.0\cdot 10^{-5}$} & 	\multirow{1.5}{*}{\small mm} & maximum distance for&\\[-0.5em]&&& anastomosis&\\
		\multirow{1.5}{*}{$\tau_{an}$} &	\multirow{1.5}{*}{ $24$} &	\multirow{1.5}{*}{\small h}& maximum capillary age &\\ [-0.5em]&&& for anastomosis&
	\end{tabular}	
\end{table}

The initial distributions of pressure, oxygen and VEGF are obtained solving respectively problems \eqref{eq1_pressure}-\eqref{eq_end_pressure}, \eqref{eq1_concentration}-\eqref{eq_end_concentration} and \eqref{eq_g}-\eqref{eq_g_fin} in steady state conditions. In the numerical simulations the initial set of vessels is supposed, at all time instants, to have a lower permeability with respect to the ones generated during angiogenesis. This is in
	accordance with the biological observation that tumour-induced vessels are highly leaky, due to a weakening of the junctions within the endothelial cells that leads to an increase in the permeability of the blood vessel \cite{Nagy2008} both to fluids and to oxygen.
In particular we suppose
\begin{equation}
\beta_p(\bm{x})=\begin{cases}\beta_p^0 &\forall \bm{x}\in \Lambda^0\\
r_p^\beta\beta_p^0 &\forall \bm{x}\in \Lambda^k\setminus\Lambda^0,~\forall k =0,...,K 
\end{cases}
\end{equation}
and 
\begin{equation}
\beta_c(\bm{x})=\begin{cases}\beta_c^0 &\forall \bm{x}\in \Lambda^0\\
r_c^\beta\beta_c^0 &\forall \bm{x}\in \Lambda^k\setminus\Lambda^0,~\forall k =0,...,K .
\end{cases}\label{betac}
\end{equation}
with $r_p^\beta,r_c^\beta\geq1$. In the default set of parameters we consider $r_p^\beta=100$ and $r_c^\beta=10$. The fact that the permeability of vessels to fluids increases by two order of magnitude in tumour-induced vessels is in agreement with \cite{cattzun0}. Different choices of the ratio $r_c^\beta$ will instead be discussed in the sensitivity analysis carried out in this section.
	Finally, we suppose that only stalk and tip cells can bind VEGF, so that
$$\tilde{\sigma}(\bm{x})=\begin{cases}0 &\forall \bm{x}\in \Lambda^0,\\
\tilde{\sigma} &\forall \bm{x}\in \Lambda^k\setminus\Lambda^0,~\forall k =0,...,K 
\end{cases}$$
\subsection{\textit{TestFace}} 
\begin{figure}
	\centering	
	\includegraphics[width=0.44\linewidth]{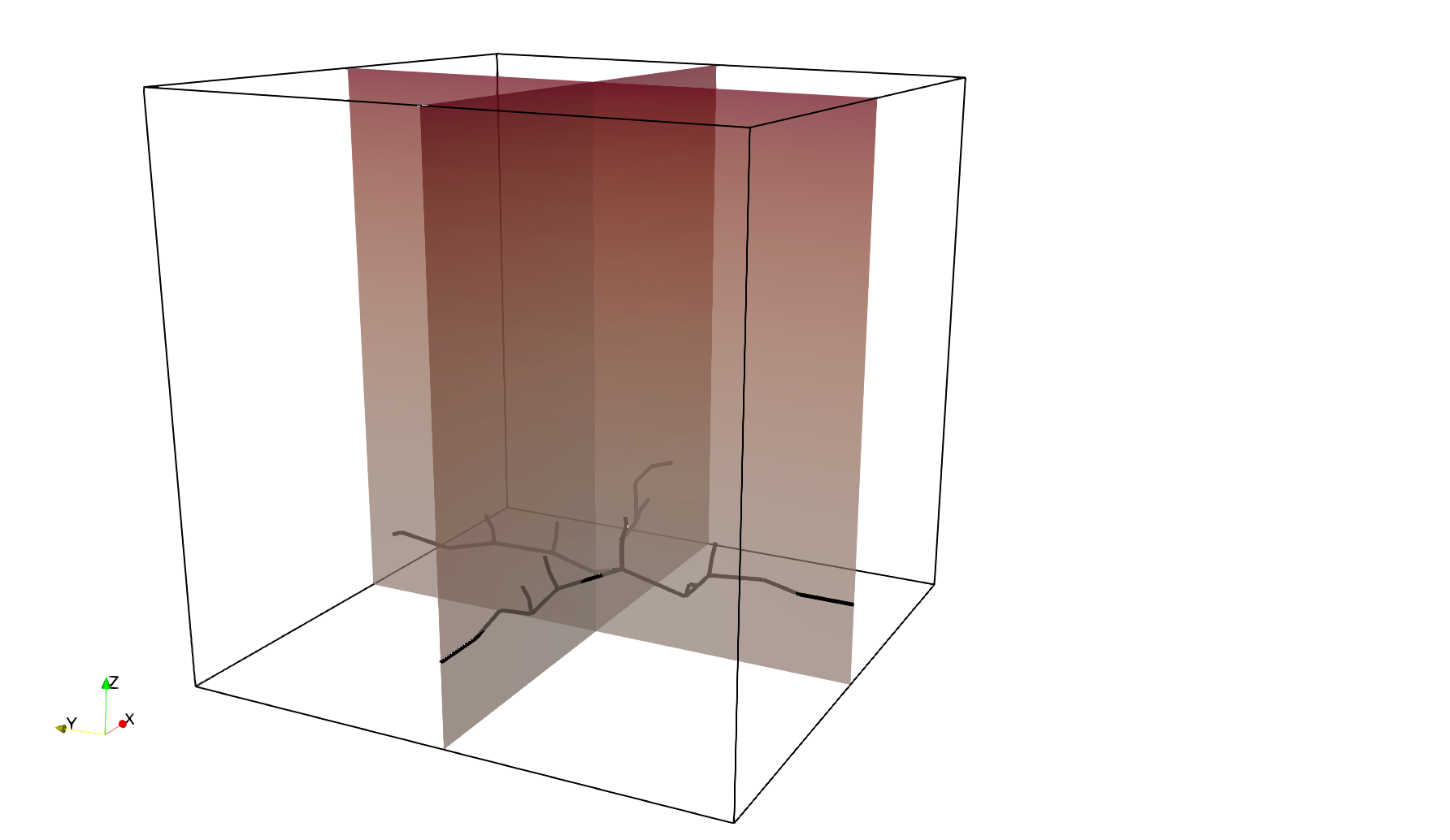}%
	\hspace{-2.4cm}\includegraphics[width=0.44\linewidth]{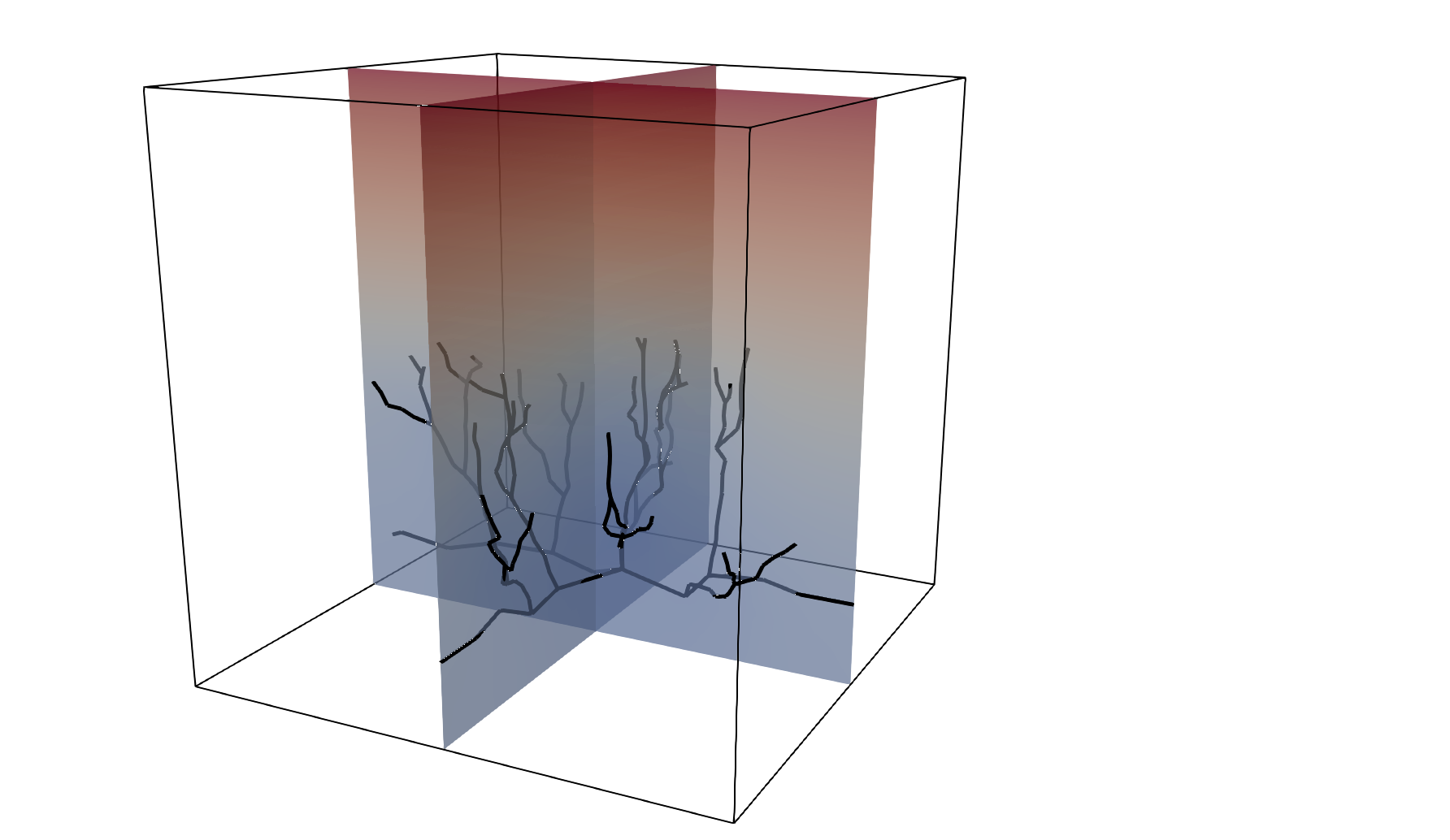}%
	\hspace{-2.4cm}\includegraphics[width=0.44\linewidth]{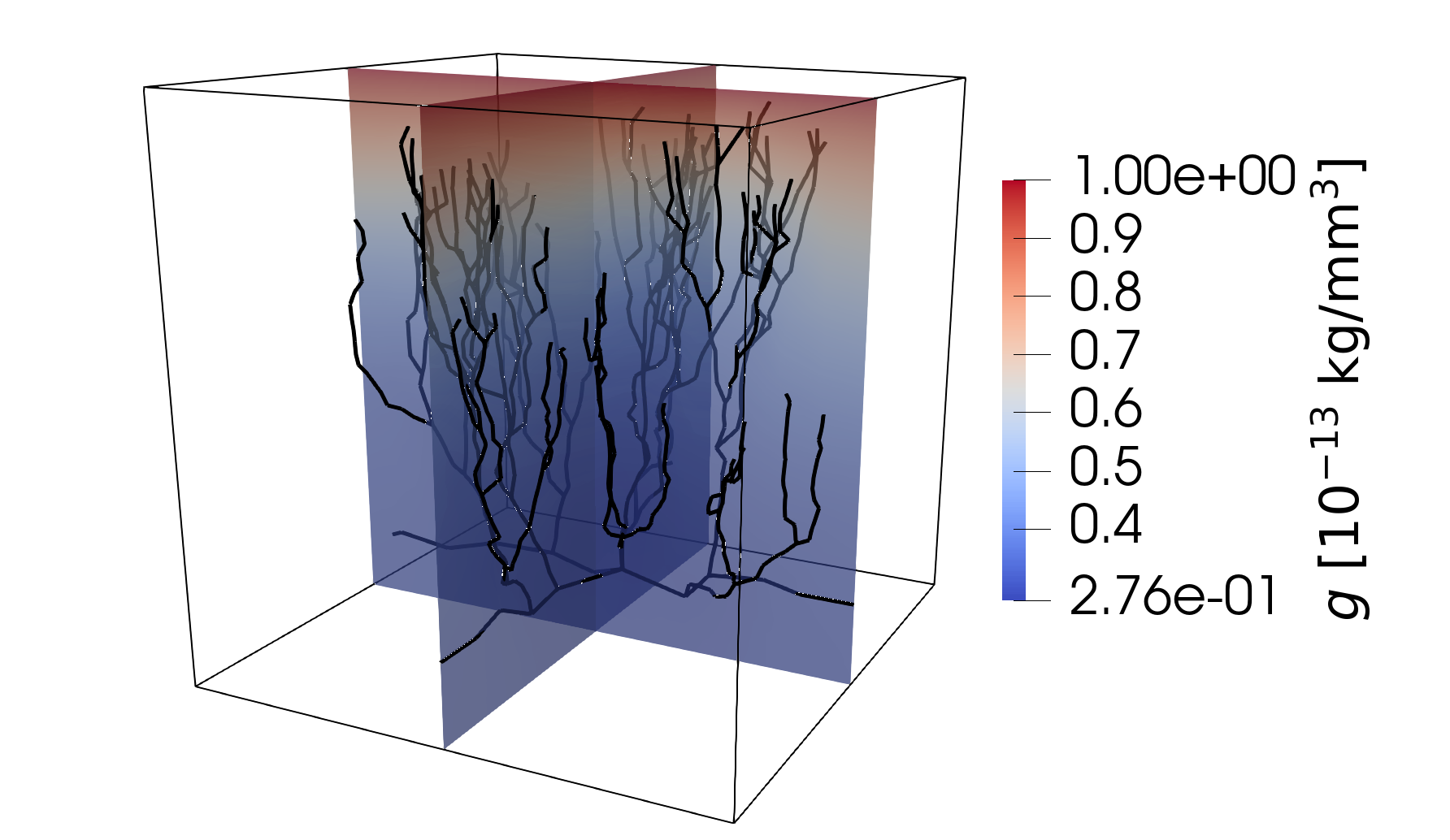}%
	\vspace{0.5cm}
	\includegraphics[width=0.44\linewidth]{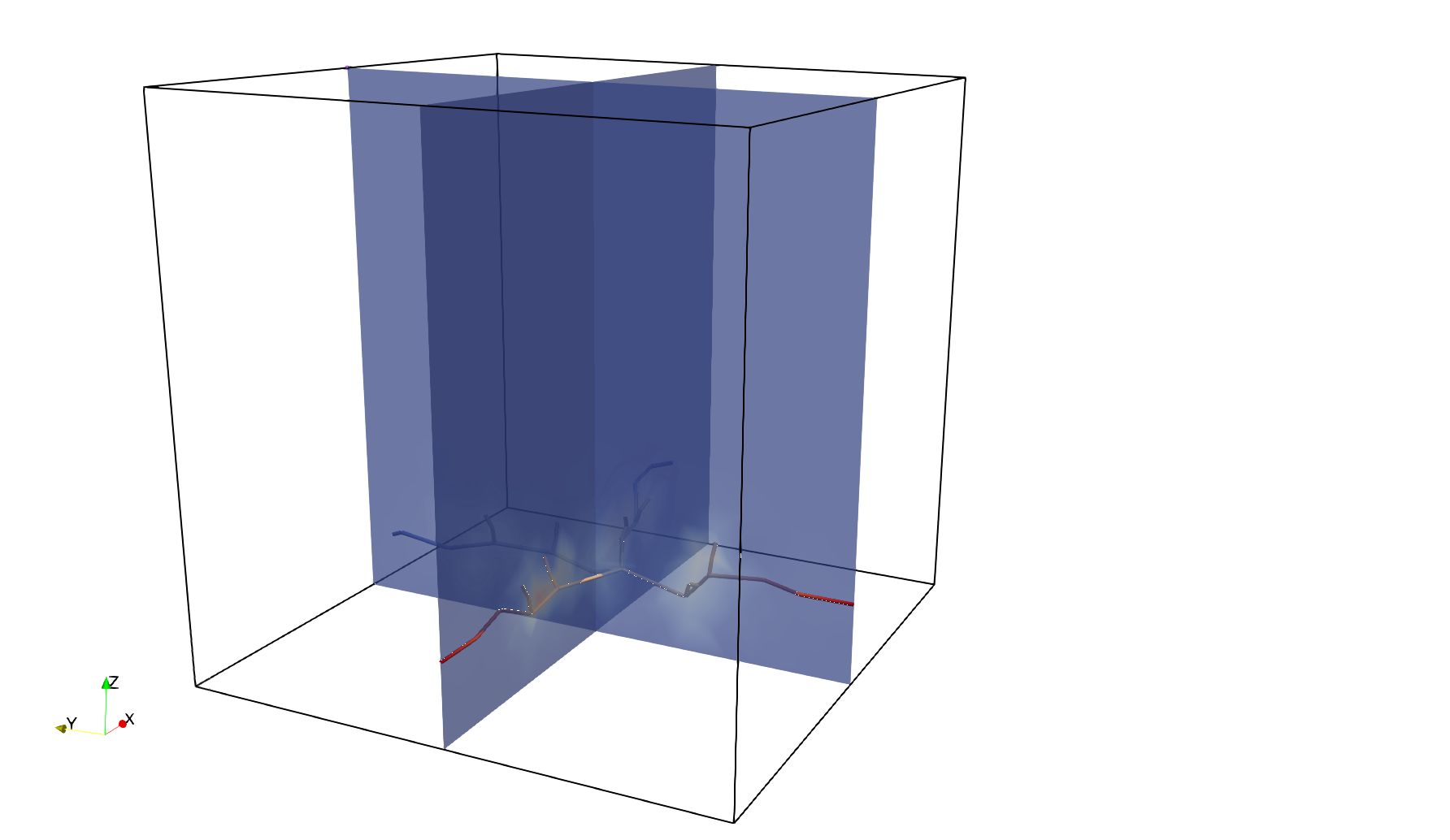}%
	\hspace{-2.4cm}\includegraphics[width=0.44\linewidth]{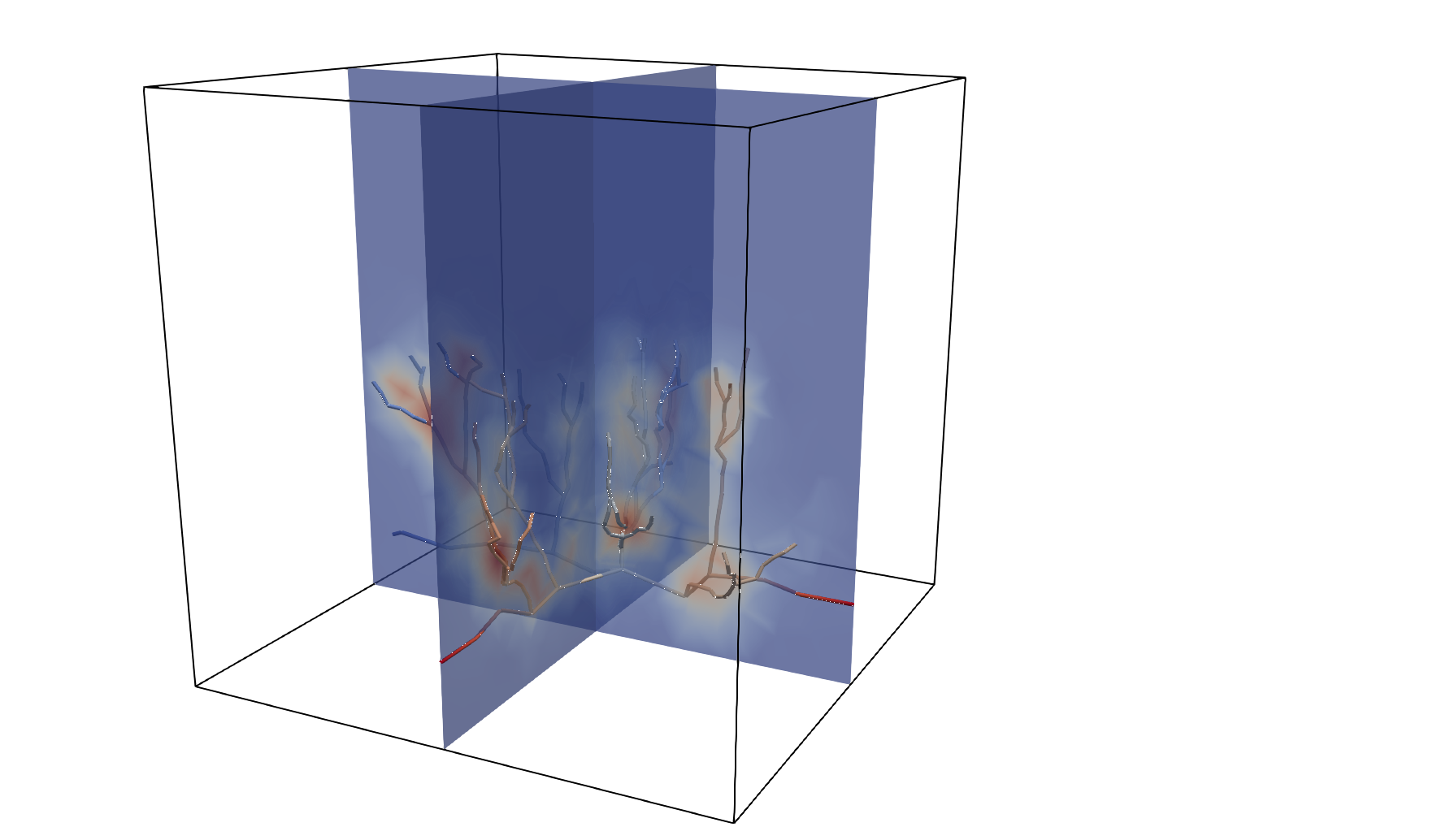}%
	\hspace{-2.4cm}\includegraphics[width=0.44\linewidth]{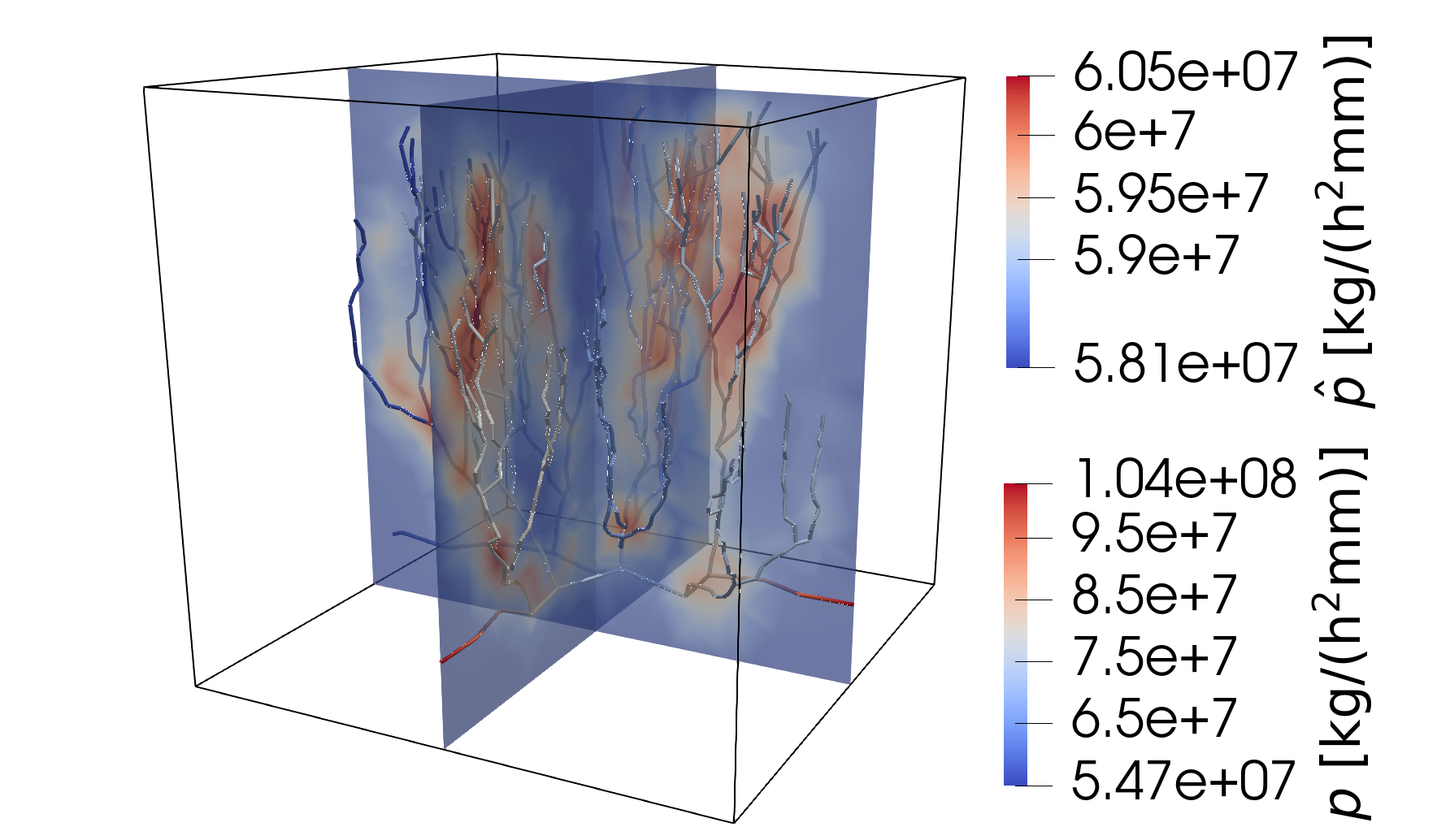}%
	\vspace{0.5cm}
	\includegraphics[width=0.44\linewidth]{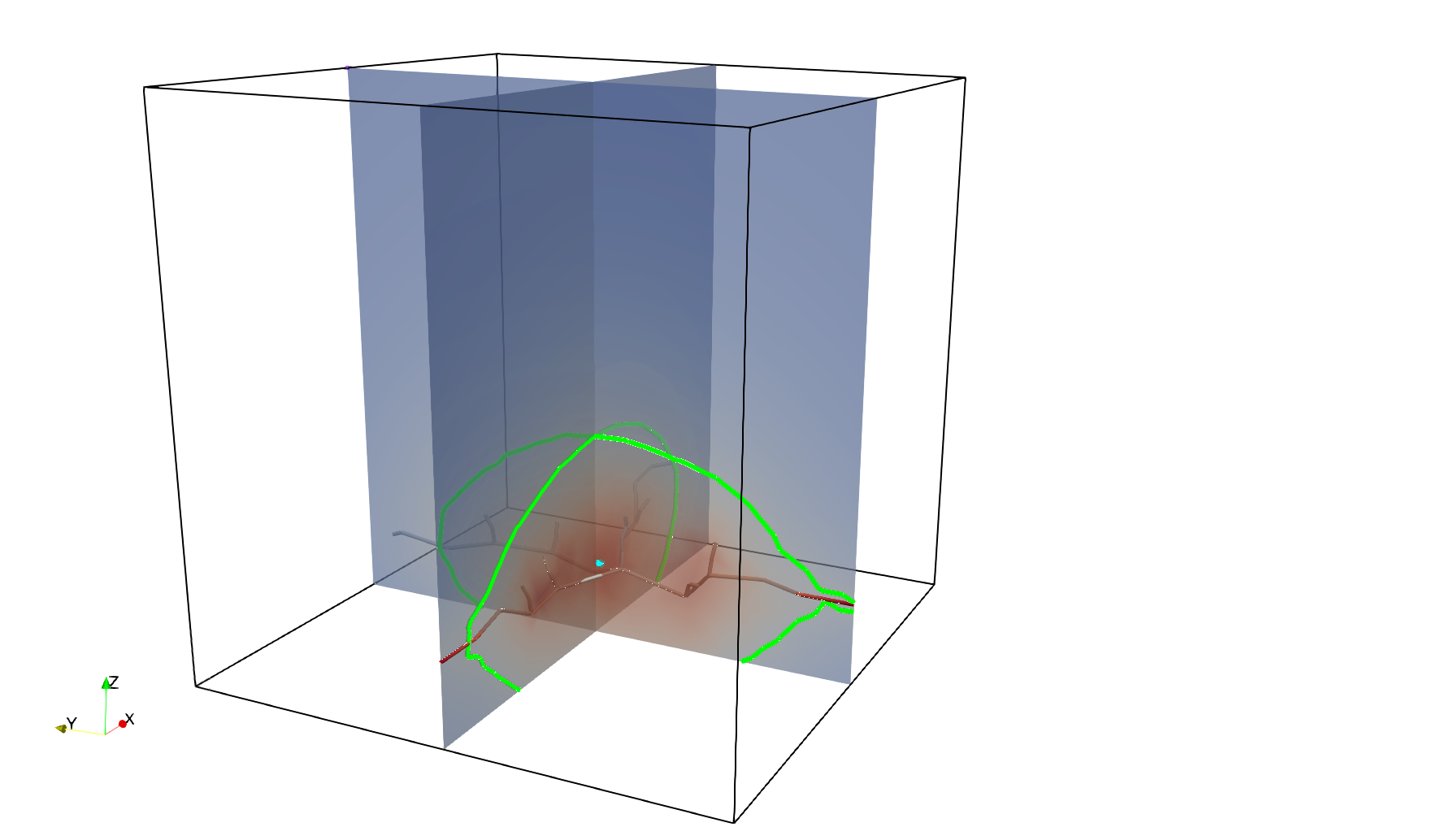}%
	\hspace{-2.4cm}\includegraphics[width=0.44\linewidth]{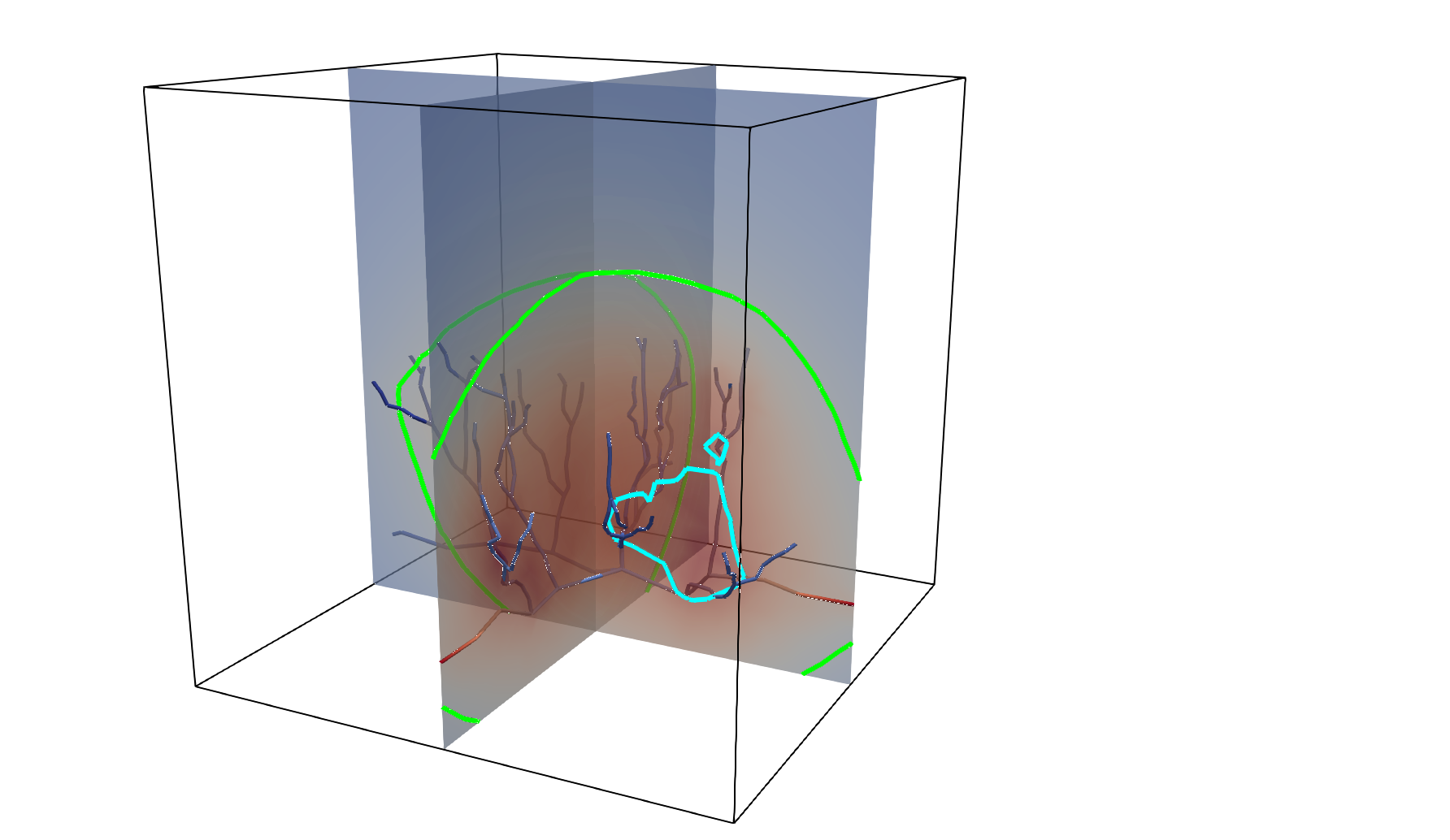}%
	\hspace{-2.4cm}\includegraphics[width=0.44\linewidth]{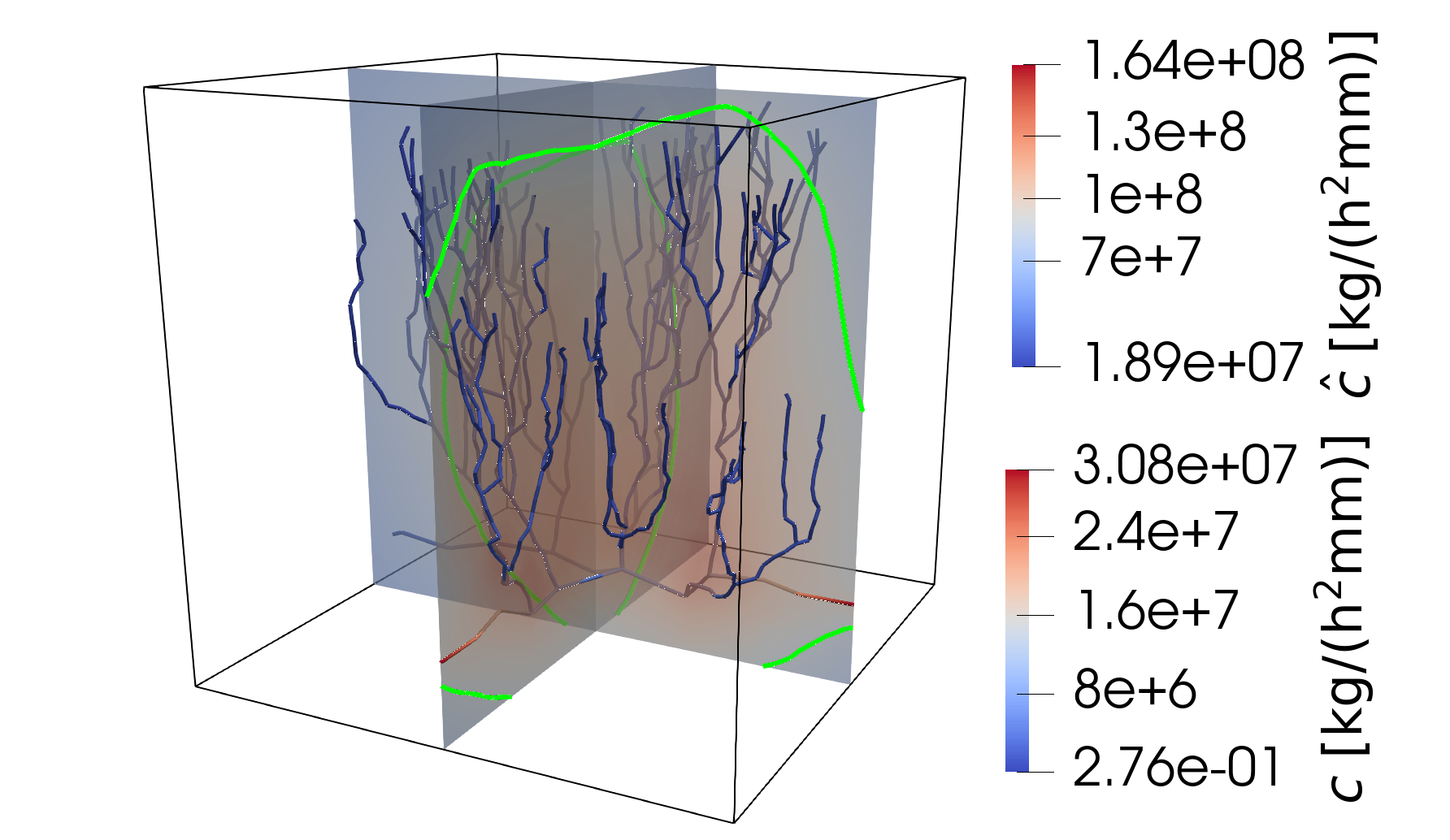}
	\caption{\textit{TestFace}: From left to right, distributions at time $t=2,7,14$ days of VEGF concentration (top), pressure (center), oxygen concentration (bottom), with highlighted isolines corresponding to $c= 8 \rm mmHg$ (green) and $c= 15 \rm mmHg$ (cyan). Parameters set from the tables.}
	\label{DefTest}
\end{figure}

Let us consider a cubic domain $\Omega=(0, L)^3$, $L=0.5 \rm mm$. The interface between the tumor and the tissue sample corresponds to the top face of $\Omega$, highlighted in red in Figure \ref{initialconfig}, while the blood inlet and outlet points on the initial network are marked in red and blue respectively.
For the time discretization we choose a step $\Delta t=\Deltaik=12 h$ $\forall k$, while a maximum volume of $1\cdot 10^{-5}~ \rm mm^3$ is considered for the tetrahedra of the space discretization. 
Figure \ref{DefTest} shows, from top to bottom, VEGF, pressure and oxygen distribution at time $t=2, ~7, ~14$ days obtained with the parameters in Tables~\ref{table_geom}, \ref{table_pr}, \ref{table_oxy} and \ref{table_VEGF}. The time scale of the numerical simulations is qualitatively in agreement with the biological observations reported in \cite{Gimbrone, Ausprunk, Muthukkaruppan}, where it takes approximately 10 to 21 days for the growing network to link the tumour to the parent vessel. For what concerns oxygen, the isolines corresponding to $c=1.38\cdot10^7 \rm kg/(h^2mm) \approx 8 \rm mmHg$ (green) and $c=2.60\cdot10^7 \rm kg/(h^2mm) \approx 15 \rm mmHg$ (cyan) are highlighted, representing, according to \cite{normoxia}, the pathological and physiological hypoxia thresholds. We observe that, starting from a condition in which the whole tissue is highly hypoxic, the formation of new vessels brings oxygen levels above the pathological hypoxia threshold. Around half the simulation time, concentrations higher than $15~\rm mmHg$ are registered. However the maximum decreases towards the end of the simulation as an effect of the exchanges through the boundaries, since an external concentration $c_{ext}=3.5\rm{~mmHg}$ is imposed. 
This choice, together with the vessel wall permeability to oxygen $\beta_c$, influences of course the results.

\begin{figure}
\centering
	\includegraphics[width=0.34\linewidth]{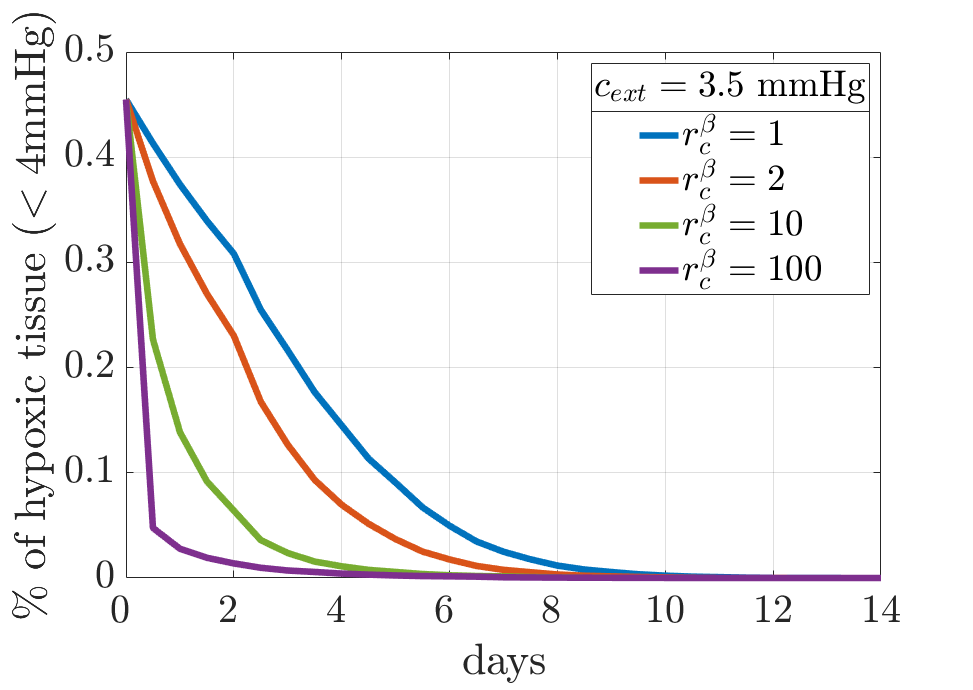}%
	\includegraphics[width=0.34\linewidth]{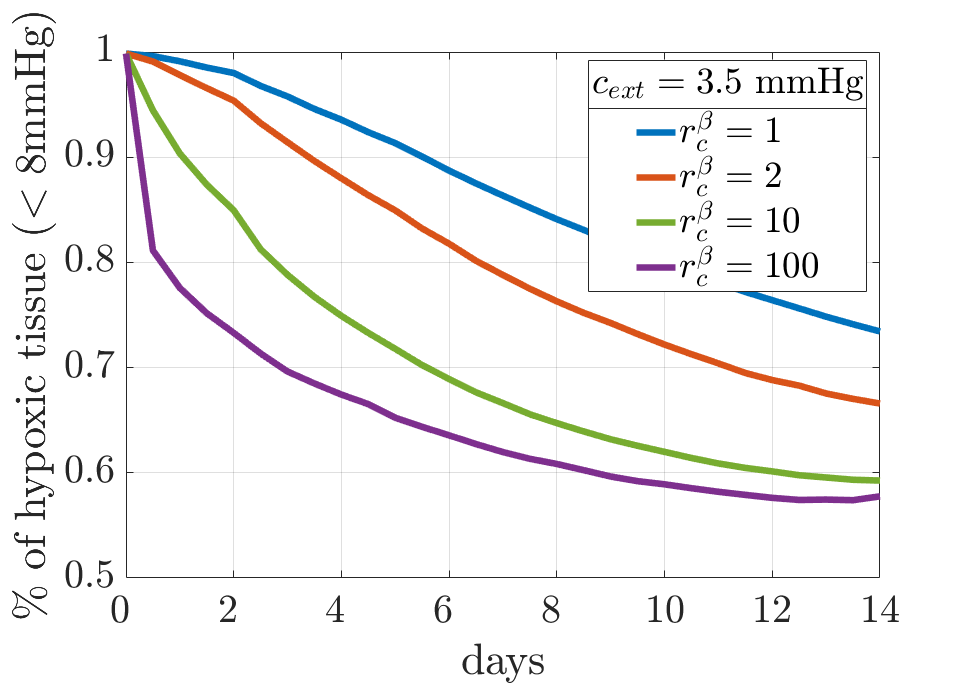}%
	\includegraphics[width=0.34\linewidth]{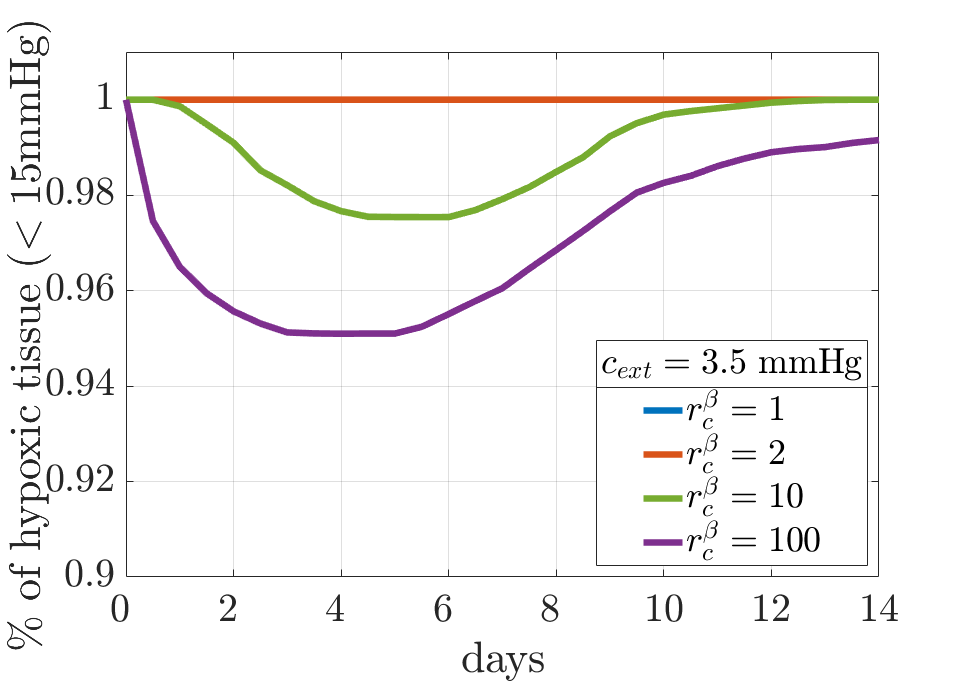}
	\caption{\textit{TestFace}: Percentage of hypoxic tissue under the variation of $r_c^\beta$ (see \eqref{betac}). From left to right hypoxia levels at $4,8,15\rm~mmHg$ respectively, $c_{ext}=6.05\cdot 10^{6} \rm kg/(h^2mm)~ (=3.5mmHg)$ for all the three cases.}
	\label{oxyperm_3_5}
\end{figure}
\begin{figure}
	\centering
	\includegraphics[width=0.34\linewidth]{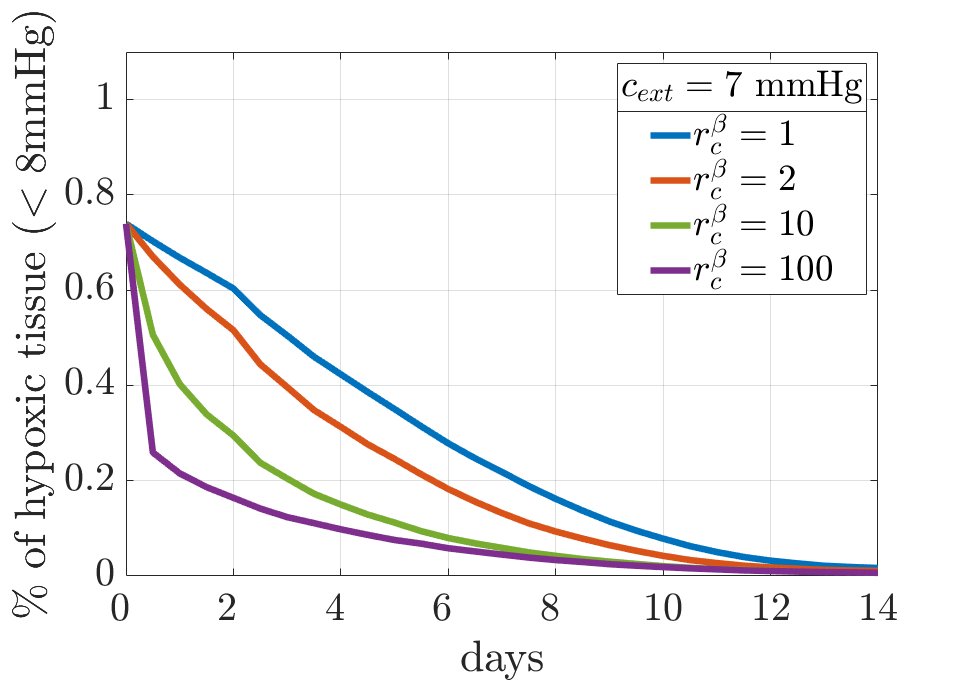}%
	\includegraphics[width=0.34\linewidth]{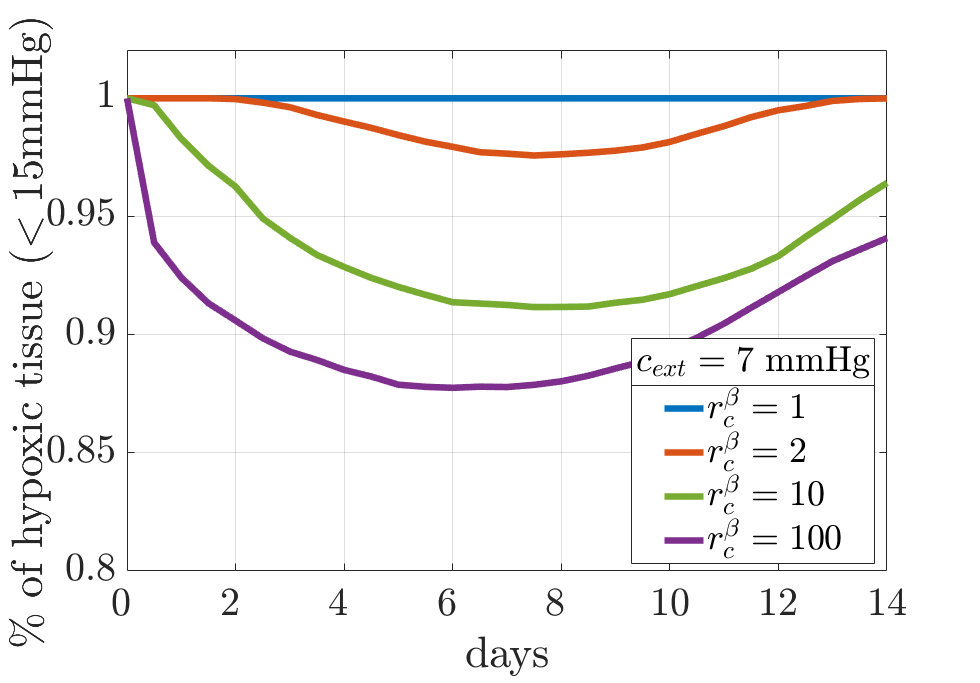}
	\caption{\textit{TestFace}: Percentage of hypoxic tissue under the variation of $r_c^\beta$ (see \eqref{betac}). On the left hypoxia level at $8\rm~mmHg$, on the right at $15\rm~mmHg$; $c_{ext}=1.21\cdot 10^{7} \rm kg/(h^2mm)~ (=7mmHg)$ in both cases.}
	\label{oxyperm_7}
\end{figure}
An analysis of the impact of such parameters is reported in Figures \ref{oxyperm_3_5} and \ref{oxyperm_7}, where the sensitivity of the percentage of hypoxic tissue to changes of $r_c^\beta$ and $c_{ext}$ is analyzed. For all the plotted curves we consider $\beta_0^c=12.6 ~\rm mm/h$ while we increment the permeability for the vessels generated by angiogenesis as $r_c^\beta\beta_c^0$. In Figure \ref{oxyperm_3_5} we set $c_{ext}=3.5~\rm mmHg$ and we analyse the percentage of tissue below 4, 8 and 15 mmHg of oxygen concentration. As expected, an increment of the permeability promotes a faster tissue oxygenation. For all the considered values of $r_c^\beta$ the tissue sample reaches an oxygen concentration higher than 4 mmHg within the simulation time (Figure \ref{oxyperm_3_5}-left), while the percentage of tissue below 8 mmHg still ranges between 57\% (for the highest value of $r_c^\beta$) and 73\% (for the lowest value of $r_c^\beta$) after 14 days (Figure \ref{oxyperm_3_5}-center). For what concerns the physiological hypoxia level (15 mmHg) we can see that for too low values of $r_c^\beta$ the threshold is never reached in the sample. As aforementioned, a small amount of tissue (3-5\%) goes above 15 mmHg if we consider higher permeability values, but the trend with time of the hypoxic portion is not monotonically decreasing as an effect of the exchanges through the boundaries. Therefore, in Figure \ref{oxyperm_7} we consider the trend with time of the percentage of tissue below 8 and 15 mmHg (left and right plots respectively) in the case $c_{ext}=7~\rm mmHg$. As expected, since the external concentration is higher, better oxygenation levels are reached into the tissue sample. Let us remark, in Figure \ref{oxyperm_7}-right, that about 3\% of the tissue sample goes above the physiological hypoxia level already with $r^\beta_c=2$. However the trend of the hypoxic portion is again non monotonically decreasing, due to the exchange of oxygen with the surrounding tissue, maintained in an hypoxic state through the imposed boundary conditions. 

\begin{figure}
	\centering
	\includegraphics[width=0.48\linewidth]{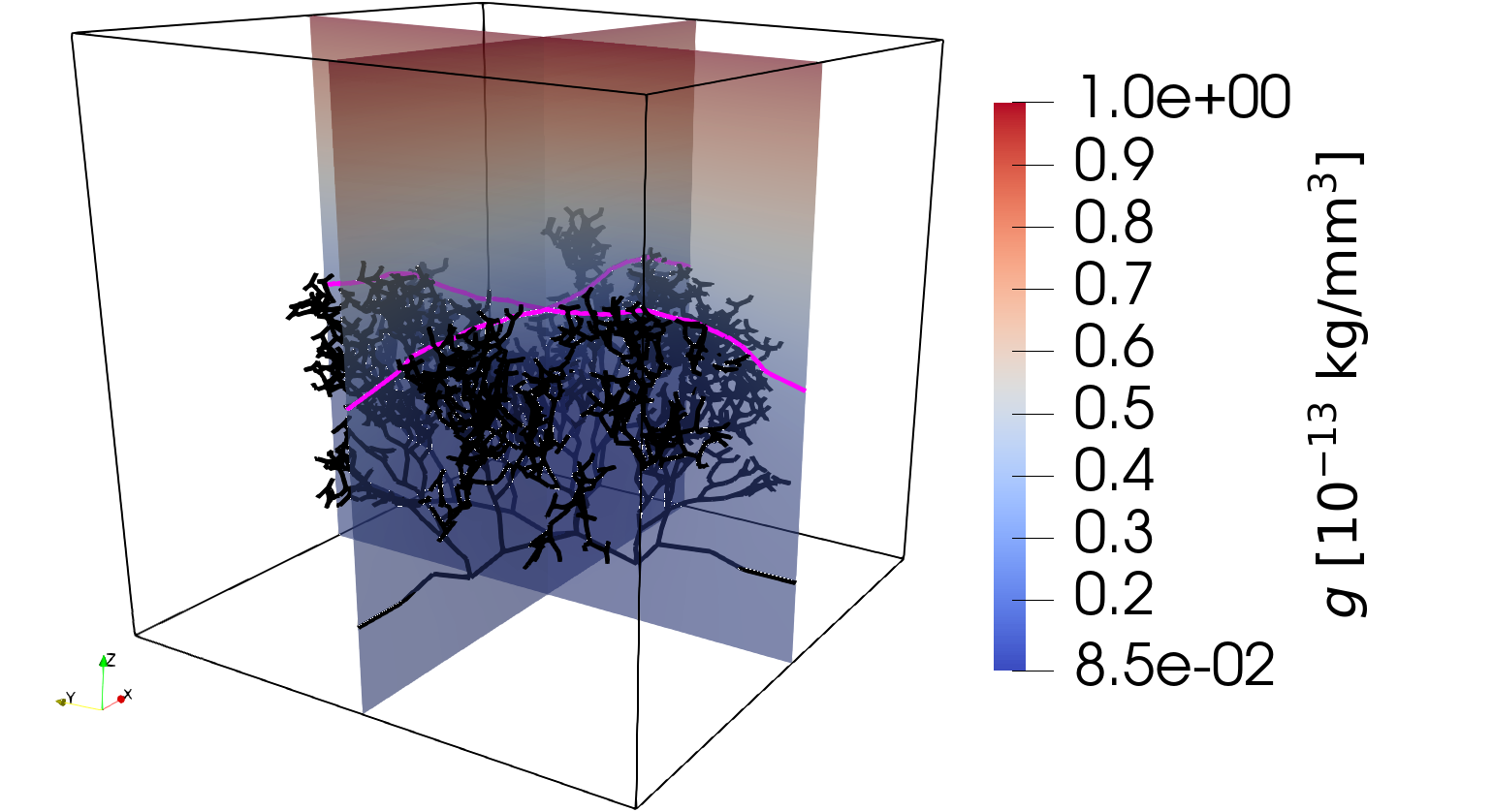}%
	\includegraphics[width=0.48\linewidth]{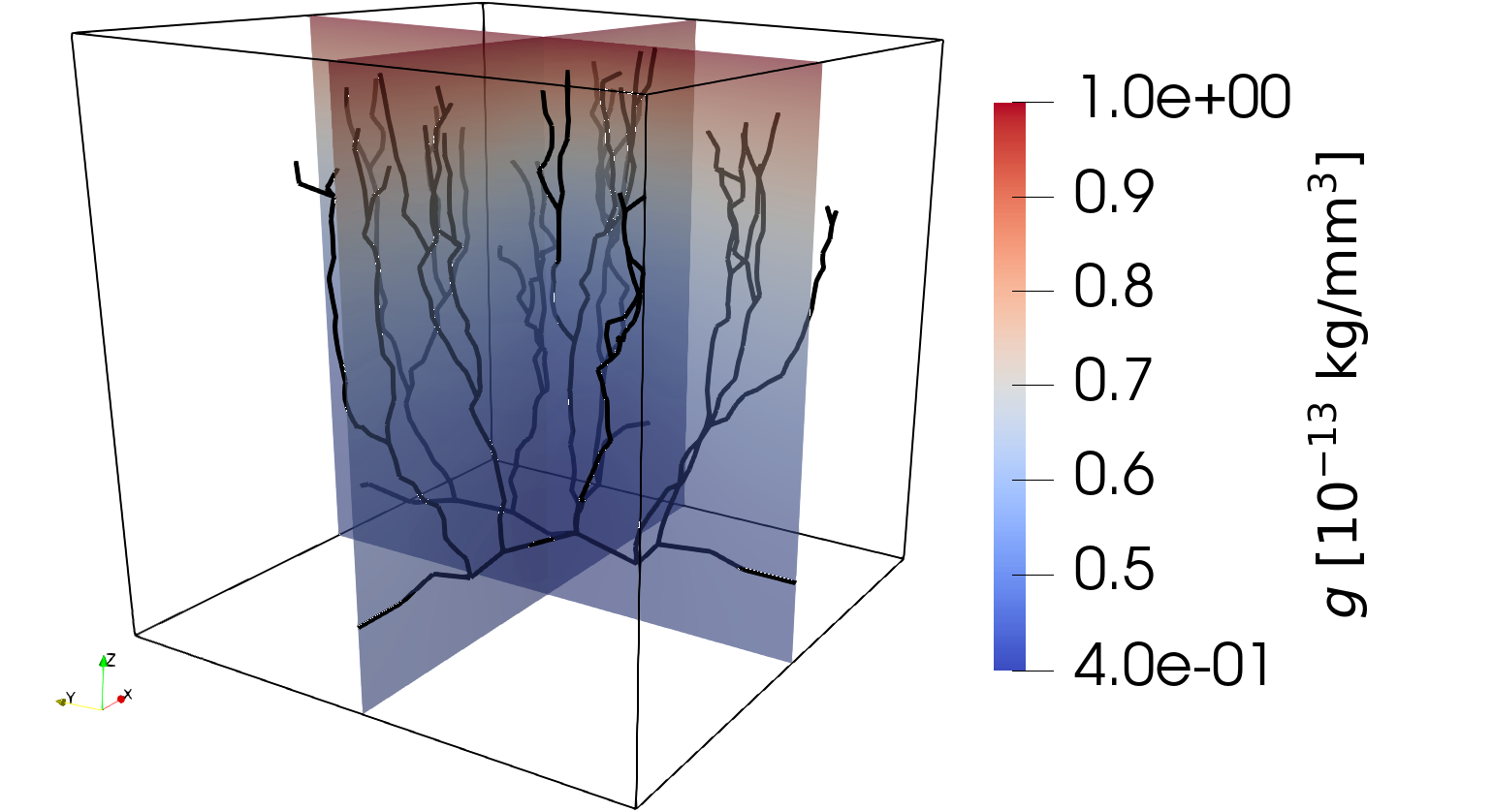}\vspace{0.1cm}
	\includegraphics[width=0.48\linewidth]{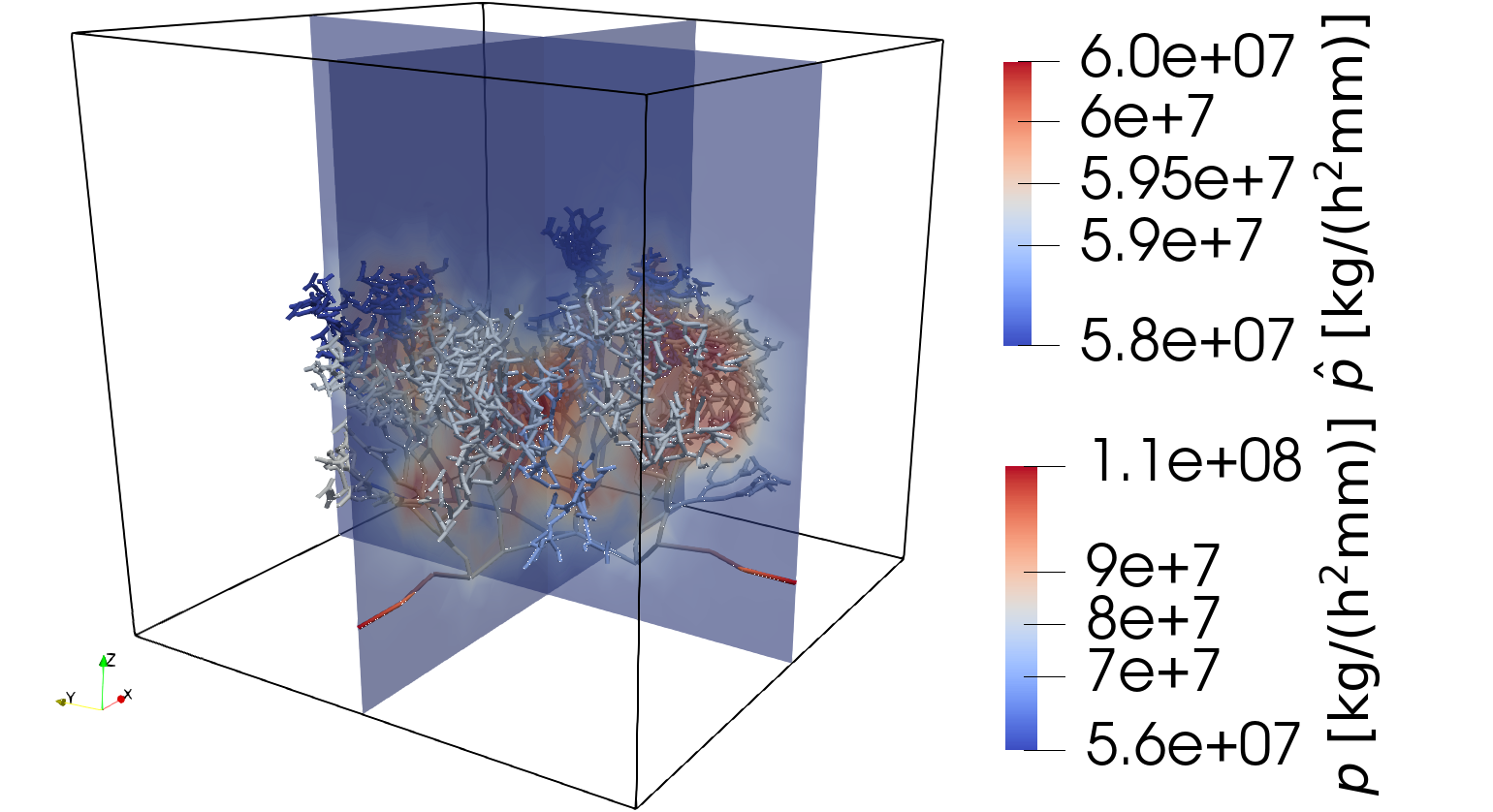}%
	\includegraphics[width=0.48\linewidth]{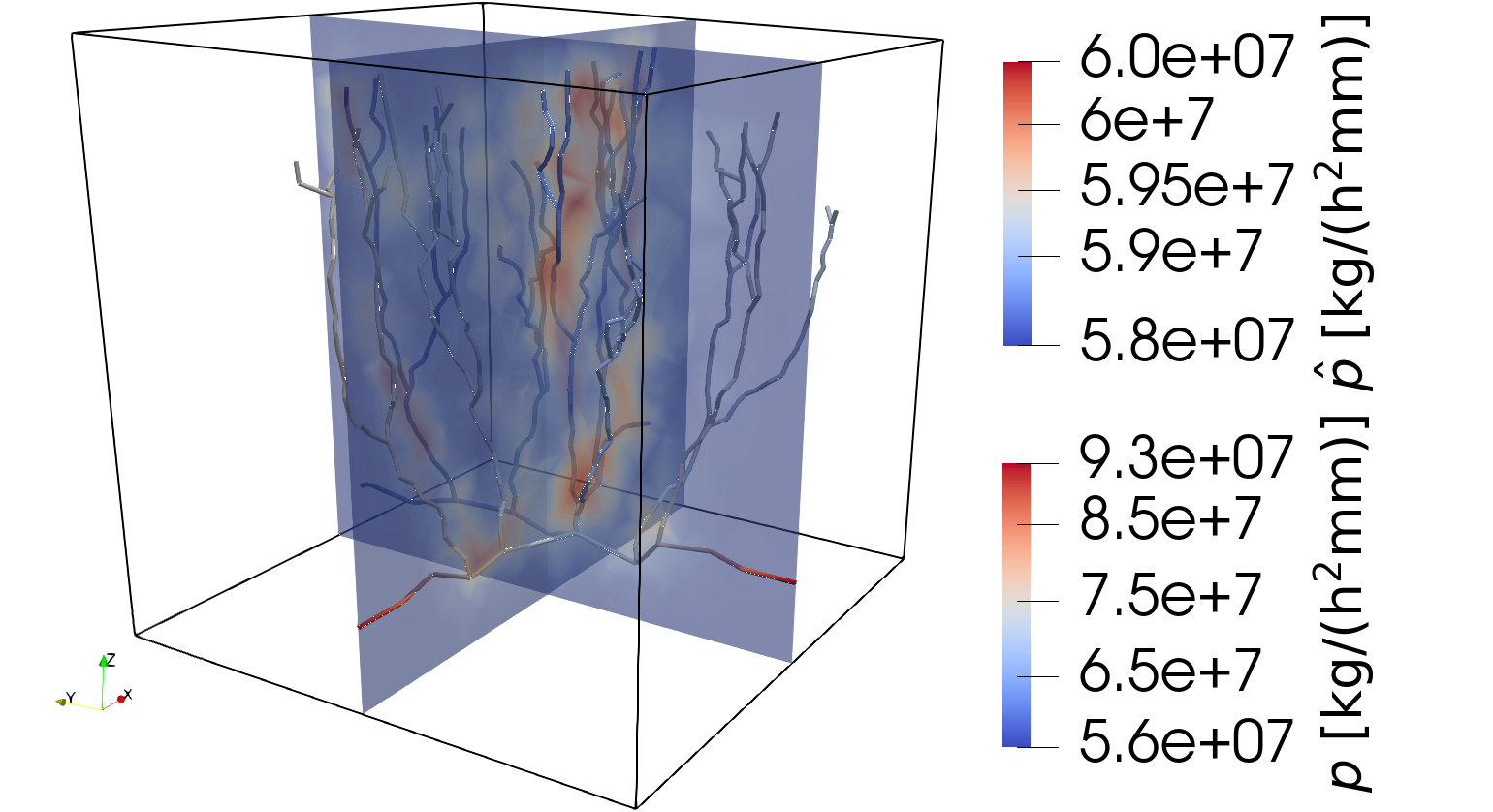}\vspace{0.1cm}
	\includegraphics[width=0.48\linewidth]{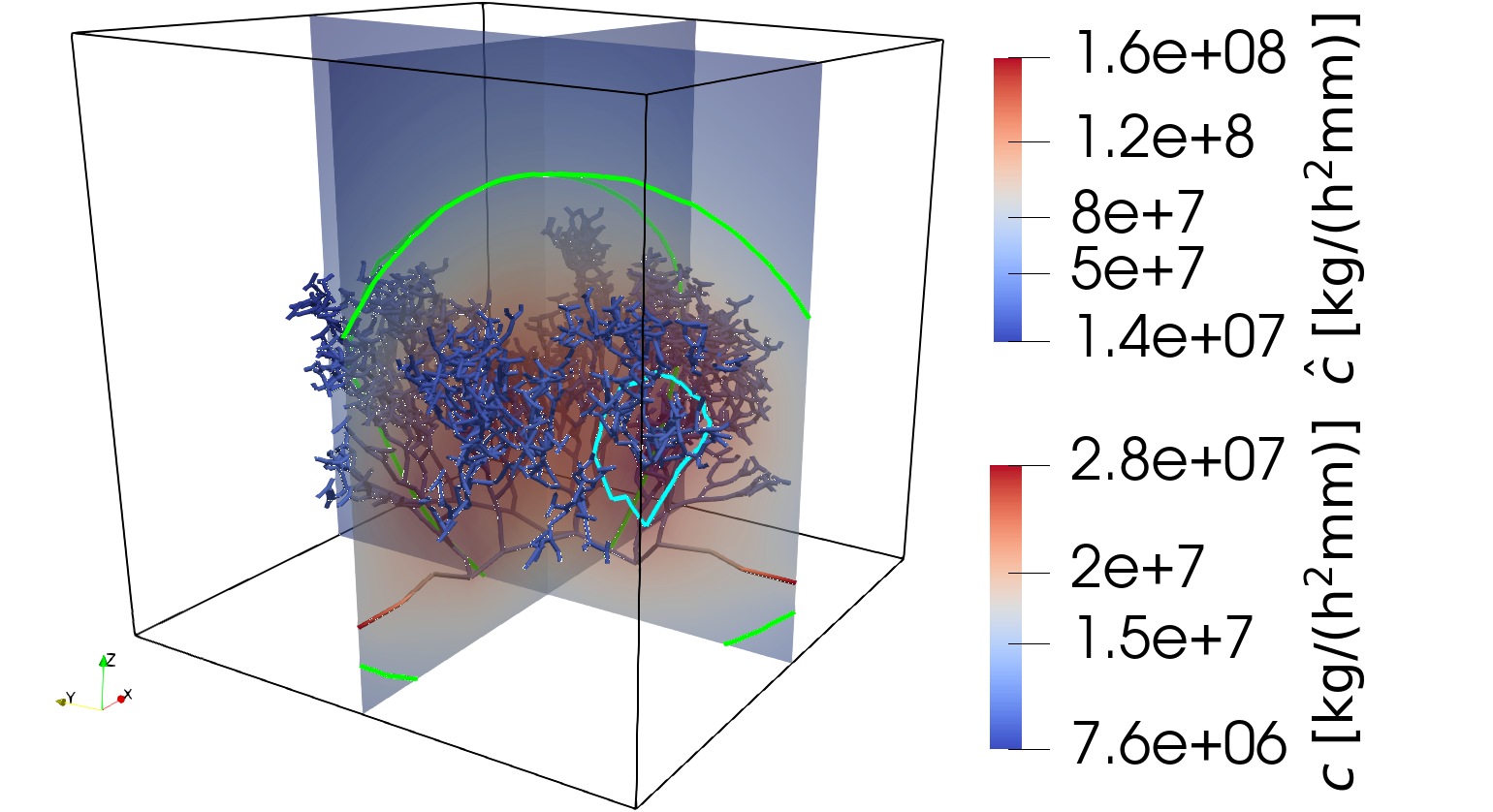}%
	\includegraphics[width=0.48\linewidth]{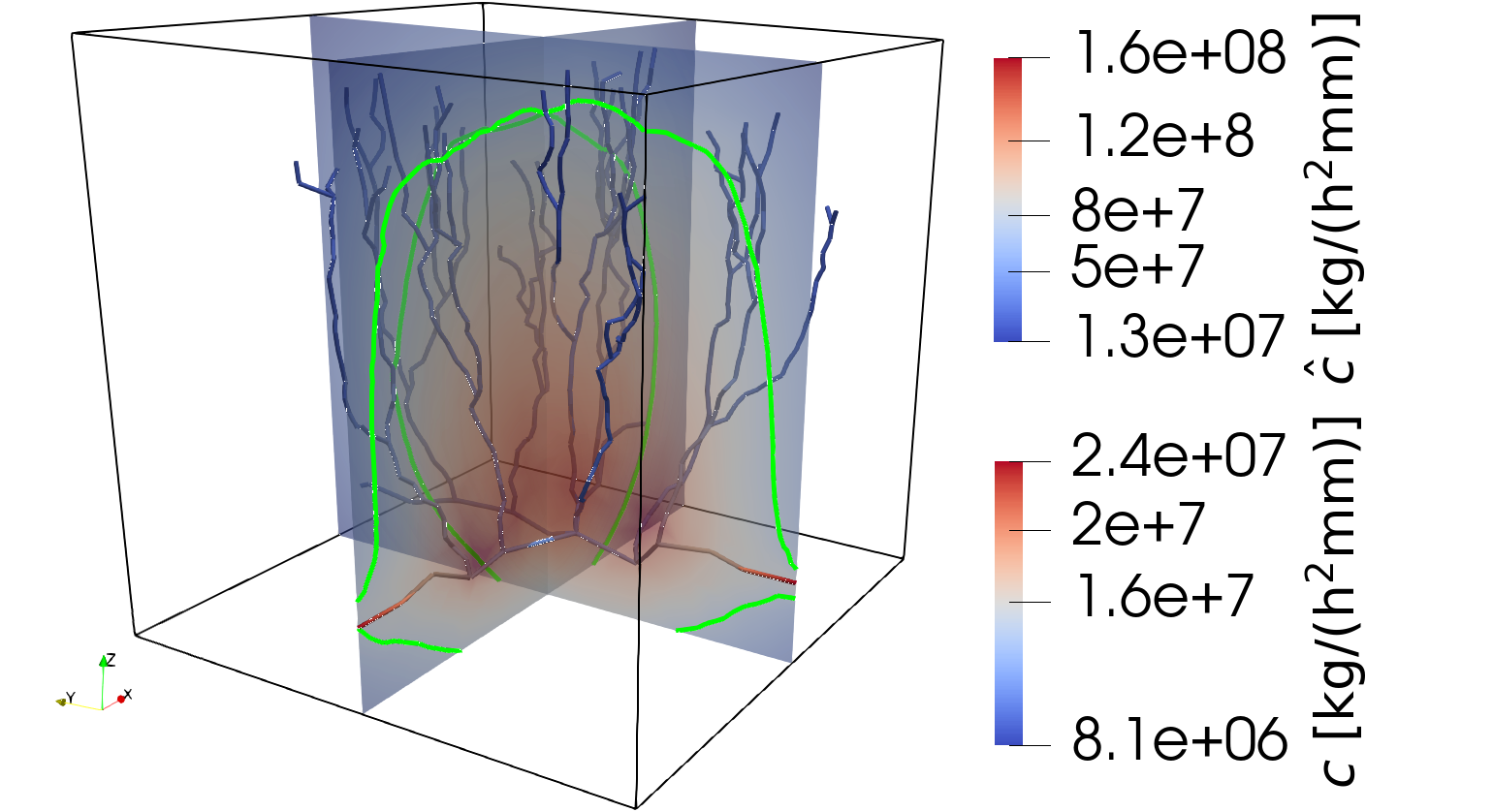}
	\caption{\textit{TestFace}: From top to bottom, VEGF concentration (in magenta isoline corresponding to $g=g_{lim}=2.5\cdot 10^{-14}~\rm\frac{kg}{mm^3}$), fluid pressure and oxygen concentration (in green and cyan isolines corresponding to $8 \rm mmHg$ and $15 \rm mmHg$, respectively) for the \emph{early branching} (left) and the \emph{late branching} (right) cases.}
	\label{br3D}
\end{figure}

The effectiveness of oxygenation can be related also to the structure of the vascular network, in particular to the presence of more or less branches and to the velocity of formation of new vessels. Specifically, we consider two opposite conditions, namely the \emph{early branching} and the \emph{late branching} cases. The early branching is obtained by forcing abnormally the fast creation of new branches also at low concentrations of VEGF, i.e. by setting $\tau_{br}=24\rm h$ and $P_{br}(g)=1,~\forall g$. The late branching setting is instead obtained by increasing the threshold age for branching to $\tau_{br}=96\rm h$, and by setting $g_{br}=2\cdot 10^{-13}\rm kg/mm^3$ in the branching probability definition \eqref{br_prob}. The morphology of the resulting vascular networks, along with the distribution of VEGF, fluid pressure and oxygen concentration are shown in Figure \ref{br3D}. The results are obtained considering $c_{ext}=3.5~\rm mmHg$ and $r^\beta_c=10$.
\begin{figure}
	\centering
	\includegraphics[width=0.4\linewidth]{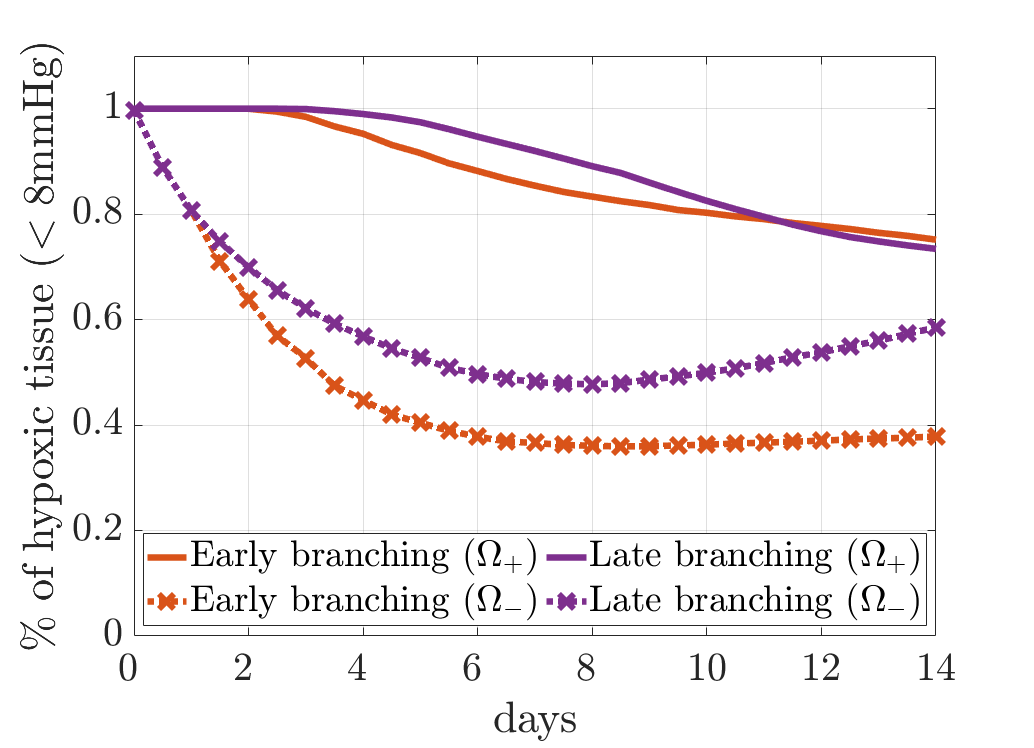}\hspace{0.5cm}%
	\includegraphics[width=0.4\linewidth]{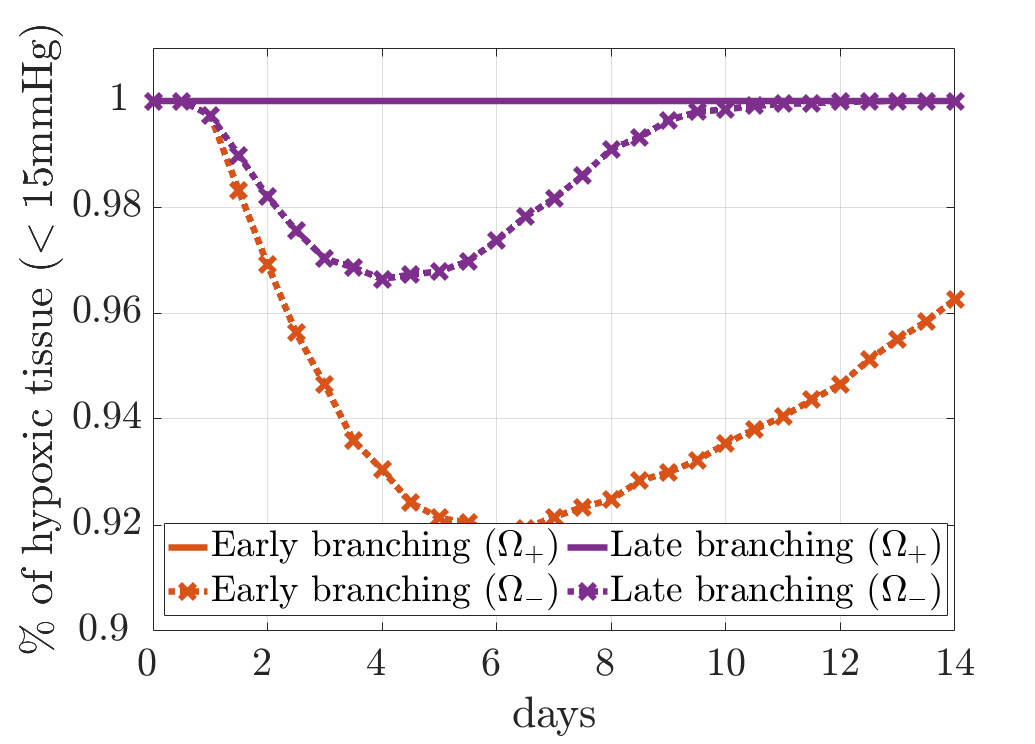}
	\caption{\textit{TestFace}: Variation in time of the percentage of hypoxic tissue in the \textit{early branching} and \textit{late branching} cases. On the left, hypoxia threshold at $8\rm mmHg$, on the right at $15\rm mmHg$.}
	\label{branching}
\end{figure}

Figure \ref{branching} shows the trend with time of the volume percentage in which the oxygen concentration is below $8 ~\rm mmHg$ (Figure \ref{branching}-left) and $15 ~\rm mmHg$ (Figure \ref{branching}-right).
For this analysis the volume has been split into two parts: $\Omega_{+}=\lbrace \bm{x}=(x,y,z)\in \Omega~\text{s.t. }z>\frac{L}{2}\rbrace$ and $\Omega_{-}=\lbrace \bm{x}=(x,y,z)\in \Omega~\text{s.t. }z\leq\frac{L}{2}\rbrace$, in order to account for oxygenation levels far and close to the tumor boundary. We can observe how fast and early branching does not produce a better oxygenation of $\Omega_{+}$, i.e. of the tissue portion which is closer to the tumor. In fact, as it can be seen in Figure \ref{br3D} on the left column, the early branching network is very dense but not much extended in the $z$ direction. This produces an efficient oxygenation of the lowest part of the domain and a consequent exchange through the boundaries, while the oxygenation level in the highest part tends to be very similar to the one obtained in the late branching case. We remark that this result is highly influenced by the boundary condition imposed at the tumour interface and future studies will certainly look at the description of tumour oxygenation. For both the configurations, the physiological hypoxia level (15 mmHg) is never reached in $\Omega_{+}$, as shown by the perfectly overlapped full lines in Figure \ref{branching}-right. 
 However, we remark that even if the hypoxic condition persists in the tumour region, it is rather a gain for the cancer. Indeed cancer cells are more resistant than healthy cells to lack of oxygen, since they can efficiently switch to an anaerobic metabolism (\emph{Warburg effect}) \cite{Gatenby} and the death of cells surrounding the tumour mass, due to hypoxia, fosters tumour cell invasion.
Finally, it may be worth underlining how the irregularity of the generated vessel network in the early branching configuration possibly hinders the transport and diffusion of anti-cancer drugs to the correct site.
\begin{figure}
	\centering
	\includegraphics[width=0.4\linewidth]{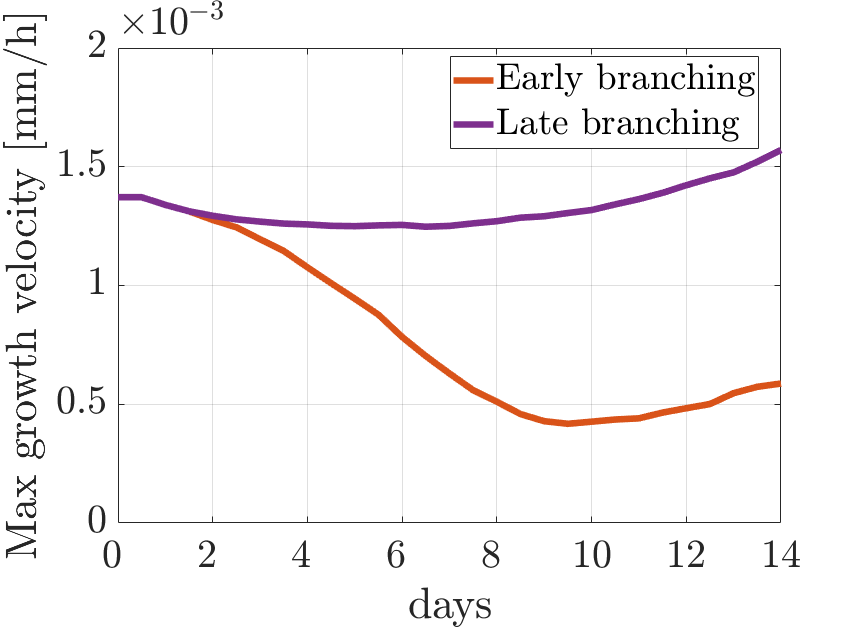}\hspace{0.5cm}%
	\includegraphics[width=0.4\linewidth]{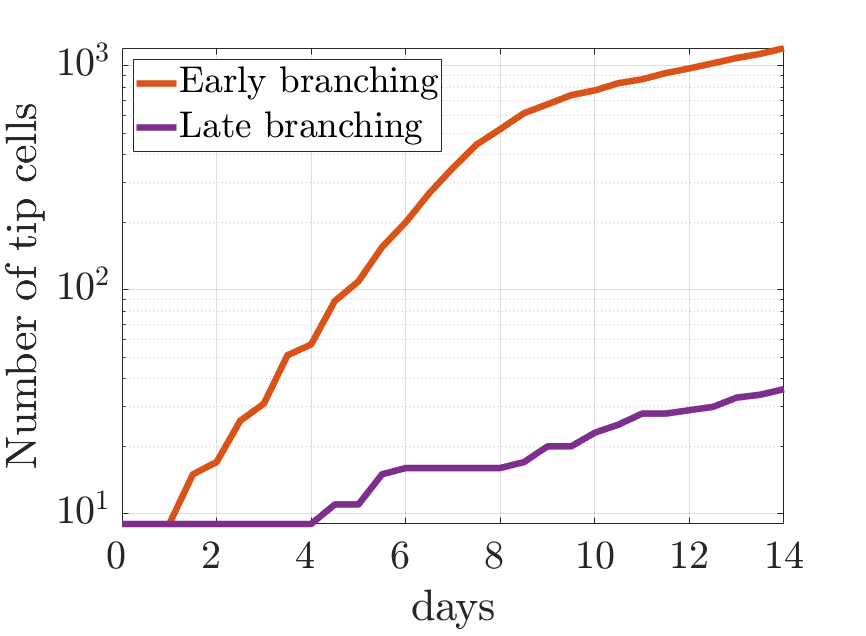}
	\caption{\textit{TestFace}: Variation in time of the maximum growth velocity (on the left) and of the number of tip cells (on the right) for an early and a late branching vascular network.}
	\label{branching2}
\end{figure}

Figure \ref{branching2} reports instead the trend with time of the maximum growth velocity and of the number of tip cells inside the domain both for the early and the late branching configuration. 
 The maximum growth velocity is obtained at each time-step by computing the maximum of the norm of $\bm{w}$ among the tip cells. Both for the early and the late branching configuration, we can observe (see Fig. \ref{branching2}-left) how the growth velocity decreases at the beginning of the simulation to increase again once the vessels approach the tumour region. Indeed, according to Equation \eqref{growth_vel}, the growth velocity is related to the VEGF concentration: in response to the VEGF binding operated by endothelial cells, such concentration goes below its initial minimum in the lowest part of the domain while it is maintained at the maximum level at the tumour boundary, thus explaining the trend of the velocity. This behaviour is particularly evident in the early branching case, where the very dense structure of the capillaries network consumes a high amount of VEGF and the region close to the tumour boundary is never reached.
 Therefore the VEGF concentration goes actually below the threshold for proliferation in a considerable part of the tissue sample, as shown, in Figure \ref{br3D}-top left, by the isoline corresponding to the minimum VEGF concentration required for EC proliferation, i.e., $g=g_{lim}=2.5\cdot 10^{-14}~\rm\frac{kg}{mm^3}$. Finally, as aforementioned, Figure \ref{branching2}-right reports also the trend with time of the number of tip cells which are contemporaneously active inside the domain. As the vascular network approaches the tumour, the number of tip cells and, consequently, the vessel density increase, in accordance with the so-called \emph{brush-border effect} experimentally observed \cite{Muthukkaruppan, Sholley}. The decrease in the rate at which this number increases in the early branching configuration is related to the tips which reach the lateral boundary of $\Omega$ and ideally leave the analyzed tissue sample. 

 \begin{figure}
 	\centering
 	\includegraphics[width=0.32\linewidth]{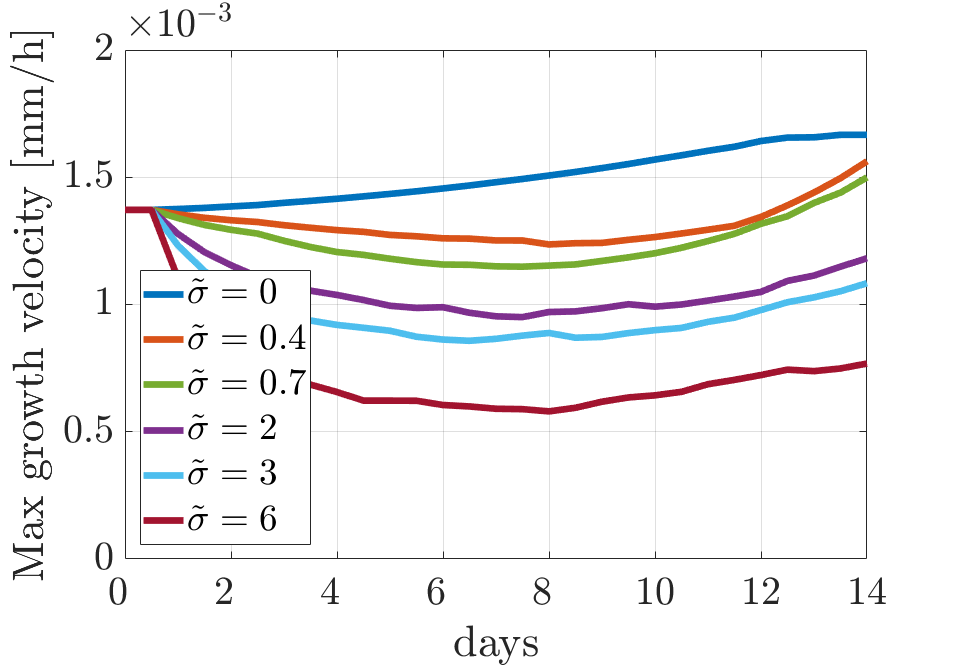}%
 	\includegraphics[width=0.32\linewidth]{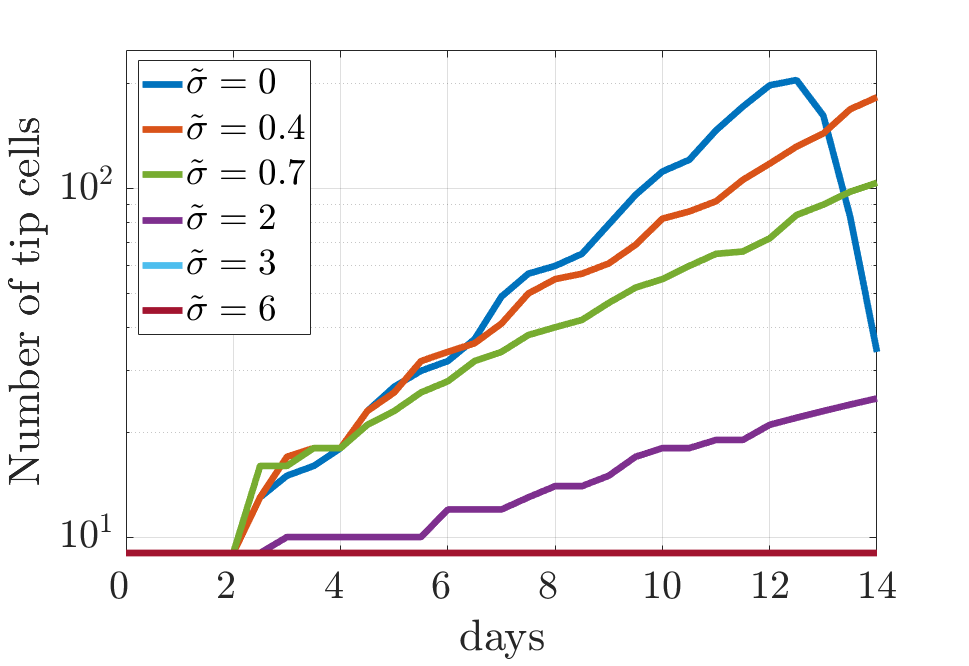}
 	\includegraphics[width=0.32\linewidth]{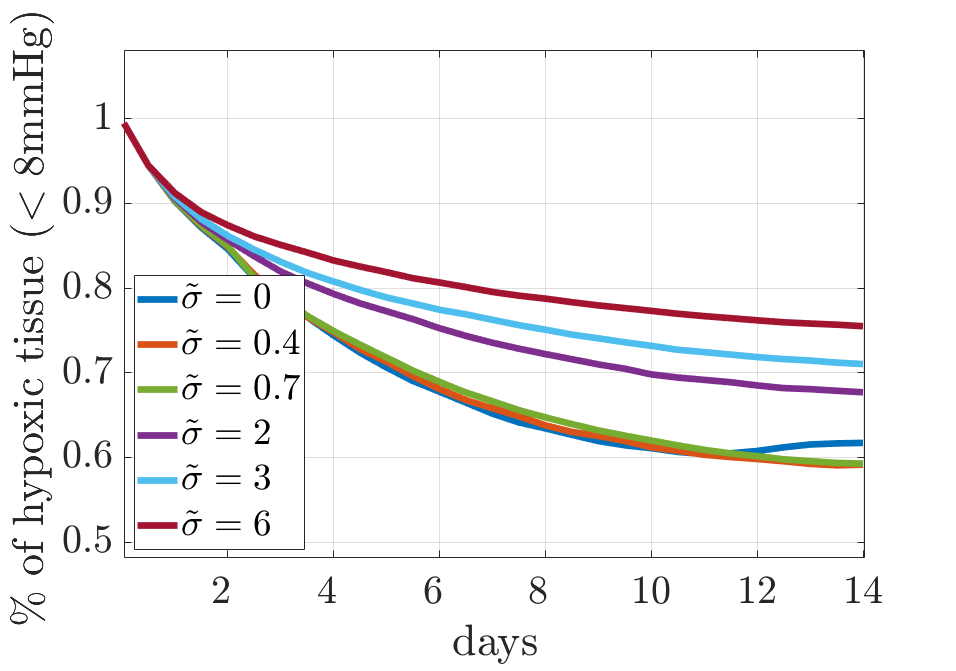}
 	\caption{\textit{TestFace}: From left to right: variation in time of the maximum growth velocity, of the number of tip cells inside the domain and of the percentage of hypoxic tissue for different values of $\tilde{\sigma}~[1/h]$. }
 	\label{consum}
 \end{figure}
 \begin{figure}
 	\centering
 	\includegraphics[width=0.42\linewidth]{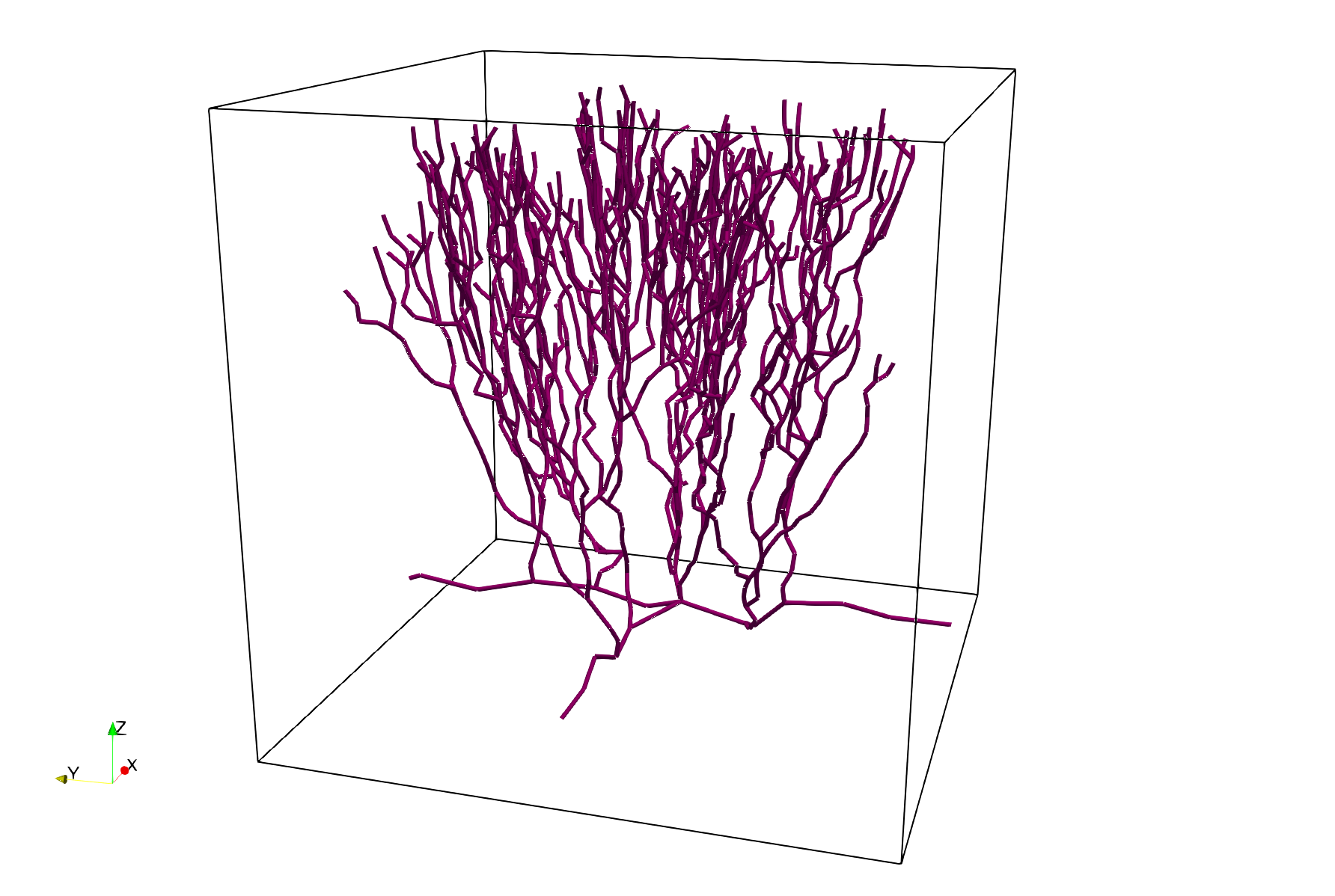}\hspace{-2.1cm}%
 	\includegraphics[width=0.42\linewidth]{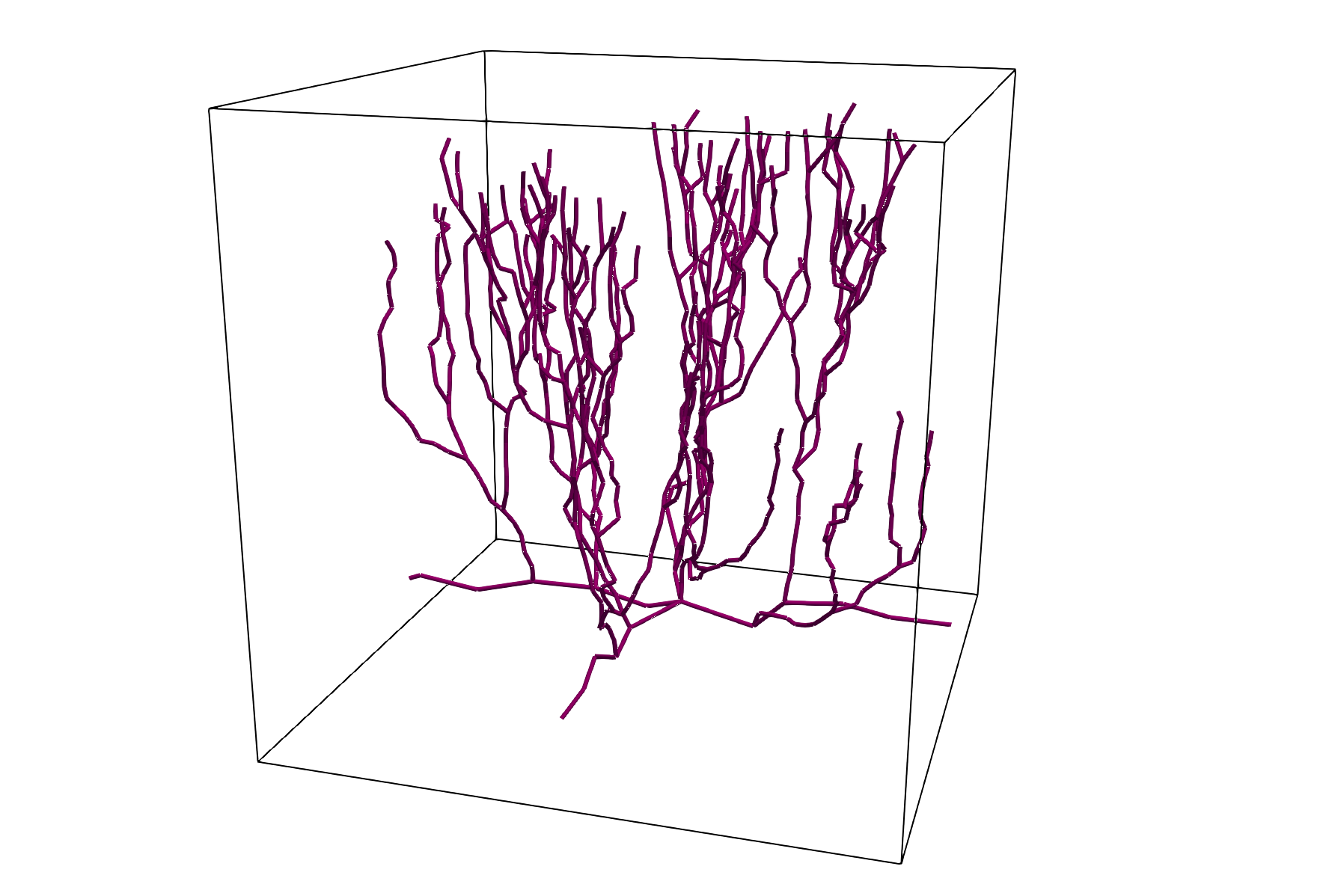}\hspace{-2.1cm}
 	\includegraphics[width=0.42\linewidth]{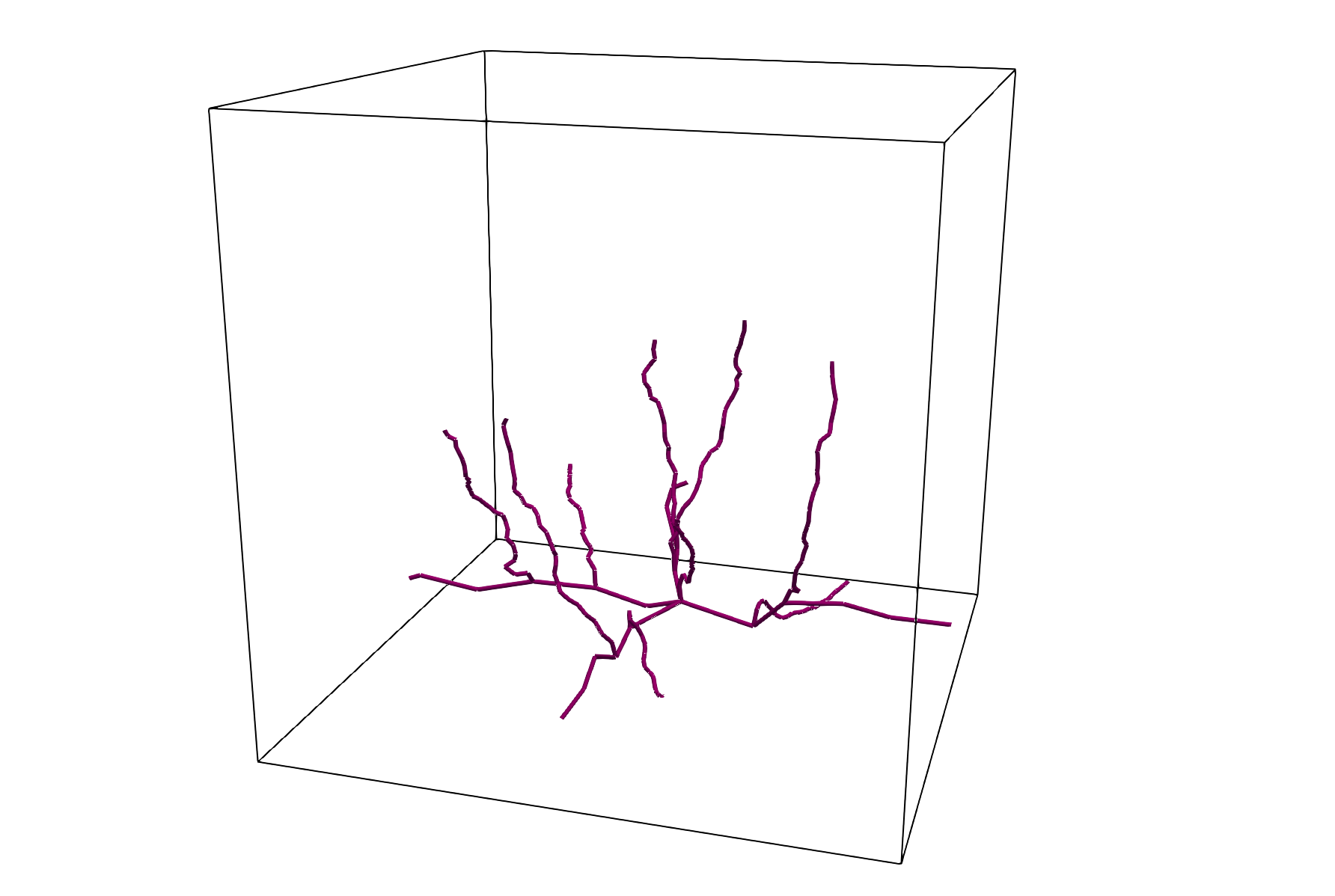}
 	\caption{\textit{TestFace}: Comparison of the vascular networks obtained after 14 days for a low, a medium and a high VEGF consumption rate. From left to right: $\tilde{\sigma}=0.4~\rm{h^{-1}}$, $\tilde{\sigma}=0.7~\rm{h^{-1}}$ and $\tilde{\sigma}=6~\rm h^{-1}$ }
 	\label{net_comparison}
 \end{figure}

In order to better analyze the impact that the rate of consumption of VEGF by the endothelial cells has on the vascular network structure and performances let us refer to Figure \ref{consum}, in which the trends in time of the maximum growth velocity, of the number of tip cells in the domain and of the percentage of hypoxic tissue are investigated for different values of the endothelial cell VEGF consumption rate, $\tilde{\sigma}$. As expected, a bigger consumption rate corresponds to lower values of $||\bm{w}||$. On the other hand, when no consumption is considered, the trend of the growth velocity with time is monotonically increasing. A higher consumption rate corresponds also to a lower number of tip cells, since branching is inhibited when VEGF concentration is low: actually for the cases $\tilde{\sigma}=3 \rm h^{-1}$ and $\tilde{\sigma}=6 \rm h^{-1}$ no branching occurs at all. The rapid final decrease in the number of tip cells in the case $\tilde{\sigma}=0 \rm h^{-1}$ is due to the fact that, with a fast growth, the network manages to reach the tumor interface within 14 days and the tips reaching the boundary are not considered in the set of tip cells anymore. For what concerns the impact of $\tilde{\sigma}$ on oxygenation we can see how, neglecting the case $\tilde{\sigma}=0 \rm h^{-1}$, the final percentage of tissue below $8~\rm mmHg$ increases as the consumption rate gets higher. Since in this case we are not abnormally forcing branching at low VEGF concentrations, an inadequate presence of angiogenic factor corresponds both to poor branching and scarce development, thus hindering oxygenation. A low consumption rate allows instead for a more branched structure, but since we are using the branching probability $P_{br}(g)$ defined in \eqref{br_prob}, the branches are not massively developing in the lowest part of the domain, thus allowing a more efficient oxygenation. The small increase registered in the hypoxic volume for $\tilde{\sigma}=0$ is related to the fact that a fast and branched development promotes oxygen exchange with the lateral boundary. The vascular networks generated for $\tilde{\sigma}=0.4 \rm h^{-1}$, $\tilde{\sigma}=0.7 \rm h^{-1}$ and $\tilde{\sigma}=6 \rm h^{-1}$ are reported in Figure \ref{net_comparison}.

\subsection{TestSphere}\label{TestSphere}
\begin{figure}
	\centering
	\includegraphics[width=0.51\linewidth]{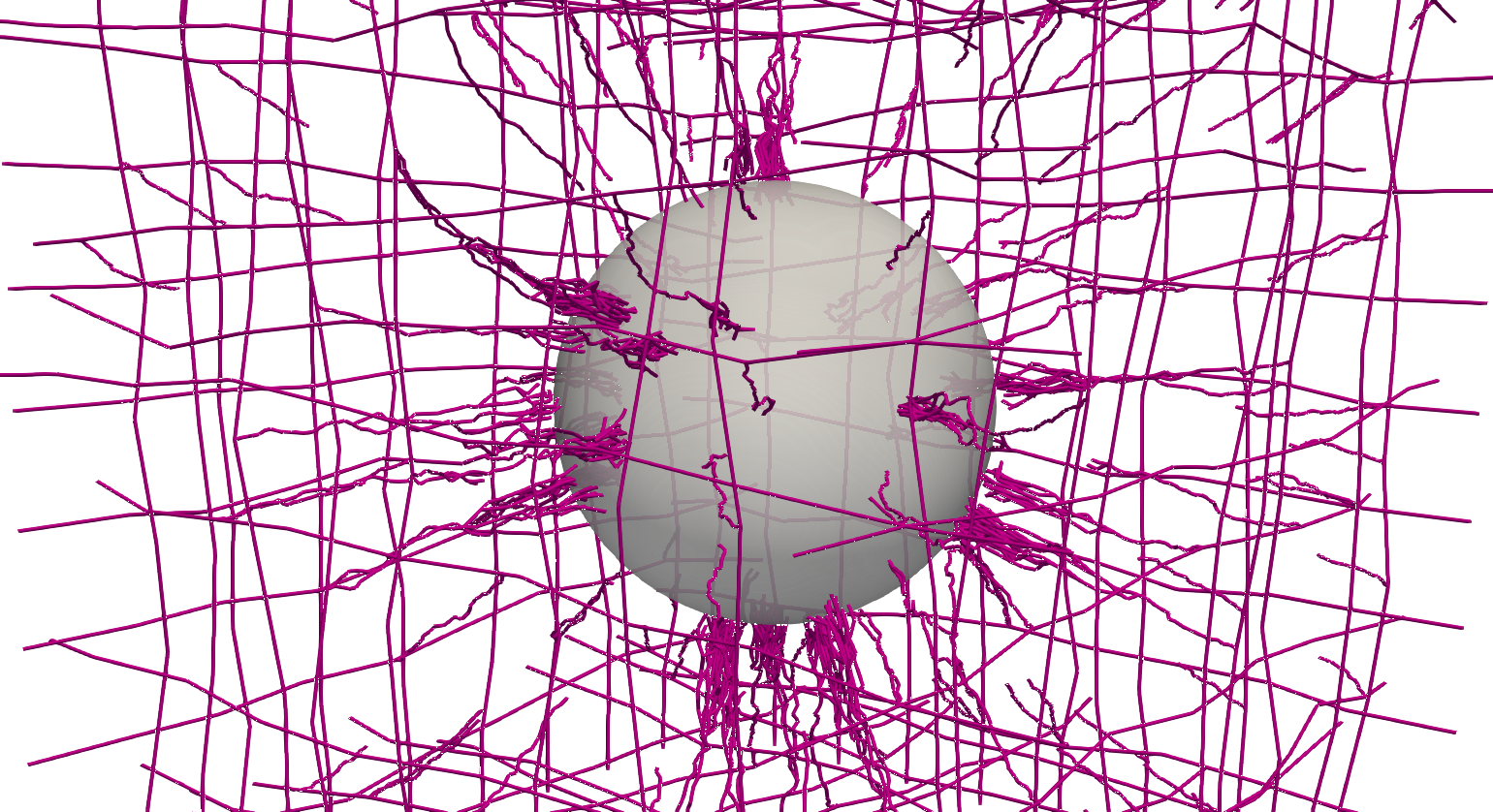}\hspace{-0.2cm}%
	\includegraphics[width=0.5\linewidth]{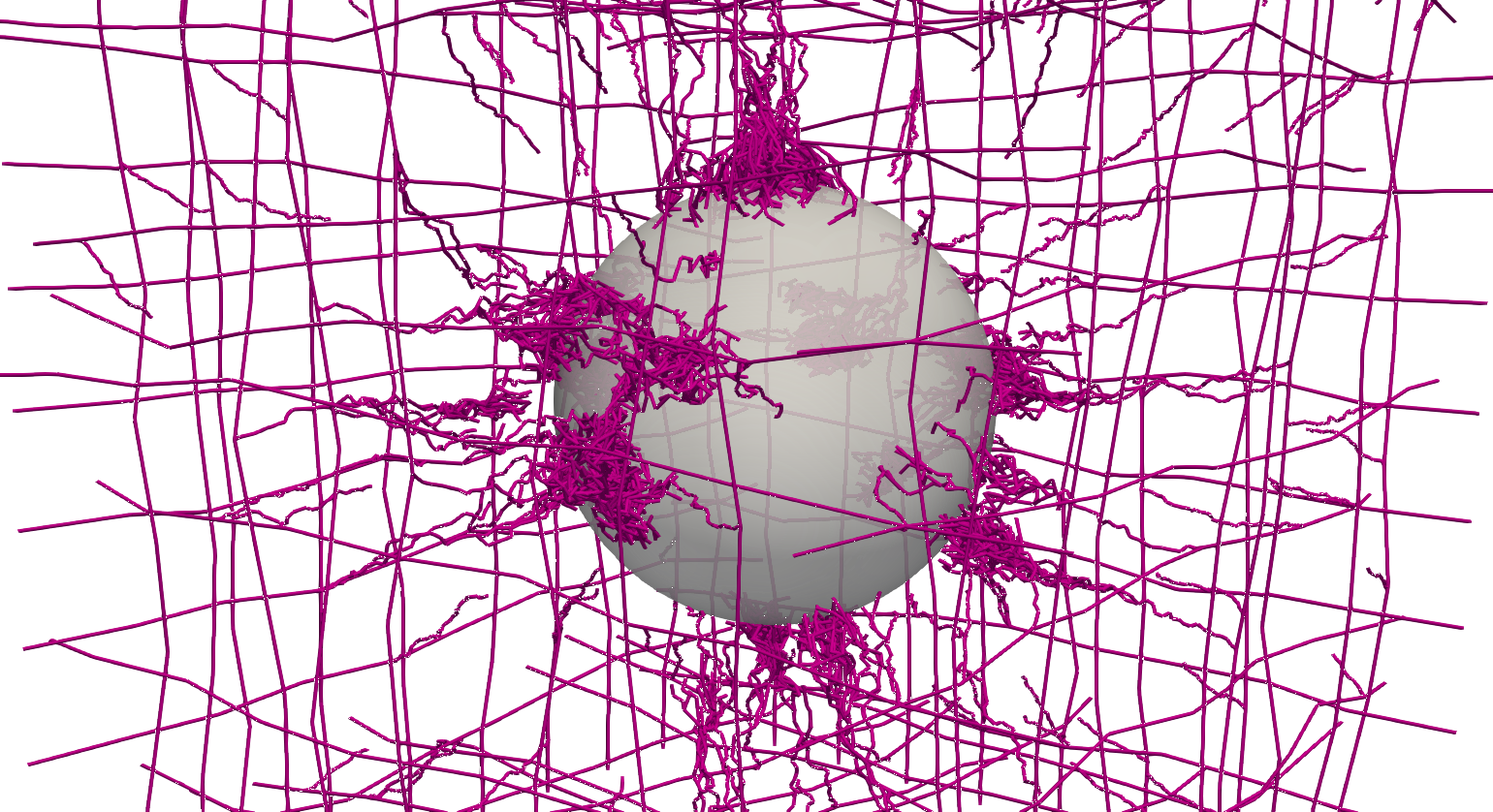}
	\caption{\textit{TestSphere:} Detail on the network configuration at time $t=40$ days. On the left $\bm{K}_{ECM}=\bm{K}_{ECM}^{rand}$ (see \eqref{kecm}), on the right $\bm{K}_{ECM}=\bm{K}_{ECM}^{\perp}$ (see \eqref{Kperp}).}
	\label{tsph_net}
\end{figure}

 As stated in the Introduction, tumor can grow in the avascular phase until a critical size of about $1-2 \, \rm mm$ is reached \cite{HillenFbio, Folkman_1987}. Above this size, the existing vasculature can no longer sustain cancer growth and the tumor stimulates new vessel formation across distances of some millimeters ($1-3 \, \rm mm$ in \cite{angio_velocity, Cavallo}). Hence, let us consider a cube of edge $L=2.5~\rm mm$ and a sphere $\mathcal{C}$ of radius $R_{\mathcal{C}}=0.5~\rm mm$ centered in the middle of the cube, as reported in Figure \ref{initialconfig} on the right. The sphere represents the tumor, while the computational domain $\Omega$ for this numerical example is given by the portion of the cube lying outside the sphere. For the pressure and oxygen concentration problems the inlet extrema are chosen as the ones lying on the planes $x=0$, $y=0$ or $z=0$, while the outlets lie on $x=2.5$, $y=2.5$ or $z=2.5$. All the parameters, except the domain edge length $L$ and the VEGF diffusivity $D_g$, have the values reported in Tables \ref{table_geom}, \ref{table_pr}, \ref{table_oxy} and \ref{table_VEGF}. The value of $D_g$ was slightly increased (but still remaining in the range proposed by \cite{Miura, Serini}) in order to have a sufficient concentration for EC proliferation also far from the center of the faces of the cube. In particular $D_g=0.36~\rm mm^2/h$ was used. For this numerical example we also consider a peculiar anistropic structure of the extracellular matrix. Specifically,
we suppose that the growth of the tumor produced a modification in the orientation of the surrounding extracellular matrix fibers, leading to concentric spherical layers around the cancer mass. 
Denoting by $\bm{x}_{\mathcal{C}}$ the center of the tumor (corresponding in this case with the center of the domain) and by $\bm{e}_r(\bm{x})$ the unit vector in the direction $\bm{x}_{\mathcal{C}}-\bm{x}$ we define
\begin{equation}
\bm{K}_{ECM}^{\perp}=\big(\bm{I}+(\varepsilon(\bm{x})-1)\bm{e}_r\otimes \bm{e}_r \big)\bm{K}_{ECM}^{rand}\label{Kperp}
\end{equation} with $\varepsilon(\bm{x})=\frac{2||\bm{x}_{\mathcal{C}}-\bm{x}||}{L\sqrt{3}}$ and $\bm{K}_{ECM}^{rand}$ defined as in \eqref{kecm}. By setting $\bm{K}_{ECM}=\bm{K}_{ECM}^{\perp}$ in \eqref{growth_vel},  we are still accounting for a random perturbation of the ECM fiber direction, but we are also enhancing the circumferential direction as the ECM approaches the tumour boundary. Figure \ref{tsph_net} shows the vascular network generated after $40$ days, on the left when a $\bm{K}_{ECM}^{rand}$ is considered (\emph{spatially random anisotropicity}), and on the right when the displacement in the radial direction is penalized, exploiting $\bm{K}_{ECM}^\perp$ (\emph{circumferential anisotropicity}). For the discretization a tetrahedral mesh with maximum element volume of $2\cdot 10^{-3} ~\rm mm^3$ is considered, while a time step $\Delta t=24~\rm h$ is used to reach a final time of $40$ days. The position of the initial tip cells is chosen randomly on the initial network. In particular in this case we consider 165 initial tips. 
As expected, sprouts that generate closer to the tumor region, i.e. next to the center of the cube faces, grow more rapidly and branch at a higher rate, since a higher concentration of VEGF is available. Conversely, tip cells located far from the tumour region do not either sprout or progress.
 From the morphology reported in Figure \ref{tsph_net} it is evident the effect of the imposed directional anisotropy: when the anisotropy is spatially random, the new vessels only slightly deviate from the radial direction, on the other hand when the circumferential anisotropicity is imposed the endothelial cell tends to follow the ECM fibers, while moving toward the source of VEGF. Therefore, in this latter case, the vascular network develops also in the direction transversal to the radial one, ending up in wider gatherings on the tumor surface.
\begin{figure}
	\centering
	\includegraphics[width=0.48\linewidth]{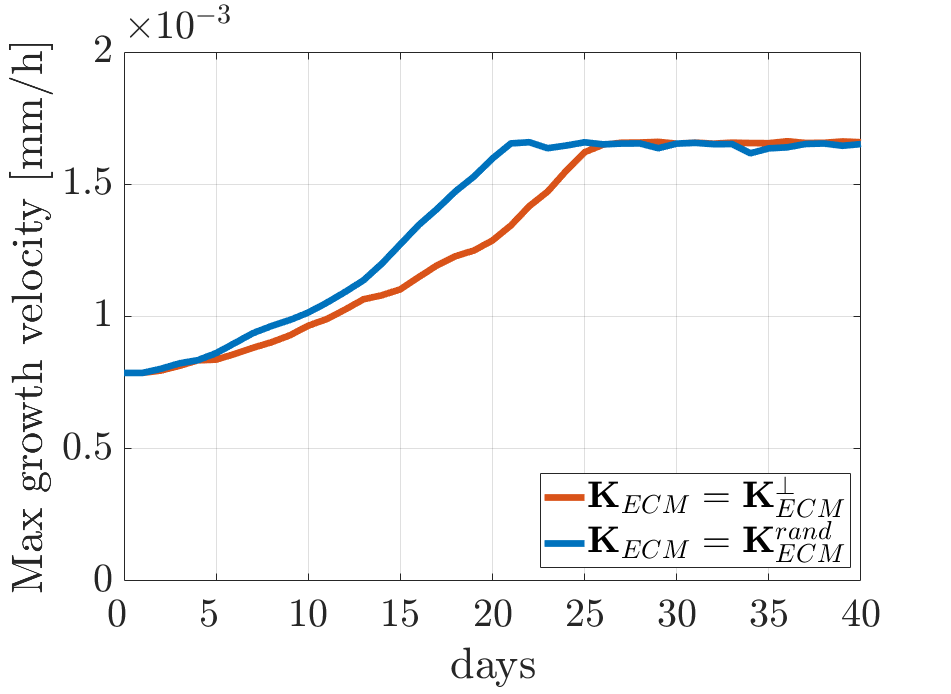}%
	\caption{\textit{TestSphere}: comparison of the maximum growth velocity in the circumferential anistropicity case ($\bm{K}_{ECM}=\bm{K}_{ECM}^{\perp}$) and in the spatially random anisotropicity case ($\bm{K}_{ECM}=\bm{K}_{ECM}^{rand}$).}
	\label{fig:testSphere_comp_vel}
\end{figure}
\begin{figure}
	\centering
	\includegraphics[width=0.48\linewidth]{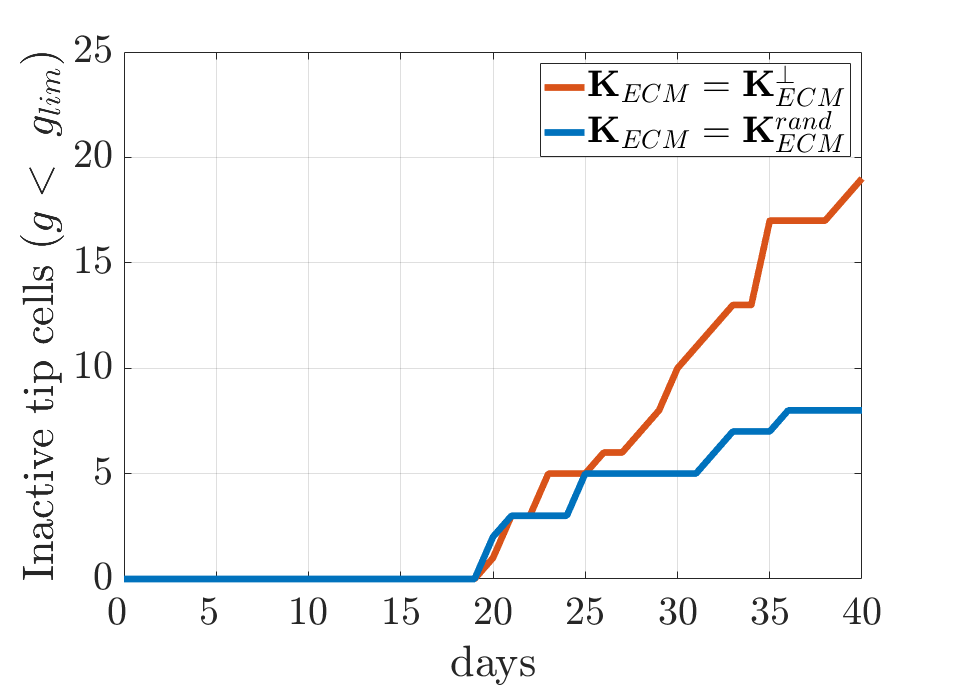}%
	\includegraphics[width=0.48\linewidth]{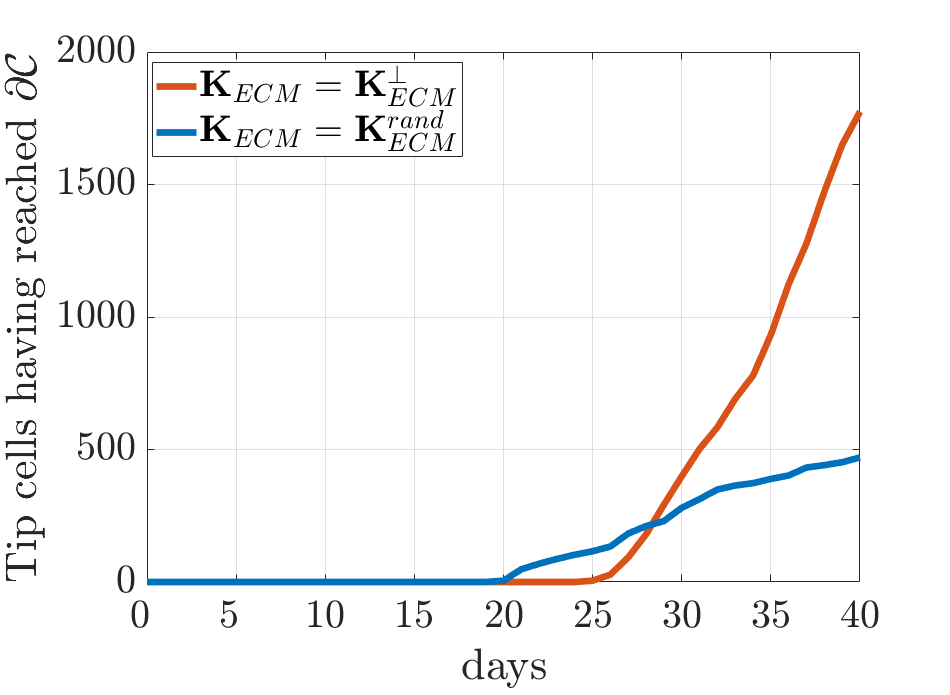}
	\caption{\textit{TestSphere}: On the left, trend with time of the the number of inactive tip cells (see text for the definition), on the right trend with time of the cumulative sum of tip cells having reached the boundary.  Comparisons between the circumferential anistropicity case ($\bm{K}_{ECM}=\bm{K}_{ECM}^{\perp}$) and the spatially random anisotropicity case ($\bm{K}_{ECM}=\bm{K}_{ECM}^{rand}$).}
	\label{fig:testSphere_comp}
\end{figure}

To quantitatively compare the two cases, we consider the growth velocity of the network and some data on the number of tip cells. Figure \ref{fig:testSphere_comp_vel} reports the trend with time of the maximum growth velocity for the spatially random and the circumferential anisotropicity cases. We can observe how, since the tumour surface is reached in both ECM scenarios, the same maximum growth velocity is registered at the end of the simulation. We recall, indeed, that a constant Dirichlet boundary condition is imposed for the VEGF at the tumour interface. The velocity increase with time is however faster in the spatially random anisotropic case, i.e., when $\bm{K}_{ECM}^{rand}$ is used. This is related to the different structure of the network: with the same total distance to cover from the starting tip cells to the tumor surface, the capillaries are actually longer and more branched in the circumferential anisotropic case, since radial displacement is inhibited. This ends up in a stronger VEGF consumption, thus leading to a lower growth velocity. 

Figure \ref{fig:testSphere_comp} compares instead, for the two ECM configurations, the trends with time of the number of inactive tip cells and of the total number of tip cells having reached the tumour boundary. We define an inactive tip cell at time $t^*$ as a tip $\bm{x}_P$ such that $g(\bm{x}_P,t^*)<2.5 \cdot 10^{-14} \rm kg/mm^3$, i.e. a tip cell in a position where the concentration of VEGF at time $t^*$ is lower than the minimum required for proliferation. As it can be observed in Figure \ref{fig:testSphere_comp}-left, the number of tips suffering of a too low VEGF concentration is higher in the $\bm{K}_{ECM}^{\perp}$ case, coherently with the previous considerations on VEGF consumption. For what concerns the tip cells having reached the tumor boundary, their number is of course higher in the $\bm{K}_{ECM}^\perp$ case, since the network is much more branched (3189 branching events were observed, versus 944 in the $\bm{K}_{ECM}^{rand}$ case). In the random anisotropicity case the first tip cells reach $\partial \mathcal{C}$ after around $t=20$ days, while it takes 25 days in the circumferential anisotropicity case to reach the tumour. This is coherent with the time at which the maximum growth velocity becomes constant (Figure \ref{fig:testSphere_comp_vel}). Let us finish the comparison by mentioning that the maximum number of tip cells reached inside the domain is 729 in the $\bm{K}_{ECM}^\perp$ case and 270 in the $\bm{K}_{ECM}^{rand}$ one. This number does not account for the tips that leave the domain through the tumour boundary.

\begin{figure}
	\centering
	\includegraphics[width=0.52\linewidth]{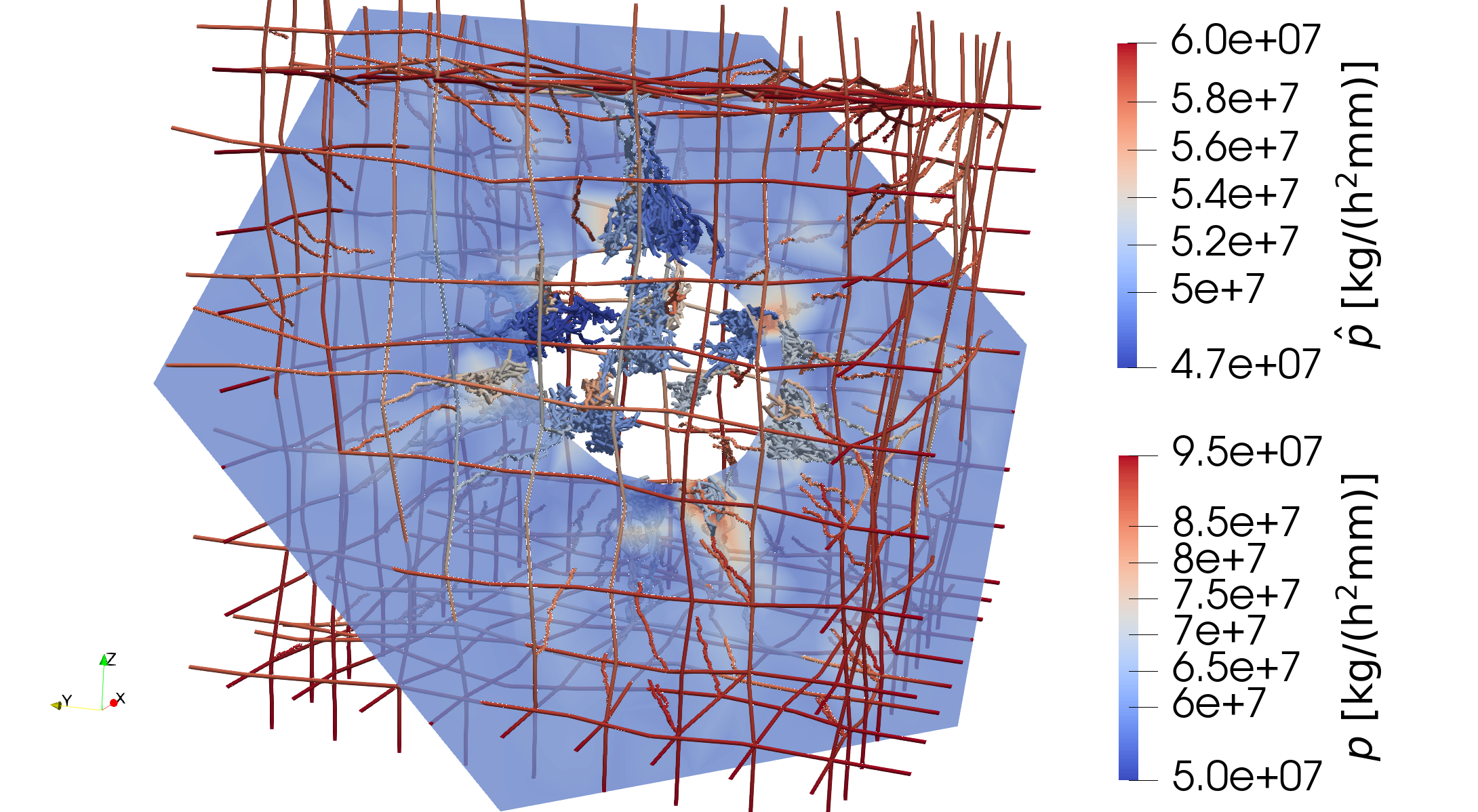}\hspace{-1cm}%
		\includegraphics[width=0.52\linewidth]{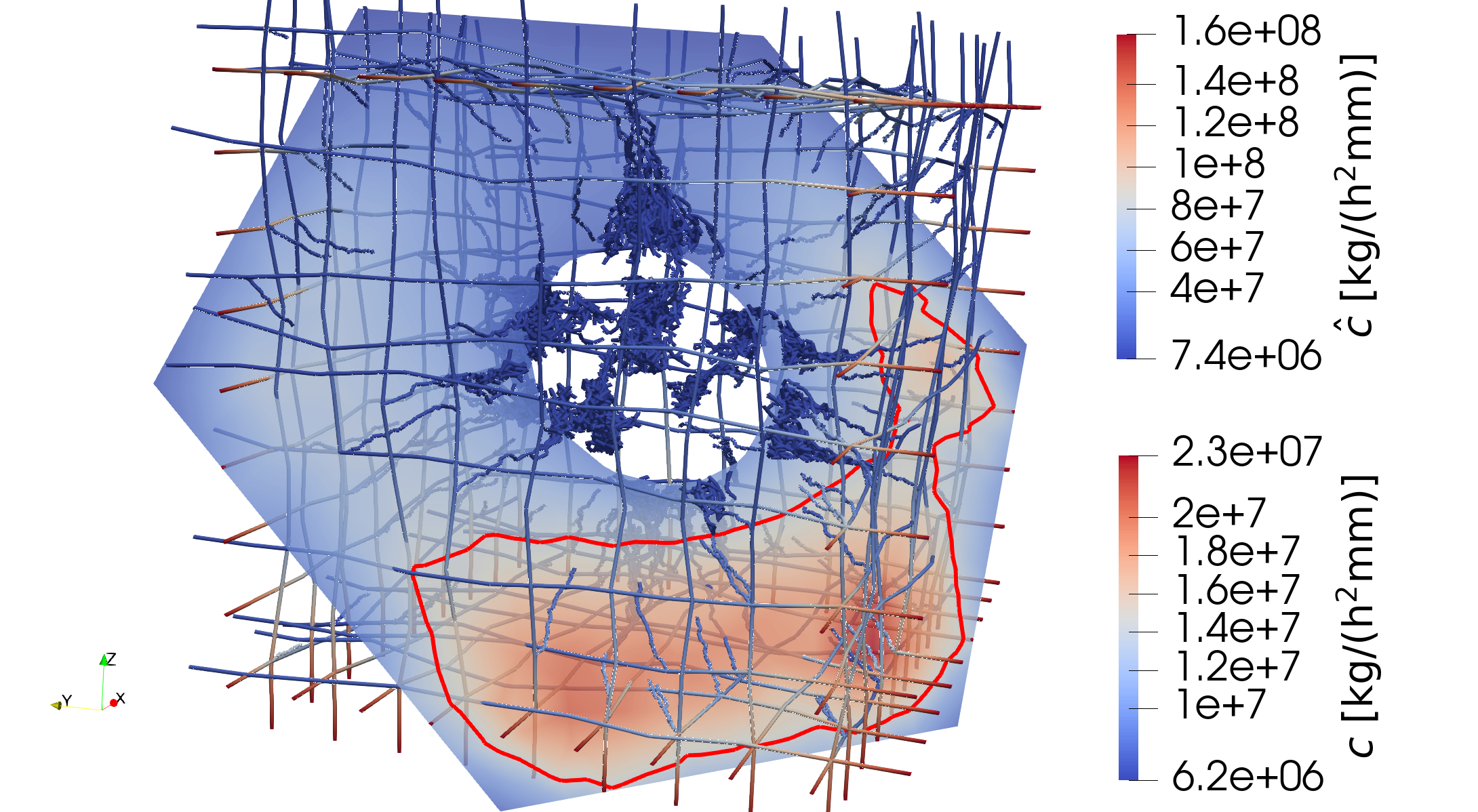}
	\caption{\textit{TestSphere:} pressure distribution (on the left) and oxygen concentration (on the right) for $\bm{K}_{ECM}=\bm{K}_{ECM}^\perp$ and $t=40$ days. Isoline corresponding to $c=8~\rm mmHg$ highlighted in red in the right figure.}
	\label{tsph_oxy}
\end{figure}

\begin{figure}
	\centering
	\includegraphics[width=0.52\linewidth]{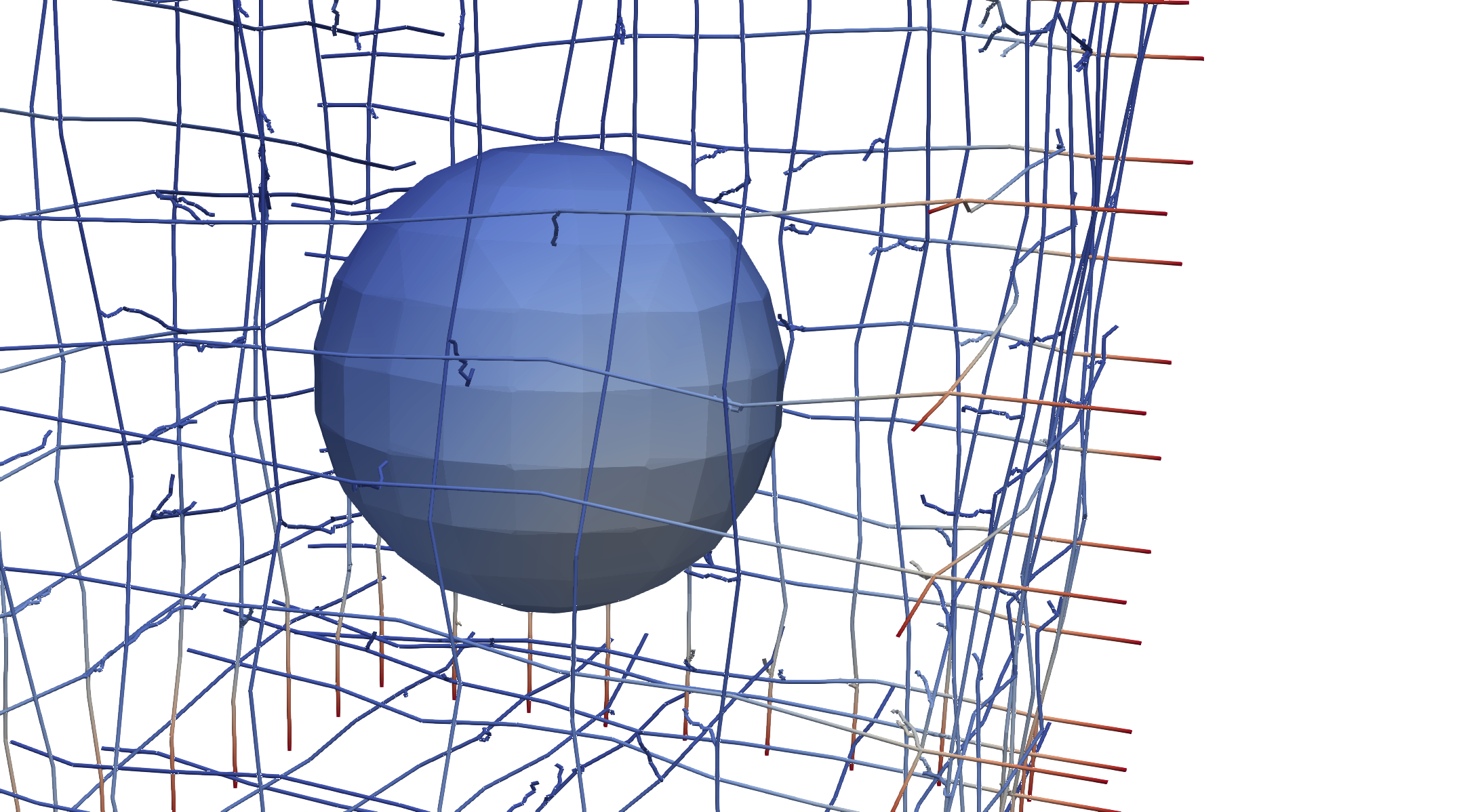}\hspace{-1cm}%
	\includegraphics[width=0.52\linewidth]{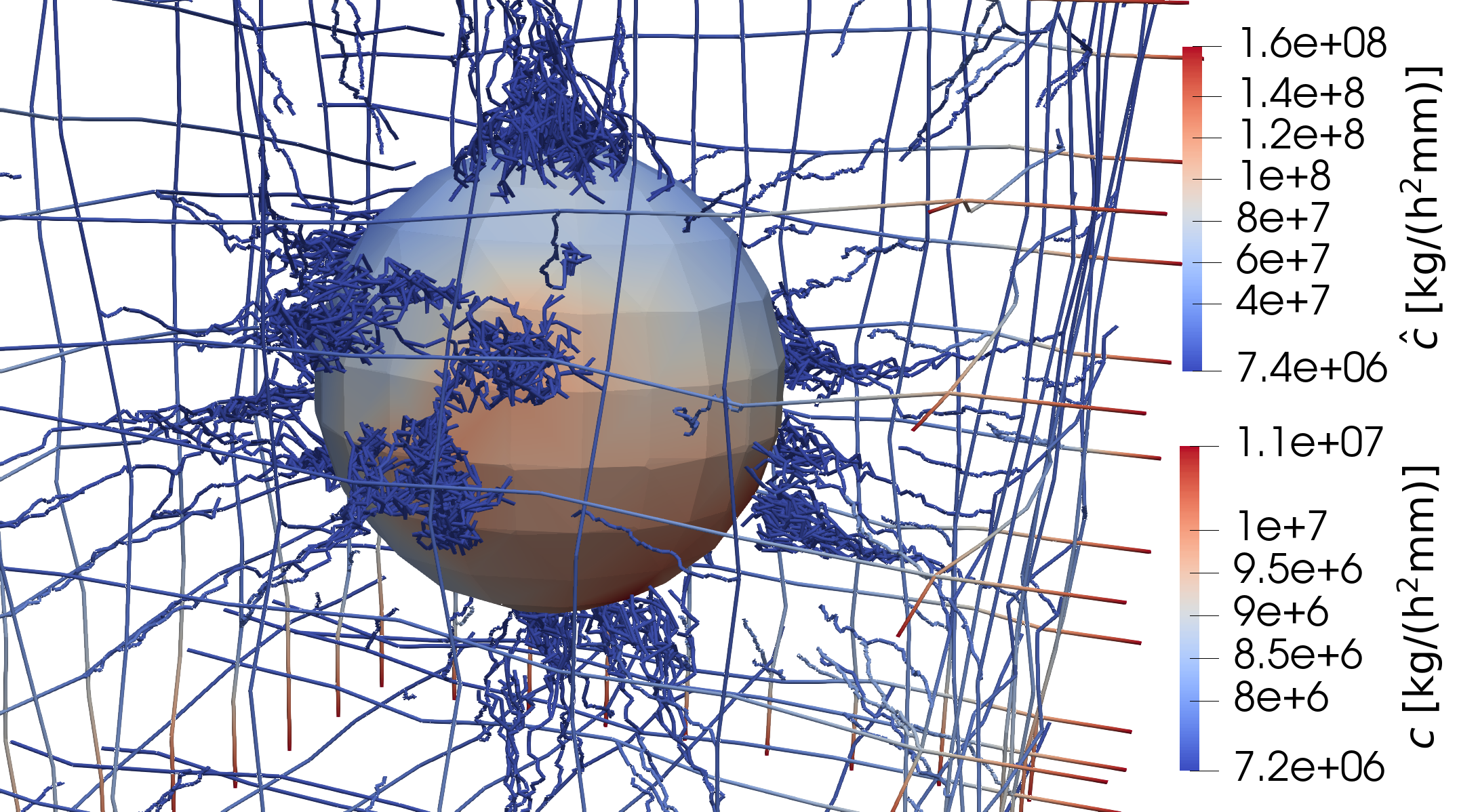}
	\caption{\textit{TestSphere:} oxygen concentration on the tumour surface for $\bm{K}_{ECM}=\bm{K}_{ECM}^\perp$. On the left $t=10$ days, on the right $t=40$ days. }
	\label{fig:tsph_oxyball}
\end{figure}

Referring to the circumferential anisotropic case, Table \ref{table_sphere} reports the maximum growth velocity and the percentage of tissue below the pathological hypoxia level at different instants of time (namely $t=1,10, 20\text{ and } 40$ days). We observe that, even though the hypoxic region reduces with time thanks to the formation of new vessels, the imposed boundary conditions and the exchange with the surrounding tissue, highly affects the supply of oxygen in the domain. Figure \ref{tsph_oxy} reports the pressure distribution and the oxygen concentration in the capillary network and on a tissue slice at $t=40$ days. Figure \ref{fig:tsph_oxyball} focuses instead on the oxygen concentration at the tumour boundary, for $t=10$ days and $t=40$ days , left and right respectively.

\begin{table}
\centering 
\caption{\textit{TestSphere}: maximum sprout growth velocity and percentage of hypoxic tissue for different time instants. $\bm{K}_{ECM}=\bm{K}_{ECM}^{\perp}$ (see \eqref{Kperp}).}
\label{table_sphere}
\begin{tabular}{c|cc}
	\hline
	\textbf{Time} &\textbf{Max Growth Velocity} &	\multirow{2}{*}{\textbf{\% Vol < 8 mmHg $\bm{ O_2}$}} \\(days)& (mm/h) &\\
	\hline
	$1$ & $7.9\cdot 10^{-4}$ &92\%  \\
	$10$ & $9.7\cdot 10^{-4}$ &87\%\\		
	$20$ & $1.3\cdot 10^{-3}$  &83\% \\		
	$40 $ & $1.7\cdot 10^{-3}$ &80\% \\		
\end{tabular}
\end{table}

\section{Conclusions}\label{sec:conclusion}
In the present work, the application of an optimization based 3D-1D coupling to the modeling of tissue perfusion and oxygen delivery during tumor-induced angiogenesis is presented. A hybrid model is adopted: fluid and chemicals distributions are represented in a continuous manner, while a discrete tip-tracking model is chosen for the vascular network growth, together with some proper rules for branching and anastomosis. The evolution in time and space of three quantities is analyzed, namely interstitial fluid/blood pressure, oxygen concentration and VEGF concentration, the first two being the actual 3D-1D coupled problems. The optimization based solving strategy fits particularly well this kind of application since it has no mesh conformity requirements: this means that, although the vascular network is growing, we never need to re-mesh our geometry according to the new capillary configuration. The method is tested on two different numerical examples, providing some parameter sensitivity analysis both for what concerns the geometry of the vascular network and the efficiency of oxygenation. The effect of the orientation of the extracellular matrix is also investigated. 
The modeling of the expansion of the tumor and of the re-modulation of the VEGF production in response to oxygen and nutrients absorption along with the possible inclusion of drug delivery is left to a forthcoming work. 

\section*{Acknowledgements}
This work is supported by the MIUR project ``Dipartimenti di Eccellenza 2018-2022'' (CUP E11G18000350001).

\bibliographystyle{elsarticle-num}
\bibliography{3D1DAngio}

\end{document}